\documentclass[12pt]{article} 
\usepackage{fullpage}
\usepackage{array,epsfig,fancyhdr,rotating,graphicx}
\usepackage[]{hyperref}  
\usepackage{sectsty, secdot}
\sectionfont{\fontsize{12}{14pt plus.8pt minus .6pt}\selectfont}
\subsectionfont{\fontsize{12}{14pt plus.8pt minus .6pt}\selectfont}

\usepackage{amsmath}
\usepackage{graphicx,psfrag,epsf}
\usepackage{enumerate,multirow}
\usepackage{bbm}
\usepackage{amsmath, amssymb,theorem,verbatim}
\usepackage[round]{natbib} 
\usepackage{setspace}
\usepackage{hyperref}
\usepackage{caption}
\usepackage{textcomp}

\usepackage{algorithm}
\usepackage{algpseudocode}
\usepackage{rotating} 
\usepackage{pdflscape}
\usepackage{tikz}  
\usepackage[caption=false]{subfig}
\usepackage{booktabs} 
\usepackage{siunitx}
\usepackage{enumitem}
\usepackage{url}

\DeclareMathOperator{\E}{E}

\DeclareMathOperator{\argmax}{argmax}
\DeclareMathOperator{\argmin}{argmin}

\newcommand{\norm}[1]{\left\|#1\right\|}

\newcommand{\Pb}[1]{\ensuremath{\mathbb{P}\left(#1\right)}}
\newcommand{\cont}[4]{\mathcal{C}_{#1,#2}^{#3}\left(#4\right)}

\newcommand{\ip}[2]{\left\langle #1, #2 \right\rangle}
\newcommand{\Exp}[1]{\exp\left(#1\right)}

\newcommand{\Z}{\mathbb{Z}}
\newcommand{\N}{\mathbb{N}}
\newcommand{\R}{\mathbb{R}}
\newcommand{\C}{\mathbb{C}}

\newcommand{\T}{\mathbb{T}}

\newcommand{\Yb}{\mathbf{Y}}
\newcommand{\Bb}{\mathbf{B}}
\newcommand{\Mb}{\mathbf{M}}
\newcommand{\Ib}{\mathbf{I}}
\newcommand{\Ab}{\mathbf{A}}
\newcommand{\Cb}{\mathbf{C}}
\newcommand{\Db}{\mathbf{D}}

\newcommand{\vb}{\mathbf{v}}
\newcommand{\ub}{\mathbf{u}}

\newcommand{\psib}{\boldsymbol{\psi}}

\newcommand{\betab}{\boldsymbol\beta}

\newcommand{\betahb}{\hat{\boldsymbol\beta}}

\newcommand{\Mc}{\mathcal{M}}
\newcommand{\Nc}{\mathcal{N}}

\newcommand{\Sc}{\mathcal{S}}
\newcommand{\Oc}{\mathcal{O}}
\newcommand{\Ic}{\mathcal{I}}
\newcommand{\Ec}{\mathcal{E}}

\newtheorem{Theorem}{Theorem}[section] 
\newtheorem{Lemma}{Lemma}[section] 

\theoremstyle{plain}

\newtheorem{prop}{Proposition}[section]

\newcommand{\cl}{\underline{c}}
\newcommand{\all}{\underline{\alpha}}
\newcommand{\bl}{\underline{b}}
\newcommand{\bu}{\overline{b}}

\def \tables_path{}
\def \tikz_path{}
\newif\ifplot
\plottrue 

\newcommand{\pkg}[1]{{\normalfont\fontseries{b}\selectfont #1}}
\let\proglang=\textsf

\newcommand{\mzar}[1]{AMAR($#1$)}
\newcommand{\ar}[1]{AR($#1$)}

\long\def\comment#1{}

\def \tables_path{Tab}
\def \tikz_path{Fig}

\usepackage{scrextend}

\begin{document}


\title{\bf Multiscale Autoregression on \\ Adaptively Detected Timescales}
\author{Rafal Baranowski \and Yining Chen \and  Piotr Fryzlewicz\\ \\
	London School of Economics and Political Science}
\date{}
\maketitle


\begin{abstract}
We propose a multiscale approach to  time series autoregression, in which linear regressors for the process in question include features
of its own path that live on multiple timescales. We take these
multiscale features to be the recent averages of the process over multiple timescales,
whose number or spans are not
known to the analyst and are estimated from the data via a change-point detection technique.
The resulting construction, termed Adaptive Multiscale AutoRegression (AMAR) enables adaptive regularisation
of linear autoregressions of large orders. The AMAR model is designed to offer simplicity and interpretability
on the one hand, and modelling flexibility on the other. Our theory permits the longest timescale to increase with the sample size. A simulation study is presented to show the usefulness of our approach. Some possible extensions are also discussed, including the Adaptive Multiscale Vector AutoRegressive model (AMVAR) for multivariate time series, which demonstrates promising performance in the 
data example on UK and US unemployment rates.  
The \proglang{R} package 
\pkg{amar} \citep{baranowski2016mzar} 
provides an efficient implementation of the AMAR framework.
\end{abstract}

\vspace{9pt}
\par

\def\thefigure{\arabic{figure}}
\def\thetable{\arabic{table}}

\section{Introduction}
\label{mzar:sec:introduction}
\subsection{Motivation and main idea}
Autoregression in time series modelling is arguably the most frequently used device to characterise temporal dependence in data.
The classical linear autoregressive model of order $p$, known as AR($p$), for univariate time series $X_t$ assumes
that $X_t$ is a linear but otherwise unconstrained function of its own past values $X_{t-1}, \ldots, X_{t-p}$,
plus white-noise-like innovation $\varepsilon_t$, that is
$X_t = \beta_1 X_{t-1} + \ldots + \beta_p X_{t-p} + \varepsilon_t$ for $ t = 1, \ldots, T$.
However, in situations where the application of this model yields a large or even moderate $p$,
either in absolute terms or relative to $T$ (perhaps in an attempt to reflect long-range dependence in
$X_t$), it may be tempting to consider instead an alternative approach, in which
$X_t$ is regressed explicitly on some other features of its own past, rather than directly on the individual
variables $X_{t-1}, \ldots, X_{t-p}$.

Motivated by this, we propose what we call a multiscale approach to
time series autoregression, in which we include features of the path
$X_1, \ldots, X_{t-1}$ that live on multiple timescales as linear regressors for $X_t$. To fix ideas, here we take these
multiscale features to be the recent averages of $X_t$ over multiple time spans (N.B. possible extensions will be discussed in Section~\ref{sec:disc}), which are not
necessarily known to the analyst a priori and need to be estimated from the data.  This leads to the following
\textbf{A}daptive \textbf{M}ultiscale \textbf{A}uto\textbf{R}egressive model of order $q$, abbreviated as AMAR($q$),  for $X_t$:
\begin{align}  
	\label{mzar:eq:mzar_model}
X_{t}=\alpha_{1}\frac{X_{t-1}+\ldots+X_{t-\tau_{1}}}{\tau_{1}}+\ldots+\alpha_{q}\frac{X_{t-1}+\ldots+X_{t-\tau_{q}}}{\tau_{q}}+\varepsilon_{t},\quad t=1,\ldots, T,
\end{align}
where the number of timescales $q$, the timescales $1\leq\tau_{1}<\tau_{2}<\ldots<\tau_{q}$ and the scale coefficients $\alpha_{1}, \ldots, \alpha_{q} \in \R \backslash \{0\}$ are unknown (NB. zero is excluded for model identifiability),  and where the innovations $\{\varepsilon_{t}\}$ follow a white noise process, which we take to mean a sequence of random variables that are uncorrelated, having zero-mean and a finite (but non-zero) variance. The number of scales $q$ can possibly be much smaller than the largest timescale $\tau_{q}$. 
Here we use the term ``adaptive" to reflect the fact that the timescales in the AMAR model automatically
adapt to the data in the sense of being selected in a data-driven way, rather than being known a priori.

In essence, the AMAR($q$) model is a  multiscale, sparsely parameterised, version of the AR($\tau_q$) process, which permits the longest timescale $\tau_q$ to be large in practice. 
These properties make the AMAR framework particularly suitable for the exploratory analysis of processes in which a seasonal component may be suspected, or for the modelling of time series which exhibit low-frequency trends (which may give them a non-stationary appearance) accompanied by higher-frequency oscillations. We shall illustrate these claims in Section~\ref{sec:specialcases} of the supplementary materials.

\subsection{Literature review}
\label{Sec:lr}
We now provide an overview of other related literature. 
\cite{reinsel1983some} considers a model in which the current time series variable depends linearly on a small number of index variables which are linear combinations of its
own past values; in contrast to our setting, these index variables are assumed to be known a priori. Reduced-rank time series multivariate autoregression, which provides a way of reducing the parameterisation for multivariate time series via the use of automatically chosen index variables, is considered in
\cite{vrw86} and \cite{ahn1988nested}, but this approach is not explicitly designed to be multiscale or to be 
 able to cope with autoregressions of large orders.

\cite{ferreira2006multi} introduce a class of bi-scale univariate time series models that consist of two main building blocks: $Y_{t}$, $t=1,\ldots, mT $, the fine-level process, where the integer $m>1$ is known, and the coarse-level aggregate process $X_{t} = m^{-1} \sum_{j=1}^{m} Y_{t m -j} + \varepsilon_{t}$ for $t=1, \ldots, T$, 
where the noise term $\varepsilon_{t} \sim \mathcal{N}(0,\sigma^{2})$ is independently and identically distributed (i.i.d.) and independent of $Y_t$.
 \cite{ferreira2006multi} recommend choosing a simple model for $Y_t$, e.g. AR(1), and show with this choice, $X_t$ can emulate long-memory behaviour.
In contrast to this framework, AMAR assumes that the timescales are not known a priori, and uses coarse-level information for fine-level modelling, rather than vice versa.
 
\cite{ghysels2004midas} propose MIxed DAta Sampling (MIDAS) regression, in which 
time series observed at finer scales are used to model one observed at a lower frequency. In the notation of the previous paragraph, the MIDAS model is defined as
$X_{t} =  \beta_{0} + \sum_{i=1}^{p} b_{i}(Y_{tm -i}; \betab) +\varepsilon_{t}$,
where $b_{1}(\cdot; \betab), \ldots, b_{p}(\cdot;\betab)$ are given functions of the lagged observations recorded at a higher frequency and of a low-dimensional vector of unknown parameters $\betab=(\beta_{1},\ldots,\beta_{q})^T$, and where $\varepsilon_{t}$ is random noise. For each recorded observation of $X_{t}$, $m$ values of $Y_{i}$ are sampled.
We mention one particular form of $b_{i}(\cdot; \betab)$ from \cite{forsberg2007absolute}:
$X_{t} = \beta_{0} + \sum_{j=1}^{q} \beta_{j} \sum_{i=1}^{\tau_{j}} Y_{tm -i} +  \varepsilon_{t}$,
where $1\leq \tau_{1} < \ldots < \tau_{q}$ are known integers. 
One important difference between this and the AMAR framework is that $\tau_{1}, \ldots, \tau_{q}$ in our model are unknown.

In Heterogeneous AutoRegressive (HAR) modelling \citep{c09}, the quantity of interest
is regressed on its past realised averages over given known multiple timescales.
The author shows that the model is able to imitate long-memory behaviour without, in fact, possessing the long-memory property. Numerous extensions and applications of the HAR approach have been considered; see 
\cite{car12} for a review of HAR modelling of realised volatility.

 \cite{mf19} introduce bi-scale autoregression, in which the more remote autoregressive coefficients are assumed to be sampled from a smooth function; this is done to regularise the estimation problem and thus facilitate estimation of the coefficients if the autoregression order is large. The rough and smooth regions of the AR coefficient space are identified through a technique akin to change-point detection. The approach is different from AMAR
in that only two scales are present (while in AMAR the number of scales is unknown a priori and is chosen adaptively from the data), and the scales are defined by the degree of coefficient smoothness instead of their spans as in AMAR.

The Long Short-Term Memory (LSTM) model of the recurrent neural network \citep{hs97} uses a bi-scale modelling approach whereby the new hidden state at each time point combines (in a particular way that has been learned from the data) long-range ``cell state'' information with more recent information originating from the previous hidden state and instantaneous input.
The use of LSTM models in time series forecasting is less well explored and the theoretical understanding of their behaviour in the context of time series modelling is limited, but see \cite{p19} for a recent review. The complexity of LSTM models means that large samples are typically required to train them.

In addition, one could view the AMAR model as a particular instance of a linear regression model in which the coefficients have been grouped into (unknown) regions of constancy. The group LASSO
approach \citep{yl06} assumes that the groups are known and it therefore would not be suitable for AMAR. The fused LASSO
\citep{tsrzk05}, which uses a total-variation penalty on the vector of regressors, could in principle be used for the fitting of a piecewise-constant
approximation to the estimated vector of AR coefficients, but consistent detection of scales in the AMAR model is effectively a multiple change-point detection problem, and
it is known (see e.g. \cite{cf11}) that approaches based on the total variation penalty (e.g. fused LASSO) is not optimal for this task.

Finally, we note that our notion of ``multiscale autoregression'' is different from that in, for example, \cite{bbw92} or \cite{761321}, who consider statistical modelling on dyadic trees, motivated by the wavelet decomposition of data. In contrast, we are interested in the explicit multiscale modelling of the time evolution of the original process $\{X_t\}$, i.e. there is no prior multiscale transformation to speak of.

Against the background of the existing literature, the unique contributions of this work can be summarised as follows. Unlike the existing multiscale and index-based approaches to autoregression described above, the scales $\tau_1, \ldots, \tau_q$ in the AMAR model
 are not assumed to be
known by the analyst and are estimable from the data; so is their number $q$. The AMAR model is able to accommodate autoregressions of large order: the largest-scale parameter $\tau_q$ is permitted to increase with the sample size $T$ at a rate close to $T^{1/2}$. The consistent estimation of the number of scales $q$ and their spans
$\tau_1, \ldots, \tau_q$ is achieved by a change-point detection algorithm, more precisely, a ``narrowest-over-threshold''-type \citep{bcf19} adapted to the AMAR context,
and this paper both justifies this choice and shows how to overcome the significant methodological and theoretical challenges that arise in this adaptation.
Being only based on the past averages of the process but enabling data-driven selection of their number and spans, the AMAR framework is designed to offer simplicity and interpretability on the one hand, and modelling flexibility on the other. Besides, the AMAR framework can be extended in different ways to handle more complicated data structure, including multivariate time series. The promising performance of this particular extension, named Adaptive Multiscale Vector AutoRegressive model (AMVAR), is also demonstrated in this paper. 
The \proglang{R} package 
\pkg{amar} \citep{baranowski2016mzar}
provides an efficient implementation of our proposal.

\section{Methodology and theory}
\label{sec:mt}

\subsection{Model framework}

Recall that \mzar{q} is an instance of a sparsely parametrised AR model. Therefore, for any $p \ge \tau_q$, \eqref{mzar:eq:mzar_model} can be rewritten as
\begin{align}
	\label{mzar:eq:ar_p}	X_{t} &= \beta_{1} X_{t-1} + \ldots + \beta_{p} X_{t-p} + \varepsilon_{t},\quad t=1,\ldots, T,\\
	\label{mzar:eq:mzar_ar_beta} \beta_{j} &= \sum_{k: \tau_{k} \geq j} \frac{\alpha_{k}}{\tau_{k}}, \quad j=1,\ldots,p,
\end{align}
where $\{\varepsilon_{t}\}$ is a white noise process.
Here we refer to \eqref{mzar:eq:ar_p} and \eqref{mzar:eq:mzar_ar_beta} as an \ar{p} representation of the \mzar{q} process. Also note that $\beta_j = 0$ for $j = \tau_q+1, \ldots, p$.

Let $\betahb=(\hat{\beta}_1, \ldots, \hat{\beta}_{p})^T$ be the Ordinary Least Squares (OLS) estimator of $\betab=(\beta_{1}, \ldots, \beta_{p})^T$. Then $\hat{\beta}_{j}$'s can be trivially decomposed as
\begin{align}
	\label{mzar:eq:signal+noise}	\hat{\beta}_{j} = \beta_{j} + (\hat{\beta}_{j}-\beta_{j}), \quad j=1,\ldots,p.
\end{align} 
The coefficients $\beta_{1}, \ldots, \beta_{p}$ form a piecewise-constant vector with change-points at the timescales $\tau_{1}, \ldots, \tau_{q}$, and
thus the hope is that the timescales can be estimated consistently using a multiple change-point detection technique.
This observation motivates the following estimation procedure for the AMAR models.
First, we choose an adequate $p$ and find the OLS estimates of the autoregressive coefficients in the \ar{p} representation of the \mzar{q} process. Then, we estimate the timescales by identifying the change-points in the series $\beta_{1}, \ldots, \beta_{p}$, using for this purpose an adaptation of the Narrowest-Over-Threshold (NOT) approach
of \cite{bcf19}. 
Once the timescales are estimated, we estimate the scale coefficients via least squares.

Our motivation for using the NOT approach as a change-point detector in this context is that it enjoys the following change-point
isolation property: in each detection step, the NOT algorithm is
guaranteed (with high probability) to be only selecting for consideration sections of the input data (i.e. the vector $(\hat{\beta}_{1}, \ldots, \hat{\beta}_{p})$ here)
that contain at most a single change-point each. This is a key fact that makes our version of the NOT method easily amenable to a theoretical analysis in the AMAR framework.
 
In a typical application of the \mzar{q} model, we envisage that the number of timescales $q$ will be small in comparison to the maximum timescale $\tau_{q}$. For the development of our theory, we work in a framework where the number of timescales $q$, the timescales $\tau_{1}, \ldots, \tau_{q}$  and the coefficients $\alpha_{1}, \ldots, \alpha_{q}$ possibly depend on the sample size $T$ under Gaussian innovations, and are fixed under heavy-tailed innovations. However, for the economy of notation, we shall suppress the dependence of these quantities on $T$ in the remainder of the paper.

We end this section by emphasising again that the purpose of change-point detection in our context is not to find change-point in the AMAR($q$) process itself; indeed,
this paper studies stationary AMAR processes, which themselves contain no change-points. The aim of change-point detection in the AMAR context is to segment the possibly long vector
of the estimated autoregressive coefficients into regions of piecewise constancy, and thereby estimate the unknown timescales $\tau_1, \ldots, \tau_q$.

\subsection{Stationarity}

Recall that the characteristic polynomial of any AR($p$) is
$
	b(z) := b(z; \beta_1,\ldots, \beta_p) = 1-\sum_{j=1}^{p}\beta_{j}z^{j}
$
 for $z\in\C$, where $\C$ denotes the complex plane.  Also, the unit circle is denoted by $\T  = \{z\in \C: |z| = 1\}$. 

We now discuss the stationarity of the AMAR models with  $q$ and $\alpha_1,\ldots,\alpha_q$ being fixed. Since AMAR is a special form of AR, $\{X_t\}$ that follows the AMAR model has a stationary (and causal) solution if and only if the roots of $b(z; \beta_1,\ldots, \beta_{\tau_q})$ are outside $\T$, where $\beta_1,\ldots, \beta_{\tau_q}$ are defined in Equation~\eqref{mzar:eq:mzar_ar_beta}. Here any $z \in  \C$ is outside $\T$ if and only if $|z| > 1$. Furthermore, a  simplified sufficient condition for stationarity is given in the following result.

\begin{prop}
\label{Prop:stationary}
Given $\{X_t\}$ follows the AMAR($q$) model in Equation~\eqref{mzar:eq:mzar_model} with $\alpha_1,\ldots,\alpha_q$. If $\sum_{j=1}^q |\alpha_j|<1$, then $\{X_t\}$ has a causal stationary solution. Suppose all the $\alpha_j$'s are non-negative, then the converse is true.

\end{prop}
We remark that  when all the $\alpha_j$'s are non-negative, previous observations have non-negative effects on the current one. In this case, $\sum_{j=1}^q \alpha_j<1$ would be a sufficient (but not necessary) condition for stationarity. 
Furthermore, the above proposition holds even when $q = \infty$. 
When $q$ (or $\tau_q$) increases with $T$, the stationarity property of AMAR would need to be discussed in a setup that involves triangular arrays. These details are omitted for notational convenience. 

\begin{figure}[!htbp]
\centering
\begin{minipage}{.25\textwidth}
  \centering
  \includegraphics[width=\linewidth]{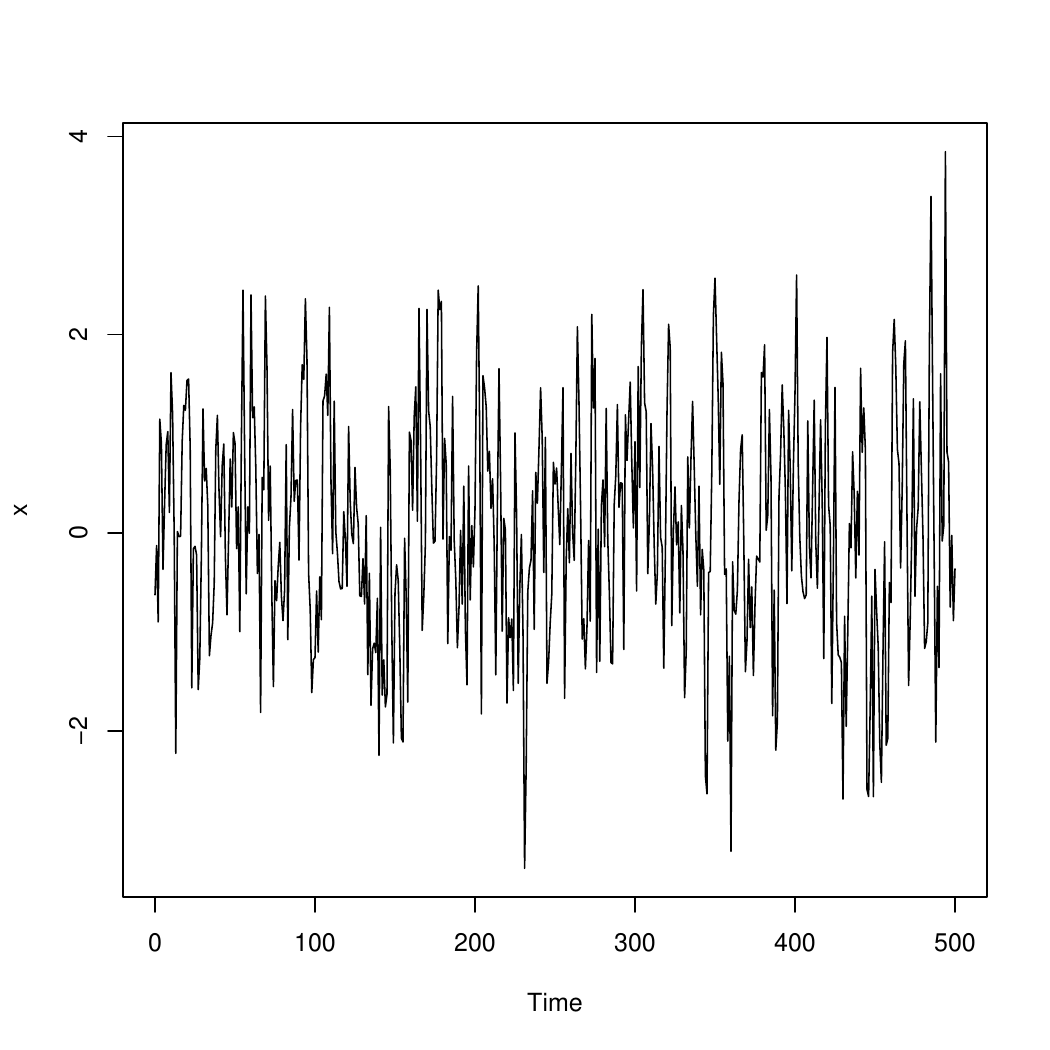}
\end{minipage}%
\begin{minipage}{.25\textwidth}
  \centering
  \includegraphics[width=\linewidth]{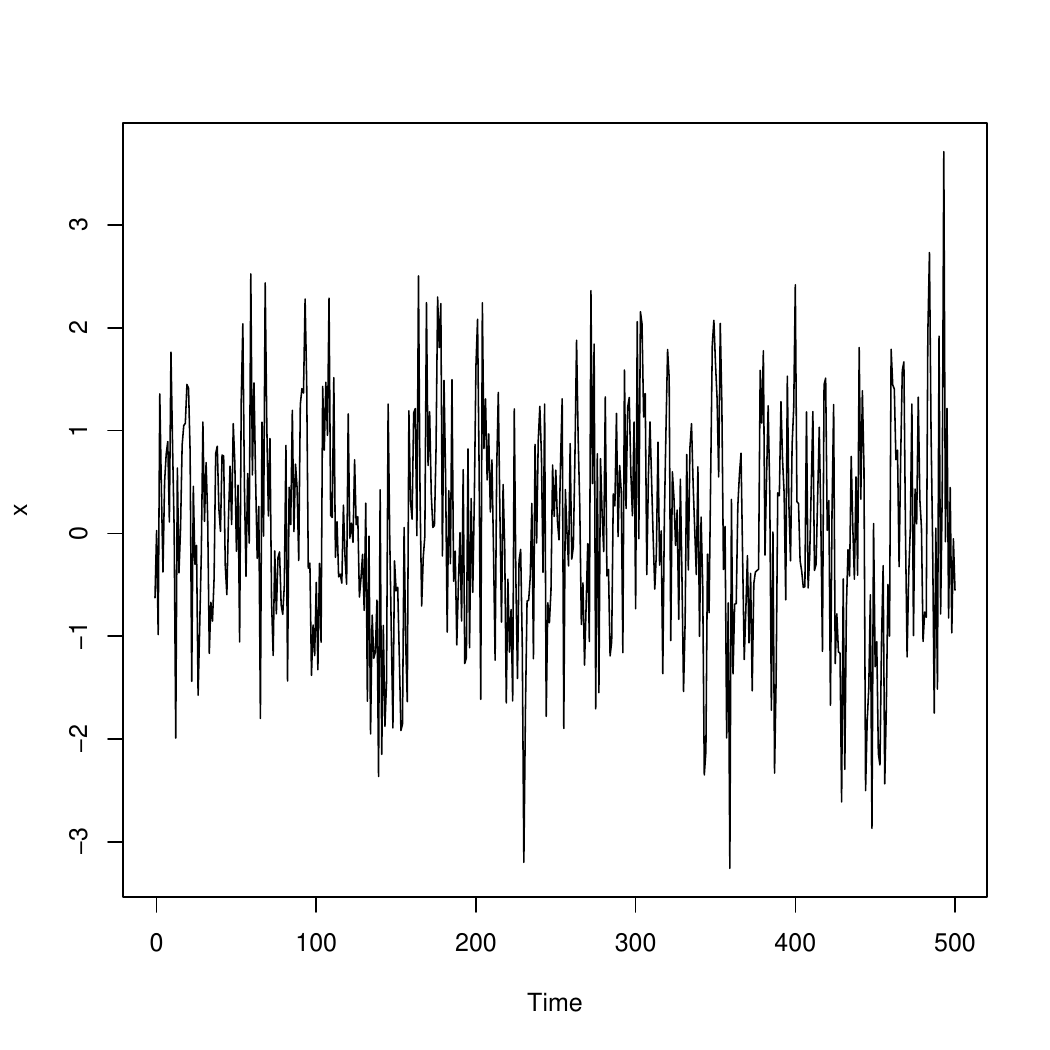}
\end{minipage}%
\begin{minipage}{.25\textwidth}
  \centering
  \includegraphics[width=\linewidth]{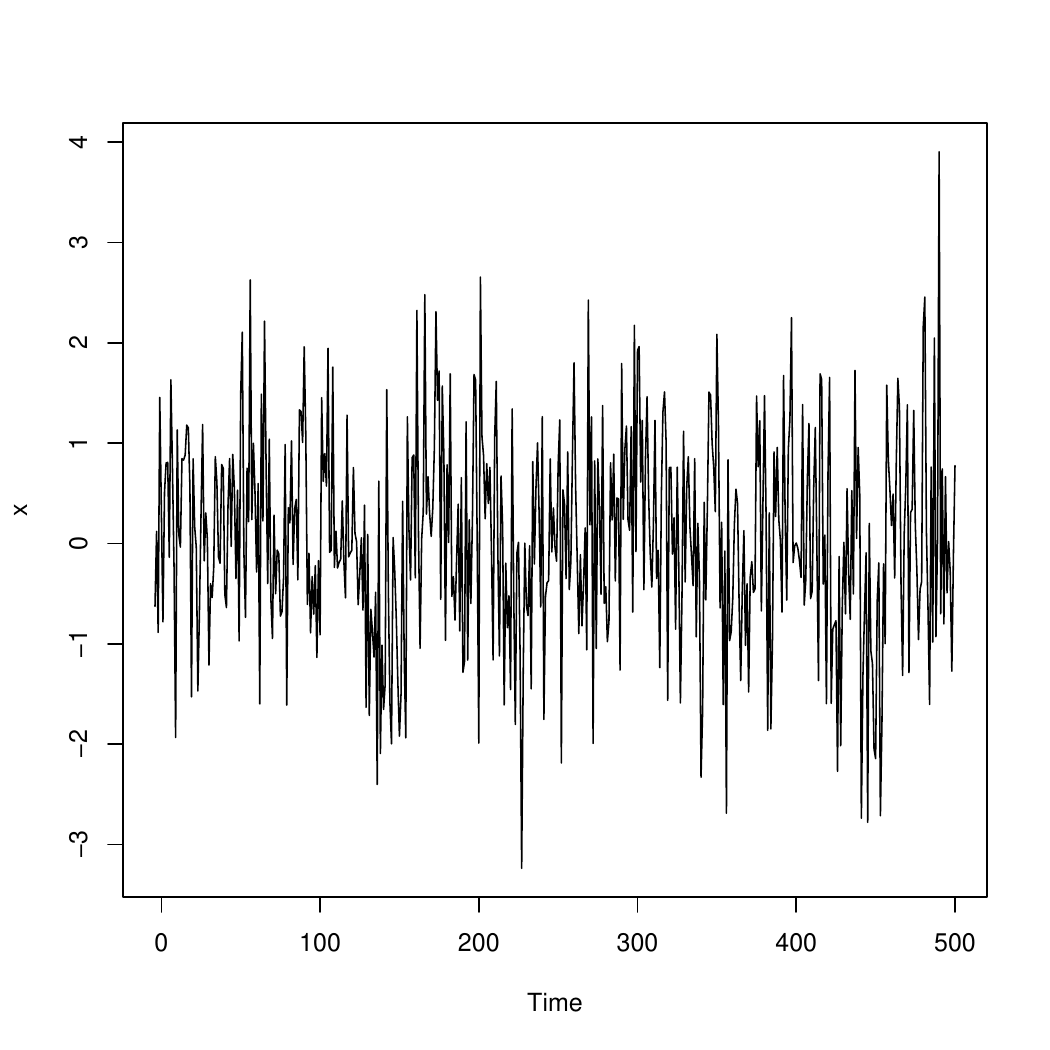}
\end{minipage}%
\begin{minipage}{.25\textwidth}
  \centering
  \includegraphics[width=\linewidth]{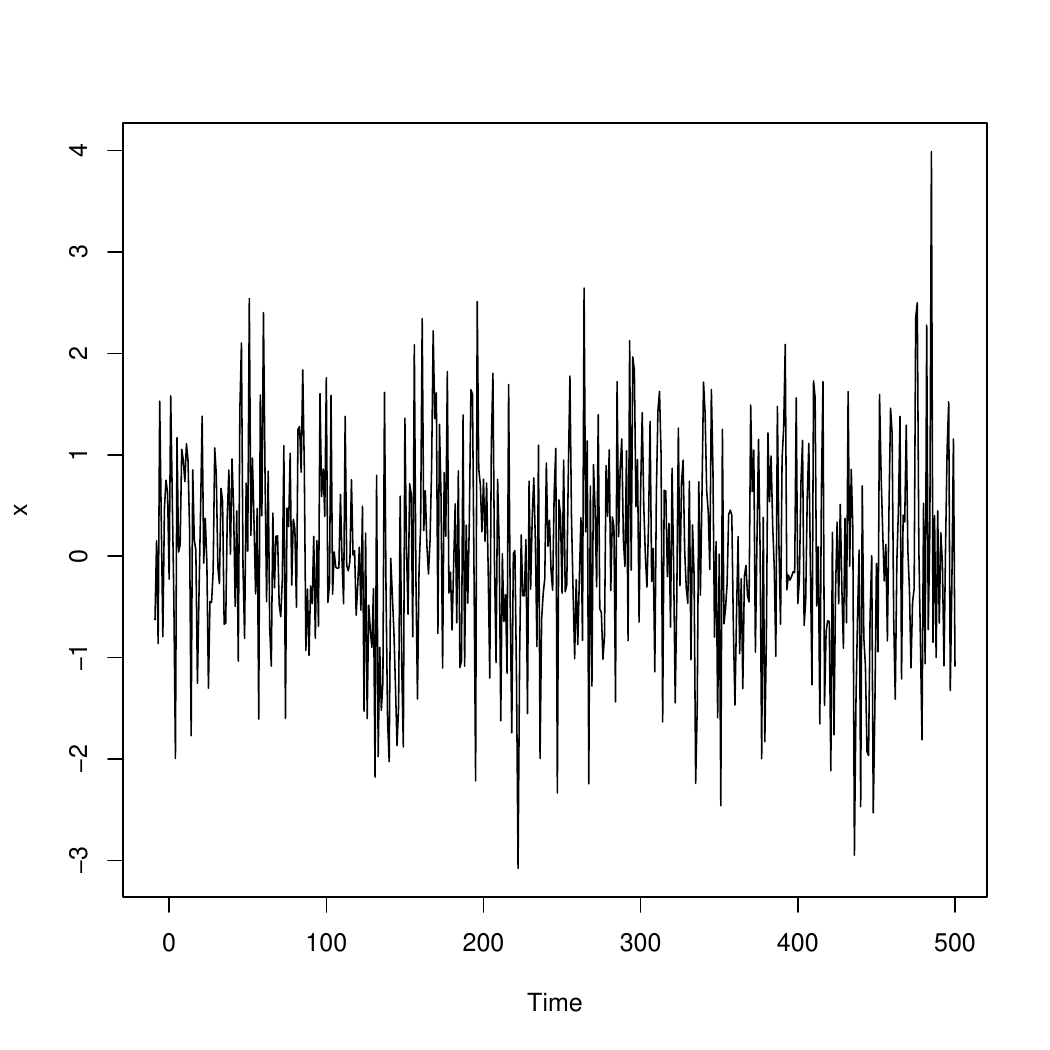}
\end{minipage}%
\\
\begin{minipage}{.25\textwidth}
  \includegraphics[width=\linewidth]{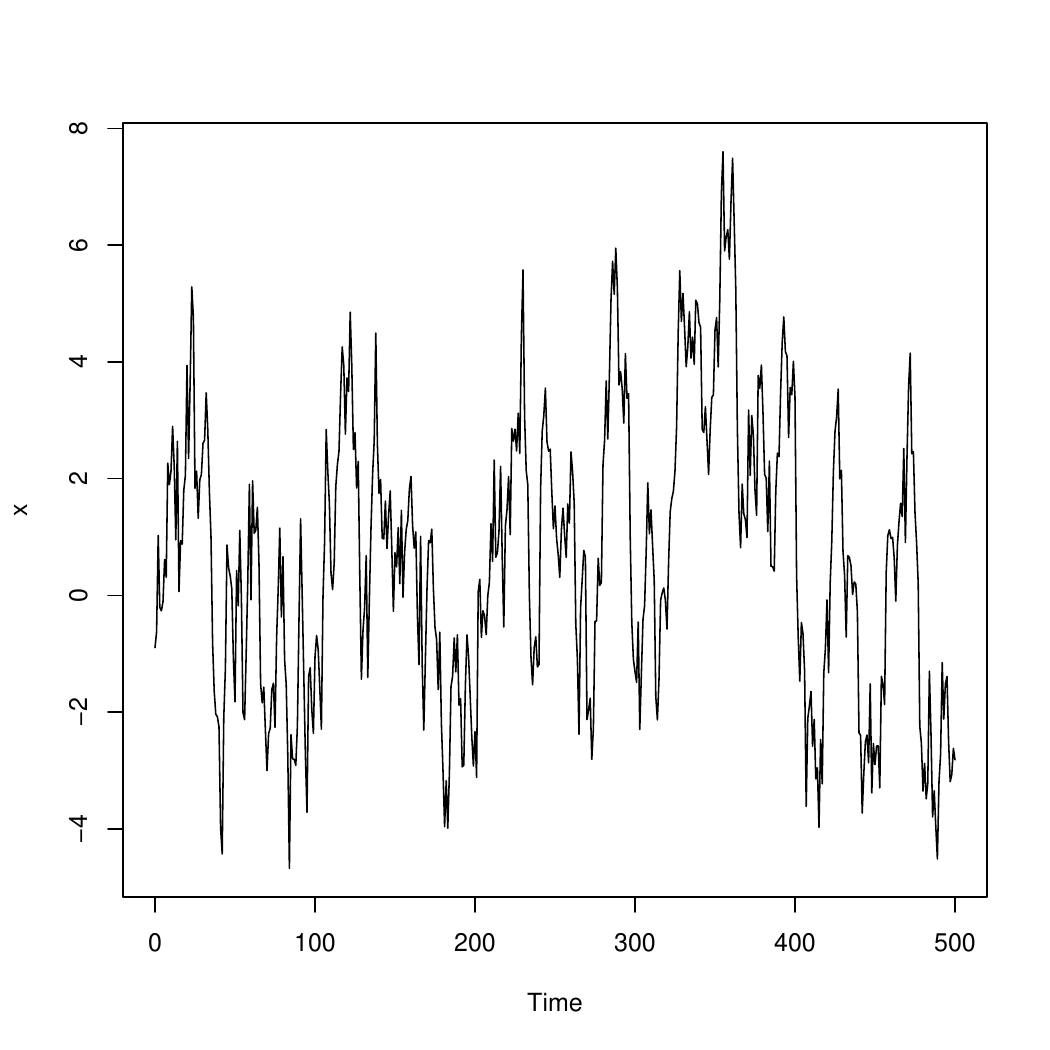}
\end{minipage}%
\begin{minipage}{.25\textwidth}
  \centering
  \includegraphics[width=\linewidth]{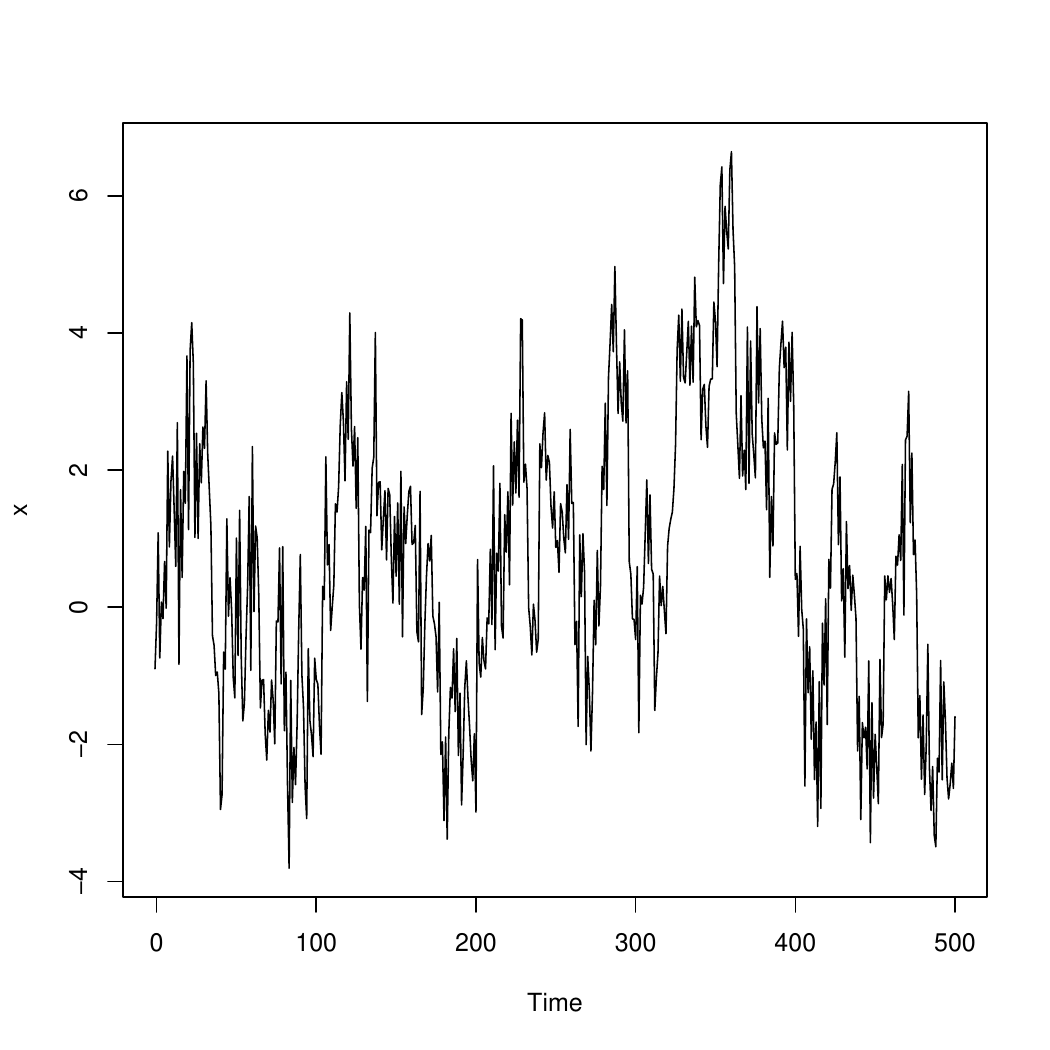}
\end{minipage}%
\begin{minipage}{.25\textwidth}
  \centering
  \includegraphics[width=\linewidth]{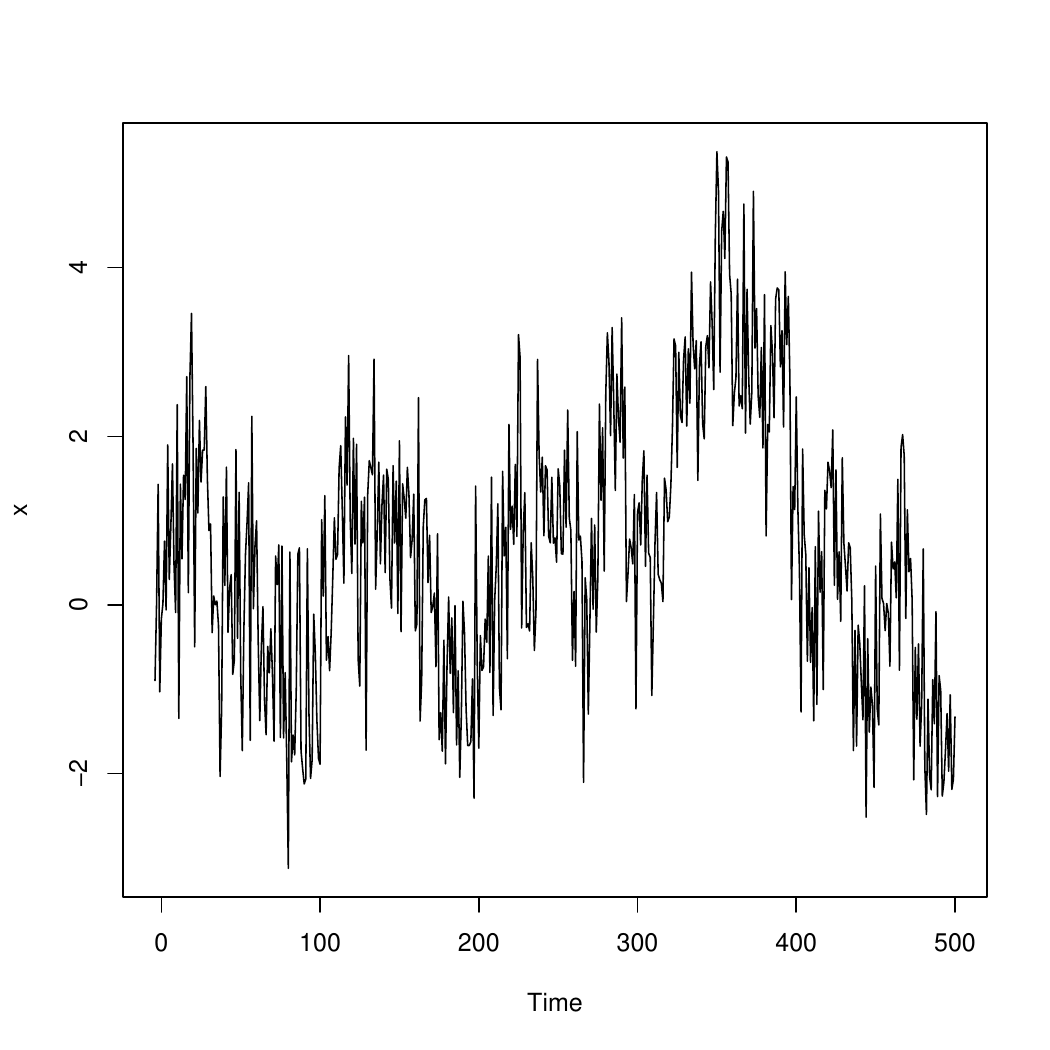}
\end{minipage}%
\begin{minipage}{.25\textwidth}
  \centering
  \includegraphics[width=\linewidth]{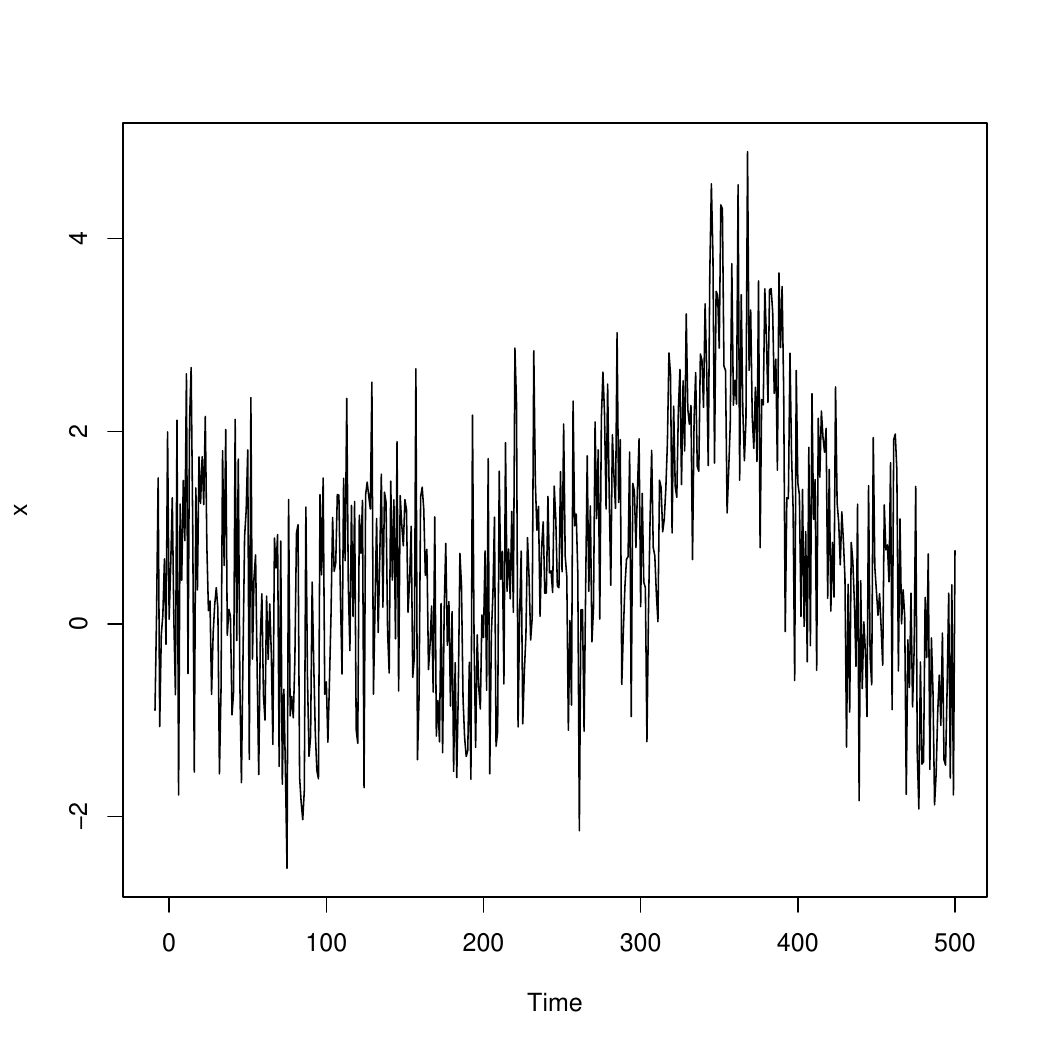}
\end{minipage}%
\\
\begin{minipage}{.25\textwidth}
  \includegraphics[width=\linewidth]{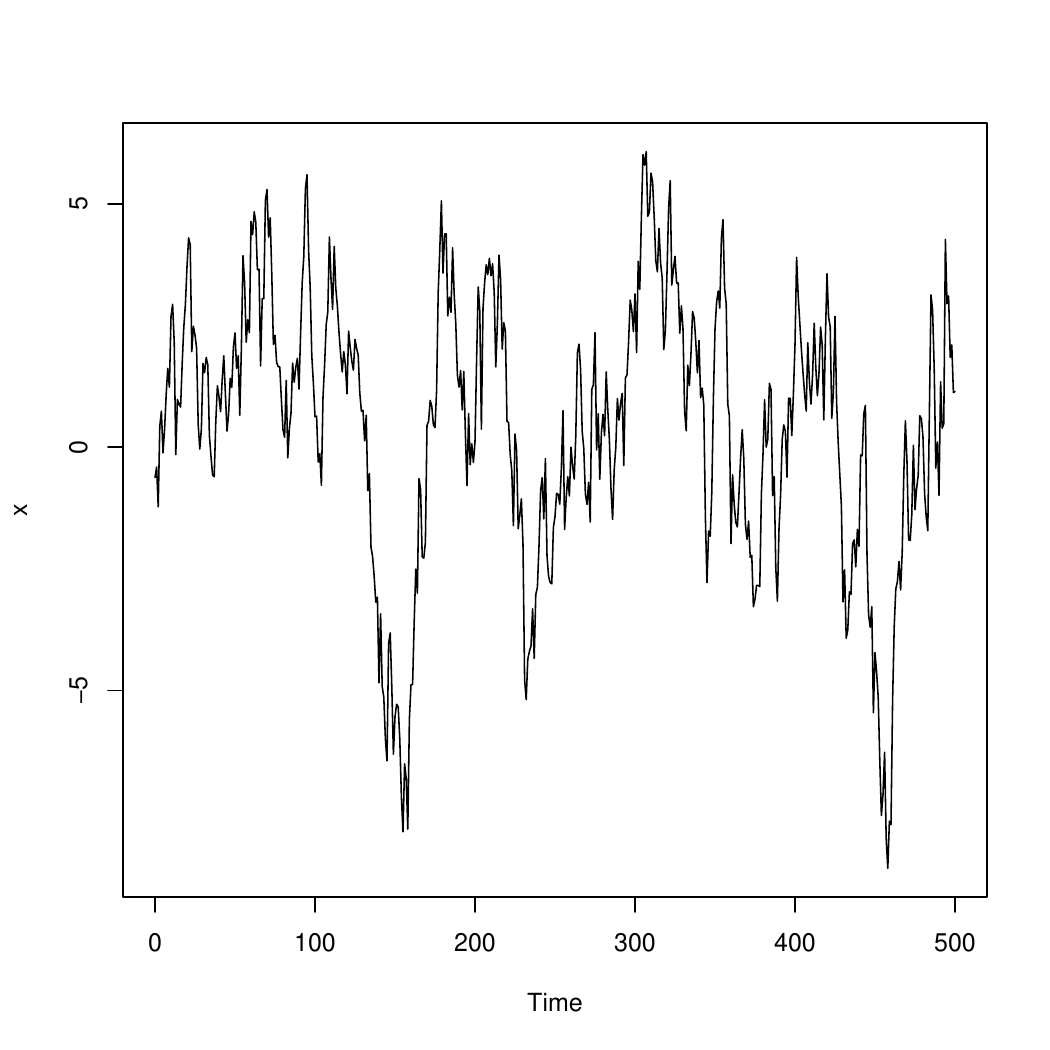}
\end{minipage}%
\begin{minipage}{.25\textwidth}
  \centering
  \includegraphics[width=\linewidth]{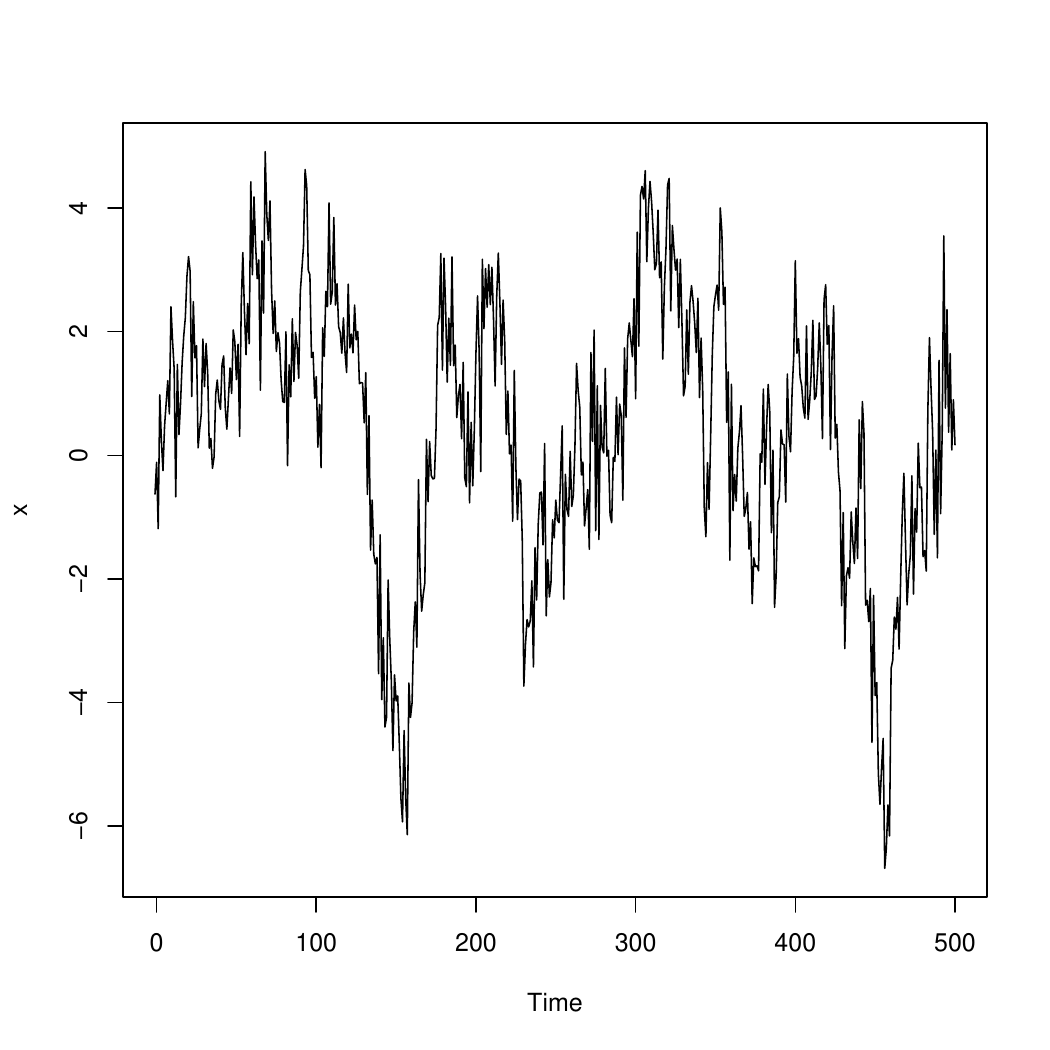}
\end{minipage}%
\begin{minipage}{.25\textwidth}
  \centering
  \includegraphics[width=\linewidth]{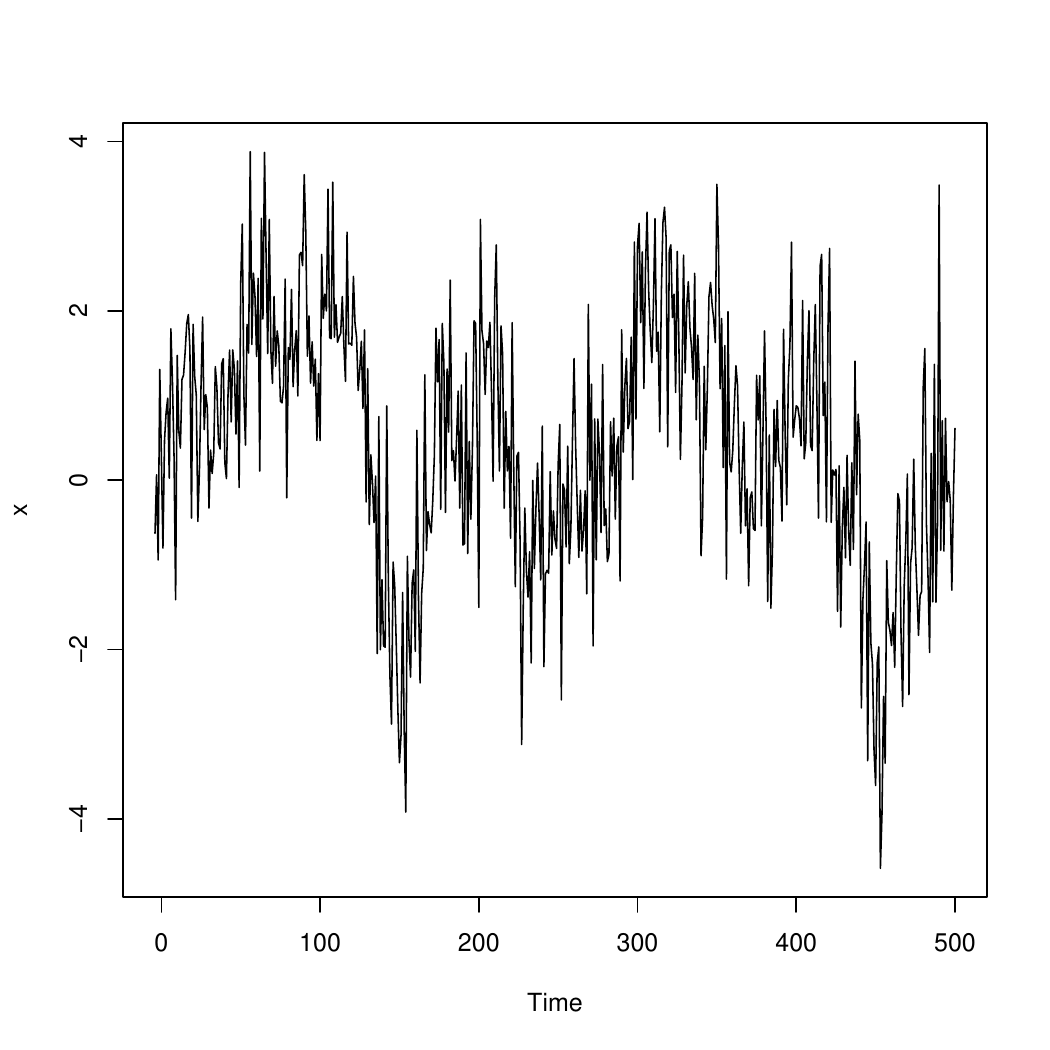}
\end{minipage}%
\begin{minipage}{.25\textwidth}
  \centering
  \includegraphics[width=\linewidth]{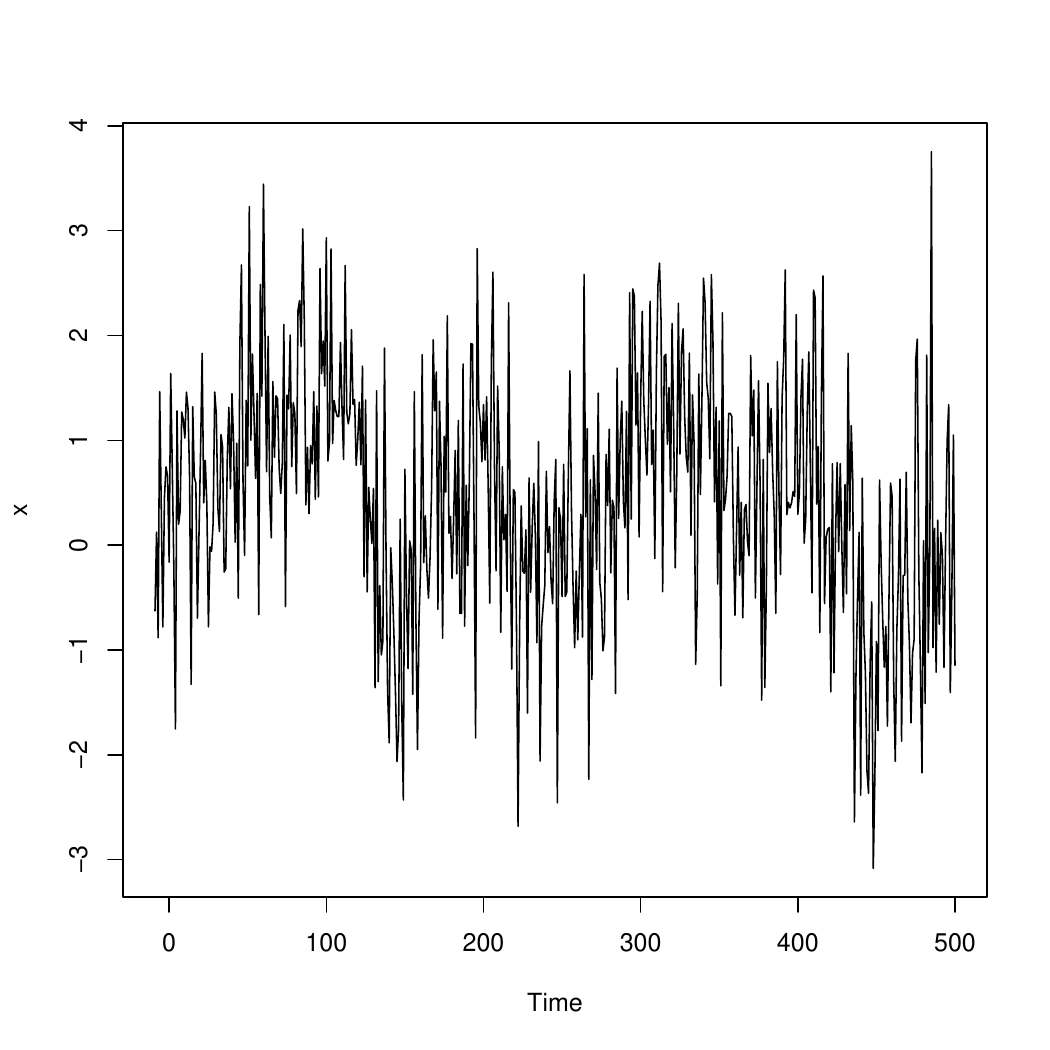}
\end{minipage}
\caption{Simulated sample paths, of length $T=500$, from the single-scale AMAR model $X_t = \alpha_1 \frac{X_{t-1} + \ldots + X_{t-\tau_1}}{\tau_1} + \varepsilon_t$, for $t = 1, \ldots, T$, with $\alpha_1 = 0.5,0.9,0.95$ (respectively from top to bottom) and $\tau_1 = 1, 2, 5, 10$ (respectively from left to right). The same random seed is used to generate path for each row. \label{fig:amar_1}}
\end{figure}

Finally, we offer visual insights into the behaviour of AMAR with a single scale in Figure~\ref{fig:amar_1}. Here realisations for different values of $\alpha_1$ (from 0.5 to 0.95, the latter corresponds to series that are near unit-root) and $\tau_1$ (from 1 to 10) with standard Gaussian noise are plotted. It appears that the longer the scale, the noisier the appearance; the overall shape (driven by the low frequencies) is preserved, but the details (driven by the high frequencies) are increasingly obscured by noise. In addition, even though all the series plotted in Figure~\ref{fig:amar_1} are weakly stationary, some of them exhibit behaviour that mimics non-stationarity, at least visually, when $\tau_1$ if large, even for a moderate $\alpha_1$. This hints at the usefulness of AMAR in the modelling of near unit-root or certain non-stationary series. Regarding this, see also additional numerical results in the supplements. Moreover, insights into other more complex special cases can be found in Section~\ref{sec:specialcases} of the supplementary materials.

\subsection{Large deviations for the OLS estimator in  \texorpdfstring{\ar{p}}{AR(p)}}
\label{mzar:sec:ar_large_deviations}

As a prelude to the study of the behaviour of our proposed AMAR scale and coefficient estimation procedure,
we obtain a tail probability bound on the Euclidean norm of the difference between the OLS estimator $\betahb$ of the autoregressive parameters $\betab$ in model \eqref{mzar:eq:ar_p}, with all bounds explicitly depending on $T$, $p$ and the other parameters of the \ar{p} process. For any vector $\vb=(v_1, \ldots, v_{k})^T\in\R^{k}$,  the Euclidean norm is denoted by $\norm{\vb} = \sqrt{\sum_{j=1}^{k} v_{k}^{2}}$.  The following theorem holds.
\begin{Theorem}
	\label{mzar:theorem:lse_in_ar_conistency_bound}
	Suppose that  $\{X_{t}\}_{t=1}^T$ follows the \ar{p} model \eqref{mzar:eq:ar_p} with the innovations $\varepsilon_{1},\ldots,\varepsilon_{T}$ being i.i.d. $\Nc(0,\sigma^2)$ with $\sigma > 0$. Also assume that the process is stationary and causal.
	Let $\betahb=(\hat{\beta}_{1}, \ldots, \hat{\beta}_{p})^T$ be the OLS estimator of the vector of the autoregressive coefficients $\betab=(\beta_{1}, \ldots, \beta_{p})^T$. Then there exist universal constants $\kappa_{1}$, $\kappa_{2}$, $\kappa_{3}>0$ not depending on $T$, $p$ or $\betab$ s.t. if $\sqrt{T}>\kappa_{2}p\log(T)$, then we have
	\begin{align}
		\label{mzar:eq:bound_on_lse_norm}
		\Pb{\norm{\betahb-\betab} \leq \kappa_{1}(\bl/\bu)^2 \norm{\betab}\frac{p\log(T)\sqrt{\log(T+p)}}{\sqrt{T}-\kappa_{2}p\log(T) } } \geq 1 - \frac{\kappa_{3}}{T},
	\end{align}
	where $\bl = \min_{z\in\T} |b(z)|$ and $\bu = \max_{z\in\T} |b(z)|$.
\end{Theorem} 
Theorem~\ref{mzar:theorem:lse_in_ar_conistency_bound} implies that,  with high probability,  the differences $\hat{\beta}_{j}-\beta_{j}$ in \eqref{mzar:eq:signal+noise} converge to zero with $T\to\infty$, provided that $\frac{p\log(T)\sqrt{\log(T+p)}}{\sqrt{T}-\kappa_{2}p\log(T) } \to 0$. Also note that this result holds for any $\sigma > 0$, as the OLS estimate has the ``self-normalising'' property in the current setting, i.e. it remains unchanged when we scale the entire observed series $\{X_t\}$ (and thus $\sigma$) by a constant.

We remark that in a setting where both the order $p$ and the autoregressive coefficients in model \eqref{mzar:eq:ar_p} do not depend on the sample size $T$, properties of the OLS estimators are well-established.
\cite{lai1983ar} show that, without assumptions on the roots of the characteristic polynomial $b(z)$, the OLS estimators are strongly consistent if $\{\varepsilon_{t}\}$ is a martingale difference sequence with bounds on the conditional second moments. \cite{barabanov1983strong} obtains similar results independently, under slightly stronger assumptions on the noise sequence. \cite{bercu2008exponential} give an exponential inequality for the OLS estimators in the \ar{1} model with i.i.d. Gaussian noise. 

\subsection{AMAR estimation algorithm}

\subsubsection{Timescale estimation}

To estimate the timescales $\tau_1, \ldots, \tau_q$, at which the change-points in model \eqref{mzar:eq:signal+noise} are located,
we adapt the Narrowest-Over-Threshold (NOT) approach of \cite{bcf19}, with the cumulative sum (CUSUM) contrast function
$\cont{s}{e}{b}{\cdot}$ suitable for the piecewise-constant model, defined by
\begin{equation}
\label{eq:cont}
\cont{s}{e}{b}{\vb} = \left| \sqrt{\frac{e-b}{(e-s+1)(b-s+1)}}\sum_{t=s}^b v_t - \sqrt{\frac{b-s+1}{(e-s+1)(e-b)}}\sum_{t=b+1}^e v_t \right|.
\end{equation}
In \cite{bcf19}, NOT was shown to recover the number and locations of change-points (the latter at near-optimal rates) in the ``piecewise-constant signal + i.i.d. Gaussian noise'' model. Although it is challenging to establish the corresponding consistency and near-optimal rates in problem \eqref{mzar:eq:signal+noise} due to the complex dependence structure
in $\hat{\beta}_{j}-\beta_{j}$, we 
show in Section~\ref{mzar:sec:mzar_algorithm} that here NOT estimators enjoy properties similar to those established in the i.i.d. Gaussian setting.

Let $\zeta_{T}>0$ be a significance threshold with which to identify large CUSUM values (with its choice to be discussed in Section~\ref{mzar:sec:parameter_choice}). The NOT procedure for the estimation of the timescales in the \mzar{q} model is described in Algorithm~\ref{mzar:alg:not_algorithm_for_mzar}, which serves as a key ingredient of the AMAR estimation algorithm, given in Section~\ref{mzar:sec:mzar_algorithm}. Core to this approach is a particular blend of global and local treatment of the data $\hat{\betab}$ in the
search for the multiple scales that may be present in the true $\betab_0$. At the global stage, we look at the behaviour of $\hat{\betab}$ over a large number of subintervals (either drawn randomly or systematically),  $(\hat{\beta}_s, \ldots,  \hat{\beta}_e)$, where $1 \le s < e \le p$. On each subinterval, we assume, possibly erroneously,
that only one feature (i.e. scale) is present and use a contrast function (in this setting, CUSUM-based) to find the most likely location of
the scale. We retain those subsamples for which the contrast exceeds a certain specified
threshold, and discard the others. Amongst the retained subsamples, we search for the one drawn
on the narrowest interval, i.e. one for which $e - s$ is the smallest. The focus on the narrowest
interval constitutes the local part of the method, which ensures that with high probability, at most (and also at least, with appropriate choice of threshold) one scale is present in the selected interval. Having detected the first scale, our Algorithm~\ref{mzar:alg:not_algorithm_for_mzar} then proceeds recursively to the left and to the
right of it, and stops, on any current subinterval, if no contrasts can be found that exceed the threshold. More details regarding the intuitions and the construction of the contrast function under different settings can be found in \cite{bcf19}.

\begin{algorithm}[!htbp]
	\caption{NOT algorithm for estimation of timescales in AMAR models\label{mzar:alg:not_algorithm_for_mzar}}
		\begin{algorithmic}
		\Require Estimates $\betahb=(\hat{\beta}_{1},\ldots, \hat{\beta}_{p})^T$;  $s$ and $e$ are the start- and end-points of an interval of interest; $F_{T}^{M}$ is a set of intervals within $[1,p]$; and a given threshold $\zeta_T$ and $\Sc=\emptyset$ (as an initiation).
		\Ensure Set of estimated timescales $\Sc =\{\hat{\tau}_1, \ldots, \hat{\tau}_{\hat{q}}\}\subset\{1,\ldots,p\}$, where $\hat{\tau}_1, \ldots, \hat{\tau}_{\hat{q}}$ are in increasing order.
		
		\Procedure{NOT}{$\betahb, s,e,F_T^M,\zeta_{T}$}	
		\If{$e=s$} STOP
		\Else
		\State $\Mc_{s,e}:=\left\{m:[s_{m},e_{m}]\in F_{T}^{M}, [s_{m},e_{m}]\subset[s,e]\right\}$
		\If{$\Mc_{s,e}=\emptyset$} STOP
		\Else
		\State $\Oc_{s,e}:=\left\{m\in\Mc_{s,e}: \max_{b \in \{s_m,\ldots, e_m-1\}} \cont{s_m}{e_m}{b}{\betahb}  >  \zeta_{T}\right\}$
		\If{$\Oc_{s,e}=\emptyset$} STOP
		\Else
		\State $m^{*}:\in\argmin_{m\in\Oc_{s,e}}|e_{m}-s_{m}+1|$ 
		\State $b^{*}:=\argmax_{b \in \{s_{m^*},\ldots, e_{m^*}-1\}} \cont{s_{m^*}}{e_{m^*}}{b}{\betahb}$
		\State $\Sc:=\Sc\cup\{b^{*}\}$
		\State \Call{NOT}{$\betahb, s,b^{*},\zeta_{T}$}
		\State \Call{NOT}{$\betahb, b^{*}+1,e,\zeta_{T}$}
		\EndIf
		\EndIf
 		\EndIf
		\EndProcedure
	\end{algorithmic}
\end{algorithm}

\begin{algorithm}[!ht]
\caption{AMAR algorithm}
\label{mzar:alg:mzar_algorithm} 
\begin{algorithmic}[1]
\algrenewcommand{\alglinenumber}[1]{\small\textbf{\hspace{\algorithmicindent}Step #1}}
\Require Data $X_{1},\ldots, X_{T}$, $p$;  threshold $\zeta_{T}$, and $M$ (needed only if $p > 500$).
\Ensure Estimates of the relevant scales $\hat{\tau}_{1}, \ldots, \hat{\tau}_{\hat{q}}$ and the corresponding AMAR coefficients $\hat{\alpha}_{1}, \ldots, \hat{\alpha}_{\hat{q}}$.

\Statex{\hspace{-\algorithmicindent}{\bf procedure} \Call{AMAR}{$\{X_{1},\ldots, X_{T}\}$, $p$, $\zeta_{T}$}}
	\State{\label{mzar:alg:mzar_ar_fit}} Find $\betahb=(\hat{\beta}_{1}, \ldots, \hat{\beta}_{p})^T$, the OLS estimates of the autoregressive coefficients in the \ar{p} representation of \mzar{q}.
	\State{\label{mzar:alg:gen_intervals}}
	Let $F_{T}^{M}$ be a set of all $M = p(p-1)/2$ intervals within $[1,p]$ (i.e. $[1,2],\ldots, [1,p], [2,3],\ldots, [2,p], \ldots, [p-1,p]$). If $p$ is large (e.g. $>500$), we take $F_{T}^{M}$ to be a set of $M$ intervals whose start- and end-points have been drawn independently and uniformly from $\{1,\ldots,p\}$ with replacement.
	\State{\label{mzar:alg:mzar_cpt_est}} Call \Call{NOT}{$\betahb, 1, p,F_{T}^{M}, \zeta_{T}$} from Algorithm~\ref{mzar:alg:not_algorithm_for_mzar} to find the estimates of the timescales;
	Sort them in increasing order to obtain  $\hat{\tau}_{1},\ldots,\hat{\tau}_{\hat{q}}$.
	\State{\label{mzar:alg:mzar_mzar_fit}} With the timescales in \eqref{mzar:eq:mzar_model} set to $\{\hat{\tau}_{1},\ldots,\hat{\tau}_{\hat{q}}\}$, find $\hat{\alpha}_{1}, \ldots, \hat{\alpha}_{\hat{q}}$, the OLS estimates of the scale coefficients $\alpha_{1}, \ldots, \alpha_{q}$.
\Statex{\hspace{-\algorithmicindent}\bf end procedure}
\end{algorithmic}
\end{algorithm}

\subsubsection{Parameter estimation}
\label{mzar:sec:mzar_algorithm}

We now introduce our proposed estimation procedure for the parameters of the AMAR model. We refer to it as the AMAR algorithm, and its steps are described in Algorithm~\ref{mzar:alg:mzar_algorithm}. An efficient implementation of the procedure is available in the \proglang{R} package 
\pkg{amar} \citep{baranowski2016mzar}.
The choice of all the input parameters is discussed in Section~\ref{mzar:sec:parameter_choice}.
As a remark, we note that in Step 4, finding the AMAR coefficients via OLS amounts to the same procedure as refitting the OLS estimates of the AR coefficients (e.g. $\hat{\betab}$) subject to equality constraints of having the coefficients to be the same from the $(\tau_k+1)$-th to the $\tau_{k+1}$-th time-lag for all $k$ .

\subsection{Theoretical properties}
\label{Sec:theory}
\subsubsection{Gaussian innovations}
The following two quantities will together measure the difficulty of our change-point problem detection problem (with the convention $\tau_0 = 0$ and $\tau_{q+1} = p$):
\begin{align}
	\delta_{T} &:= \min_{j=1,\ldots,q+1} |\tau_{j}-\tau_{j-1}| \label{mzar:eq:delta_T},\\
	\all_{T} &:= \min_{j=1,\ldots,q} |\beta_{\tau_{j}+1}-\beta_{\tau_{j}}|=  \min_{j=1,\ldots,q}|\alpha_{j}|\tau_{j}^{-1}. \label{mzar:eq:alpha_T}
\end{align}

To study the theoretical properties of the timescale estimators $\hat{\tau}_{1},\ldots,\hat{\tau}_{\hat{q}}$, we make the following assumptions.

\begin{enumerate}[label=(A\arabic*)]
	\item \label{mzar:as:normal_noise} $\{X_{t}\}$ is stationary and follows the \mzar{q} model given in \eqref{mzar:eq:mzar_model} with the innovations $\varepsilon_{t}$ being i.i.d. $\Nc(0,\sigma^2)$ for some $\sigma > 0$.

	\item \label{mzar:as:maximum_changepoint} $p>\tau_{q}$ and there exist constants $\theta < \frac{1}{2}$ and $c_{1}>0$ such that $p<c_{1} T^{\theta}$ for all $T$.
	
	\item \label{mzar:as:strict_stationarity} The roots of the characteristic polynomial $b(z)$  lie outside the unit circle  $\T$. Furthermore, there exists constants $\cl_{2},\bar{c}_2 > 0$ such that 	$\cl_{2} \le
	\min_{z\in \T}|b(z)| \le \max_{z\in \T}|b(z)| 
 \le \bar{c}_2$ uniformly in $T$.

	\item \label{mzar:as:changepoints_spacing}  
	$\delta_{T}^{1/2} \all_{T} \succ T^{\theta-\frac{1}{2}} (\log(T))^{3/2} =: \underline{\lambda}_T$, where  $\theta$ is as in \ref{mzar:as:maximum_changepoint}, and where $\delta_{T}$ and  $\all_{T}$ are given by \eqref{mzar:eq:delta_T} and \eqref{mzar:eq:alpha_T}, respectively. Here $f(T) \succ g(T)$ means that $\mathrm{liminf}_{T\rightarrow \infty} f(T)/g(T) = \infty$.

\end{enumerate}

Some comments regarding these assumptions are in order.
First, the Gaussianity assumption \ref{mzar:as:normal_noise} is made to simplify the theoretical arguments of the proof of Theorem~\ref{mzar:theorem:lse_in_ar_conistency_bound}, which is subsequently used to justify Theorem~\ref{mzar:theorem:consistency_result} below.  As is shown later, Theorem~\ref{mzar:theorem:consistency_result} could possibly be extended to cover more general distributional scenarios for the noise $\varepsilon_{t}$.

Second, Assumption \ref{mzar:as:maximum_changepoint} imposes restrictions on both $p$ and the maximum timescale $\tau_{q}$, which are allowed to increase with $T\to\infty$, but at rates slower than $T^{1/2}$. A similar condition on $p$ being the order of  \ar{p} approximations of an \ar{\infty} processes can be found in e.g. \cite{ing2005order}. Assumption \ref{mzar:as:strict_stationarity} implies that the \mzar{q} process $X_{t}$, $t=1,\ldots,T$, is uniformly stationary for all $T$: the requirement that $\min_{z\in \T}|b(z)|$ is bounded from below implies that the roots of the characteristic polynomial do not approach the unit circle $\T$ when $T\to\infty$, which in turn ensures that the $X_{t}$ process is, heuristically speaking, sufficiently far from being unit-root. Besides, the upper bound on $\max_{z\in \T}|b(z)|$ implies that $\|\betab \|$ is uniformly bounded from below, in view of the Parseval's identity (see Lemma~\ref{mzar:theorem:parsevals_identity} in the supplementary materials). 

Third, Assumption \ref{mzar:as:changepoints_spacing} controls both the minimum spacing between the timescales and the size of the jumps in \eqref{mzar:eq:mzar_ar_beta}. The quantity $\delta_{T}^{1/2}\all_{T}$ used here is well-known in the change-point detection literature and characterises the difficulty of the multiple change-point detection problem. 
 
\begin{Theorem}
	\label{mzar:theorem:consistency_result}
	Let assumptions \ref{mzar:as:normal_noise} -- \ref{mzar:as:changepoints_spacing} hold, and
	let $\hat{q}$ and $\hat{\tau}_{1},\ldots,\hat{\tau}_{\hat{q}}$ denote, respectively, the number and the locations of the timescales estimated with Algorithm~\ref{mzar:alg:mzar_algorithm}. There exist constants $C_{1},C_{2},C_{3},C_{4}>0$ such that if $C_{1}  \underline{\lambda}_{T} < \zeta_{T} < C_{2} \delta_{T}^{1/2}\all_{T}$, and $M>36T\delta_T^{-2}\log(T\delta_T^{-1})$ (only if needed by Algorithm~\ref{mzar:alg:mzar_algorithm}), then for all sufficiently large $T$ we have
	\begin{align}
		\label{mzar:eq:consistency_result}
		\Pb{\hat{q}=q,\max_{j=1,\ldots,q}|\hat{\tau}_{j}-\tau_{j}| \leq \epsilon_{T}} \geq 1- C_{4} T^{-1},
	\end{align}
	with $\epsilon_{T} = C_{3}\underline{\lambda}_{T}^{2}\all_{T}^{-2}$.
\end{Theorem}
The main conclusion of Theorem~\ref{mzar:theorem:consistency_result} is that Algorithm~\ref{mzar:alg:mzar_algorithm} estimates the number of the timescales correctly, while the corresponding locations of the estimates lie close to the true timescales, both with a high probability. Under certain circumstances,  Algorithm~\ref{mzar:alg:mzar_algorithm} recovers the exact locations of the timescales. Consider, for example, the case when both the number of scales $q$ and the scale coefficients $\alpha_{1},\ldots, \alpha_{q}$ in \eqref{mzar:eq:mzar_model} are fixed, while the timescales increase with $T$ such that $\delta_{T} \sim p \sim T^{\theta}$ (`$\sim$' means that the quantities in question grow at the same rate with $T\to\infty$).  This is a challenging setting, in which $\all_{T} \sim  T^{-\theta}$ and $\norm{\betab}\sim T^{-\theta/2}$, where the coordinates of $\betab$ are given by \eqref{mzar:eq:mzar_ar_beta}, so the signal strength decreases to $0$ when $T\to\infty$. Here $\delta_{T}^{1/2}\all_{T} \sim T^{-\theta/2}$, thus \ref{mzar:as:changepoints_spacing} can only be met if $\theta$ in  \ref{mzar:as:maximum_changepoint} satisfies the additional requirement $\theta\leq \frac{1}{3}$. The distance between the true timescales and their estimates is then not larger than $\epsilon_{T}\sim T^{4\theta - 1}(\log(T))^{3}$, which tends to zero if $\theta < \frac{1}{4}$. In this case, \eqref{mzar:eq:consistency_result} simplifies to $\Pb{\hat{q}=q,\,\hat{\tau}_{j}=\tau_{j}\,\forall{j=1,\ldots,q}} \geq 1- C_{4} T^{-1}$, when $T$ is sufficiently large. Furthermore, in the much simpler setting where all the locations of the timescales are fixed,  Theorem~\ref{mzar:theorem:consistency_result} concludes that with high probability  $\hat{q}=q$ and $\hat{\tau}_{j}=\tau_{j}$ for all $j = 1, \ldots,q$. As a consequence, one could establish further that all the estimated autoregressive and scale coefficients (i.e. $\beta_i$'s and $\alpha_i$'s) converging at the rate of $T^{-1/2}$. 
However, in general, we would expect the convergence rate of $\hat{\alpha}_i$'s to $\alpha_i$'s to be slower than $O(T^{-1/2})$ when either $q$ or $\tau_q$ (or both) increases with $T$. Due to its theoretical nature, we leave the complete characterisation of the asymptotic behaviours of $\hat{\alpha}_i$'s to future research.

\subsubsection{Heavy-tailed innovations} 
In the settings where the innovations follow more heavy-tailed distributions, consistency of our procedure could still be established. For simplicity, we shall assume that the number of scales $q$ is fixed, so that the presented results would have much simpler dependence on the tail behaviour of the innovation distributions. The assumptions we impose under this setup are given below.

\begin{enumerate}[label=(B\arabic*)]
	\item \label{mzar:as:ht_noise} $\{X_{t}\}$ is stationary and follows the \mzar{q} model given in \eqref{mzar:eq:mzar_model} with the innovations $\varepsilon_{t}$ being i.i.d. following a symmetric distribution $Z$ with regularly varying tail probabilities of index $\alpha$, such that 
 $\Pb{|Z|>z} = z^{-\alpha} L(z)$
 for any $z > 0$ with any positive $\alpha \neq 2$, and $L(\cdot)$ is a slowly varying function at $\infty$, i.e. 
$\lim _{{z\to \infty }}{\frac  {L(az)}{L(z)}}=1$ for any $a > 0$.
	
	\item \label{mzar:as:ht_maximum_changepoint} $\alpha_1,\ldots,\alpha_q$,  $\tau_1,\ldots, \tau_{q}$ and $p$ are fixed, with $p>\tau_{q}$.
 
	\item \label{mzar:as:ht_strict_stationarity} The roots of the corresponding characteristic polynomial $b(z)$  lie outside the unit circle  $\T$.

\end{enumerate}

Regarding Assumption~\ref{mzar:as:ht_noise}, we note that it covers many scenarios of heavy-tailed distributions, including generalized Pareto and Cauchy. Here a smaller $\alpha$ implies heavier tails. For instance, when $\alpha > 4$, the innovation distribution has finite fourth-moment, while a distribution with $\alpha \in (2,4)$ has finite variance. Here the case of $\alpha =2$ (i.e. the boundary of an infinite variance) is not included to simplify our analysis further. In the setting of autoregressive models,  much heavier tails (e.g. with $\alpha < 2$) actually tend to make the parameter estimation (fundamentally via autocorrelation) more accurate, which intuitively is due to the fact that observations would be more spread-out. See \cite{yohai1977}, \cite{hannan1977}, \cite{davis1985,davis1986}. Consequently, our Algorithm~\ref{mzar:alg:mzar_algorithm} would still work as intended, as established in the following result.

\begin{Theorem}
	\label{mzar:theorem:consistency_result_2}
	Let assumptions \ref{mzar:as:ht_noise} -- \ref{mzar:as:ht_strict_stationarity} hold, and
	let $\hat{q}$ and $\hat{\tau}_{1},\ldots,\hat{\tau}_{\hat{q}}$ denote, respectively, the number and the locations of the timescales estimated with Algorithm~\ref{mzar:alg:mzar_algorithm} (with $F_T^M$ taken as the set of all $p(p-1)/2$ intervals within $[1,p]$). For any sufficiently small $\epsilon > 0$, there exist constants $C_{1},C_{2}$ such that if $C_{1}  {T}^{-\max(1/2, 1/\alpha)+\epsilon} < \zeta_{T} < C_{2} \all_{T}$, then as $T\rightarrow \infty$,
	\begin{align*}
\Pb{\hat{q}=q,\max_{j=1,\ldots,q}|\hat{\tau}_{j}-\tau_{j}|  = 0 } \rightarrow 1.
	\end{align*}
\end{Theorem}


\section{Practicalities and simulated examples} 


\subsection{Parameter choice} 
\label{mzar:sec:parameter_choice}

{\em Threshold $\zeta_T$.} This threshold is one of the input parameters required in Algorithm~\ref{mzar:alg:not_algorithm_for_mzar} and Algorithm~\ref{mzar:alg:mzar_algorithm}. The minimum rate of magnitude permitted by Theorem \ref{mzar:theorem:consistency_result}, that is $\zeta_{T}=C T^{-1/2} (\log(T))^{3/2}$
 can be used (say, with $C=0.5$), though a more careful choice would be required for different setups. In practice, we advocate choosing the threshold using the Schwarz Information Criterion (SIC) as outlined below. 
 
For any $\zeta_{T}>0$ (and a fixed $p$), denote by $\hat{X}_{t}(\zeta_{T})$  the forecast of $X_{t}$ obtained via Algorithm~\ref{mzar:alg:mzar_algorithm} and by $\hat{q}(\zeta_{T})$ the number of the estimated timescales. Specifically, with the estimated timescales $\hat{\tau}_1(\zeta_{T}), \ldots, \hat{\tau}_{\hat{q}(\zeta_{T})}(\zeta_{T})$ and corresponding scales coefficients $\hat{\alpha}_1(\zeta_{T}),\ldots, \hat{\alpha}_{\hat{q}(\zeta_{T})}(\zeta_{T})$,
\[
\hat{X}_{t}(\zeta_{T}) = \hat{\alpha}_1(\zeta_{T})\frac{X_{t-1}+\ldots+X_{t-\hat{\tau}_1(\zeta_{T})}}{\hat{\tau}_1(\zeta_{T})}+\ldots+\hat{\alpha}_{\hat{q}(\zeta_{T})}(\zeta_{T})\frac{X_{t-1}+\ldots+X_{t-\hat{\tau}_{\hat{q}(\zeta_{T})}(\zeta_{T})}}{\hat{\tau}_{\hat{q}(\zeta_{T})}(\zeta_{T})},
\]
where we set the values of the unobserved  $X_{0},X_{-1},\ldots$ to be the sample mean of the series.

We then select the threshold that minimises the SIC defined as follows:
\begin{align}
	\label{mzar:eq:SIC} \mbox{SIC}(\zeta_{T}) = T \log\left(\sum_{t=1}^{T}(X_{t}-\hat{X}_{t}(\zeta_{T}))^2\right) + 2 \hat{q}(\zeta_{T}) \log(T),
\end{align}
where \eqref{mzar:eq:SIC} is minimised over $\zeta_T$ such that $\hat{q}(\zeta_{T}) \leq q_{\mathrm{max}}=10$. Unless stated otherwise, we take this as our default approach in the remainder of this article and in Section~\ref{Sec:addnum} of the supplementary materials.

\vspace{10pt}

{\em Number $M$ of random intervals.} As outlined in Algorithm~\ref{mzar:alg:mzar_algorithm}, we normally use all the intervals unless $p$ is extremely large. This would be computationally feasible for most applications. However, when $p$ is large (say $>500$), we would follow the recommendation in \cite{bcf19} by setting $M = 10000$. 
\vspace{10pt}

{\em The autoregressive order $p$.} We refrain from giving a universal recipe for the choice of $p$. In the real data example reported later, 
we choose the $p$ that corresponds to a large ``natural''  time span. 
If such choice is not obvious, then in principle, the SIC criterion \eqref{mzar:eq:SIC} can be
minimised with respect to both $\zeta_{T}$ and $p$. Here to reduce the computational burden, in practice, instead of going through all possible values of $p$ (and finding the corresponding threshold that minimises that particular SIC), one possibility would be to search for $p$ only on a grid with its elements increasing exponentially from 1 up to the order of $T^{1/2}$, e.g. $\{1,2,4,8,\ldots\}$. 

\subsection{Computational complexity of the AMAR algorithm} 
The calculation of the OLS estimates in Steps \ref{mzar:alg:mzar_ar_fit} and \ref{mzar:alg:mzar_mzar_fit}
of Algorithm~\ref{mzar:alg:mzar_algorithm} takes $O(Tp^2)$ operations.
The values of $\cont{s}{e}{b}{\cdot}$ can be computed for all $b$ in $O(e-s)$ operations, hence the complexity
of Step~\ref{mzar:alg:mzar_cpt_est} is $O(M p)$. This term is typically dominated by  $O(Tp^2)$, and therefore the usual computational complexity of the 
AMAR algorithm is $O(Tp^2)$. We make use of an efficient implementation of OLS estimation
available from the \proglang{R} package \pkg{RcppEigen} \citep{bates2013fast}.

\subsection{Simulation study} \label{mzar:sec:simulation_study}

We illustrate the finite sample behaviour and performance  Algorithm~\ref{mzar:alg:mzar_algorithm} in a comprehensive simulation study.  The data are simulated from \eqref{mzar:eq:mzar_model} for the following four scenarios. In all these scenarios, the noise $\varepsilon_{t}$ follows i.i.d. $\Nc(0, 1)$.

\begin{enumerate}[label=(M\arabic*)] 
	\item \label{mzar:model:1} Two timescales at $\tau_{1}=1$ and $\tau_{2}=3$,  with the corresponding coefficients $\alpha_{1}= 0.3$, $\alpha_{2}=0.6$ (i.e. $\betab = (0.5,0.2,0.2)^T$). 
	\item  \label{mzar:model:2}  Two timescales at $\tau_{1}=2$ and  $\tau_{2}=5$, with the corresponding coefficients $\alpha_{1}= 1.9$, $\alpha_{2}=-1$ (i.e. $\betab = (0.75,0.75,-0.2,-0.2,-0.2)^T$).
	\item  \label{mzar:model:3}  Three timescales at $\tau_{1}=1$, $\tau_{2}= 5$ and $\tau_{3}= 14$,  with the corresponding coefficients $\alpha_{1}= 0.4$, $\alpha_{2}=-1$, $\alpha_{3}=1.4$ (i.e. $\betab = (0.4, -0.1,-0.1,-0.1,-0.1,0.1,\ldots,0.1)^T$).
		\item  \label{mzar:model:4}  Seasonal model with four timescales at $\tau_{1}=1$, $\tau_{2}=6$, $\tau_{3}=7$ and $\tau_{4}=8$,  with the corresponding coefficients $\alpha_{1}= 0.5$, $\alpha_{2}=-4.8$, $\alpha_{3}=8.4$, $\alpha_{4}=-3.2$ (i.e. $\betab = (0.5,0,\ldots,0,0.8,-0.4)^T$, so $\varepsilon_{t} = (1-0.8B^7)(1-0.5B)X_t$).
  	\item  \label{mzar:model:5} A single timescale at $\tau_{1}=10$ with $\alpha_{1}= 0.9$ (i.e. $\betab = (0.09, \ldots,0.09)^T$), as illustrated in Figure~\ref{fig:amar_1}.
     \item  \label{mzar:model:6} Two timescales at $\tau_{1}=1$ and  $\tau_{2}=\lfloor T^{0.4} \rfloor$, (which increases with $T$), with the corresponding coefficients $\alpha_{1} = \alpha_{2}=0.49$ (i.e. $\betab = (0.49+0.49/\lfloor T^{0.4} \rfloor, 0.49/\lfloor T^{0.4} \rfloor \ldots,0.49/\lfloor T^{0.4} \rfloor)^T$), as illustrated in Figure~\ref{fig:amar_2} in the supplementary materials.
\end{enumerate}
These scenarios were designed to cover combinations of timescales of different lengths. Here $\betab$ is selected as such that the series are stationary but also strongly autocorrelated with $\sum_{j=1}^p \beta_j \approx 0.9$ (or more in \ref{mzar:model:6}). We believe that this is the regime where AMAR models are most useful, and is in the lines with what one would get from fitting some of the real data in practice, as shown in Section~\ref{Sec:data}.

We consider a few different aspects of the estimators obtained with Algorithm~\ref{mzar:alg:mzar_algorithm} with different numbers of observations $T=400,800,1500, 3000$. We assess the accuracy in terms of the number of the fitted timescales $\hat{q}$, the Hausdorff distance $D_H$ between the fitted timescale locations $\{\hat{\tau}_1, \ldots, \hat{\tau}_{\hat{q}}\}$ and the true ones $\{  {\tau}_1, \ldots, {\tau}_{q}\}$, as well as the Euclidean distance between the fitted parameter vector $\hat{\betab}$ and the true one $\betab$. We also compare the mean squared prediction errors (MSPE) of the fitted models with the oracles. For the sake of fair comparison, for each of the simulated series (with length $T$), after model fitting, we further draw $T^* = 100$ observations at the end of the series and use these observations solely for the purpose of out-of-sample mean squared prediction error estimation, given as $\sum_{i=1}^{T^*}(\hat{X}_{T+i}- X_{T+i})^2/T^*$, where for every $i=1,\ldots,T^*$, the predicted value of $X_{T+i}$ is given as 
\[
\hat{X}_{T+i}=\hat{\alpha}_{1}\frac{X_{T+i-1}+\ldots+X_{T+i-\hat{\tau}_{1}}}{\hat{\tau}_{1}}+\ldots+\hat{\alpha}_{q}\frac{X_{T+i-1}+\ldots+X_{T+i-\hat{\tau}_{\hat{q}}}}{\hat{\tau}_{\hat{q}}}.
\]
We then report the ratio between the out-of-sample mean squared prediction error and $\sum_{i=1}^{T^*}\varepsilon_{T+i}^2/T^*$, which is the mean squared prediction error from the oracle model.

Here for our proposed AMAR approach, we select both the threshold and $p$ via the Schwarz Information Criterion as mentioned previously with the maximum number of timescales $q_\mathrm{max} = 10$. With regard to the competitors, we also report results obtained using the fused LASSO (N.B. details can be recalled from our literature review in Section~\ref{Sec:lr}), where $\betab$ is estimated by minimising
\[
\sum_{j=p+1}^T (X_j - \beta'_1X_{j-1}-\cdots-\beta'_pX_{j-p})^2 + \lambda \Big(\sum_{j=1}^{p-1} |\beta'_{j+1} - \beta'_{j}| + |\beta'_{p}|\Big)
\]
with respect to $\betab' = (\beta_1',\ldots,\beta_p')^T \in \mathbb{R}^p$, where $\lambda$ is picked by cross-validation. Finally, we also consider the autoregressive model selected via AIC (i.e. among AR(1), $\ldots$, AR($p$)). Note that for the AIC, we do not enforce the parameters to be constant in between consecutive timescale locations; as such, only the corresponding $\|\hat{\betab} - \betab\|$ and the mean squared prediction errors are computed. All the numerical experiments are repeated 1000 times and the results are summarised in Table~\ref{Tab:Sim1} and  Table~\ref{Tab:Sim2}.

\KOMAoptions{fontsize=10pt}
\begin{table}[!htbp]

	\begin{center}
	\begin{footnotesize}
		\begin{tabular}{c|cc|cc|ccc|ccc}
			\hline\hline
			 \multicolumn{11}{c}{Model \ref{mzar:model:1}} \\ \hline
			& \multicolumn{2}{c|}{$E|\hat{q}-q|$}  &  \multicolumn{2}{c|}{$E(D_H)$}  & \multicolumn{3}{c|}{$E\|\hat{\betab} - \betab\|$}  & \multicolumn{3}{c}{$\frac{\mathrm{MSPE(fitted)}}{\mathrm{MSPE(oracle)}}- 1$}   \\
			Method & AMAR & Fused  & AMAR & Fused & AMAR & Fused & AIC & AMAR & Fused & AIC \\\hline
			$T = 400$ &  0.172 & 6.07 & 0.593 & 16.1 & 0.0159 & 0.0206 & 0.0156 & 0.0133 & 0.0226 & 0.0138\\
   & \tiny(0.014) & \tiny (0.088) & \tiny (0.047) & \tiny (0.05) & \tiny (0.0008) & \tiny (0.00049) & \tiny (0.0006) & \tiny (0.00093) & \tiny (0.0012) & \tiny (0.00093) \\
   
			$T=800$ & 0.051 & 8.65 & 0.181 & 24.1 & 0.0035 & 0.0114 & 0.00749 & 0.0046 & 0.0154 & 0.00802 \\
   & \tiny(0.0072) & \tiny (0.12) & \tiny (0.03) & \tiny (0.056) & \tiny (0.00026) & \tiny (0.00027) & \tiny (0.00028) & \tiny (0.00048) & \tiny (0.00089) & \tiny (0.00061) \\
   
			$T=1500$ &  0.018 & 12.5 & 0.085 & 34.1 & 0.00116 & 0.00613 & 0.00445 & 0.00138 & 0.00764 & 0.00393   \\
   & \tiny(0.0042) & \tiny (0.16) & \tiny (0.03) & \tiny (0.051) & \tiny (0.000088) & \tiny (0.00014) & \tiny (0.0002) & \tiny (0.00024) & \tiny (0.00062) & \tiny (0.00041) \\
   
			 $T=3000$ &  0.012 & 20.2 & 0.072 & 50.2 & 0.000546 & 0.0029 & 0.00207 & 0.000662 & 0.00429 & 0.002 \\
       & \tiny(0.0034) & \tiny (0.21) & \tiny (0.035) & \tiny (0.052) & \tiny (0.000027) & \tiny (0.000063) & \tiny (0.000088) & \tiny (0.00017) & \tiny (0.00046) & \tiny (0.00028) \\
			\hline\hline
			
						\multicolumn{11}{c}{Model \ref{mzar:model:2}} \\ \hline
			& \multicolumn{2}{c|}{$E|\hat{q}-q|$}  &  \multicolumn{2}{c|}{$E(D_H)$}  & \multicolumn{3}{c|}{$E\|\hat{\betab} - \betab\|$}  & \multicolumn{3}{c}{$\frac{\mathrm{MSPE(fitted)}}{\mathrm{MSPE(oracle)}}- 1$}    \\
			Method & AMAR & Fused  & AMAR & Fused & AMAR & Fused & AIC & AMAR & Fused & AIC \\\hline
			$T = 400$ & 
0.303 & 7.32 & 1.33 & 14.1 & 0.02 & 0.0717 & 0.124 & 0.0281 & 0.0857 & 0.0763\\
& \tiny (0.018) & \tiny (0.064) & \tiny (0.072) & \tiny (0.062) & \tiny (0.0013) & \tiny (0.0019) & \tiny (0.0032) & \tiny (0.01) & \tiny (0.0043) & \tiny (0.0032)\\

			$T=800$ & 0.194 & 9.39 & 0.764 & 21.8 & 0.00635 & 0.0595 & 0.061 & 0.00852 & 0.0615 & 0.0331 \\
   & \tiny(0.014) & \tiny (0.071) & \tiny (0.06) & \tiny (0.077) & \tiny (0.00071) & \tiny (0.001) & \tiny (0.0018) & \tiny (0.0013) & \tiny (0.0024) & \tiny (0.0016)\\
   
$T=1500$ & 0.108 & 10.9 & 0.921 & 31.6 & 0.00171 & 0.0535 & 0.0327 & 0.00666 & 0.0532 & 0.0165 \\
& \tiny (0.01) & \tiny (0.069) & \tiny (0.11) & \tiny (0.092) & \tiny (0.00038) & \tiny (0.0011) & \tiny (0.0011) & \tiny (0.0038) & \tiny (0.0022) & \tiny (0.001)\\

$T=3000$ &
0.07 & 12.8 & 0.646 & 47.2 & 0.0000979 & 0.0504 & 0.0131 & 0.000793 & 0.0444 & 0.00571\\
 & \tiny(0.0081) & \tiny (0.071) & \tiny (0.099) & \tiny (0.12) & \tiny (0.000021) & \tiny (0.0011) & \tiny (0.00048) & \tiny (0.0002) & \tiny (0.002) & \tiny (0.00056)\\
   
			\hline\hline
			
						\multicolumn{11}{c}{Model \ref{mzar:model:3}} \\ \hline
			& \multicolumn{2}{c|}{$E|\hat{q}-q|$}  &  \multicolumn{2}{c|}{$E(D_H)$}  & \multicolumn{3}{c|}{$E\|\hat{\betab} - \betab\|$}  & \multicolumn{3}{c}{$\frac{\mathrm{MSPE(fitted)}}{\mathrm{MSPE(oracle)}}- 1$}   \\
			Method & AMAR & Fused  & AMAR & Fused & AMAR & Fused & AIC & AMAR & Fused & AIC \\\hline
			$T = 400$ & 
0.711 & 5.76 & 1.37 & 5.12 & 0.0211 & 0.0204 & 0.0499 & 0.0296 & 0.0321 & 0.0567\\
 & \tiny(0.035) & \tiny (0.077) & \tiny (0.046) & \tiny (0.038) & \tiny (0.00076) & \tiny (0.00041) & \tiny (0.00073) & \tiny (0.0016) & \tiny (0.0017) & \tiny (0.0022)\\
 $T = 800$ &
0.344 & 7.83 & 0.643 & 12.6 & 0.00699 & 0.012 & 0.0244 & 0.00922 & 0.0158 & 0.0215\\
 & \tiny(0.026) & \tiny (0.11) & \tiny (0.034) & \tiny (0.068) & \tiny (0.00031) & \tiny (0.00024) & \tiny (0.00041) & \tiny (0.00075) & \tiny (0.00091) & \tiny (0.00099)\\
 $T = 1500$ &
0.083 & 10.1 & 0.31 & 22.4 & 0.00203 & 0.00704 & 0.013 & 0.0034 & 0.00984 & 0.0112\\
 & \tiny(0.011) & \tiny (0.13) & \tiny (0.043) & \tiny (0.08) & \tiny (0.00011) & \tiny (0.00013) & \tiny (0.00022) & \tiny (0.0004) & \tiny (0.00066) & \tiny (0.00072)\\
  $T = 3000$ &
0.054 & 13.5 & 0.219 & 38.2 & 0.000673 & 0.00397 & 0.00648 & 0.0015 & 0.00628 & 0.00683\\
 & \tiny(0.0082) & \tiny (0.16) & \tiny (0.045) & \tiny (0.084) & \tiny (0.000041) & \tiny (0.000074) & \tiny (0.00011) & \tiny (0.00023) & \tiny (0.00051) & \tiny (0.00048)\\
			\hline\hline
		\end{tabular}
	\end{footnotesize}
	\end{center}
	\caption{\label{Tab:Sim1} Performance of different methods under \ref{mzar:model:1} -- \ref{mzar:model:3}, with estimated errors given in the brackets. Here $\hat{q}$ is the number of the fitted timescales, $D_H$ is the Hausdorff distance between the fitted timescale locations $\{\hat{\tau}_1, \ldots, \hat{\tau}_{\hat{q}}\}$ and the true ones $\{  {\tau}_1, \ldots, {\tau}_{q}\}$, $\|\hat{\betab} - \betab\|$ is the Euclidean distance between the fitted parameter vector and the true one, and MPSE is the mean squared prediction errors of different models.}
\end{table}

\begin{table}[!htbp]
	\begin{center}

	\begin{footnotesize}
		\begin{tabular}{c|cc|cc|ccc|ccc}
			\hline\hline
			 \multicolumn{11}{c}{Model \ref{mzar:model:4}} \\ \hline
			& \multicolumn{2}{c|}{$E|\hat{q}-q|$}  &  \multicolumn{2}{c|}{$E(D_H)$}  & \multicolumn{3}{c|}{$E\|\hat{\betab} - \betab\|$}  & \multicolumn{3}{c}{$\frac{\mathrm{MSPE(fitted)}}{\mathrm{MSPE(oracle)}}- 1$}   \\
			Method & AMAR & Fused  & AMAR & Fused & AMAR & Fused & AIC & AMAR & Fused & AIC \\\hline
			$T = 400$ &  0.098 & 11.1 & 0.199 & 11.8 & 0.00892 & 0.0483 & 0.0246 & 0.0145 & 0.0492 & 0.0358\\
 & \tiny(0.012) & \tiny (0.073) & \tiny (0.027) & \tiny (0.017) & \tiny (0.00065) & \tiny (0.00086) & \tiny (0.00083) & \tiny (0.0011) & \tiny (0.0019) & \tiny (0.0017)\\
 
$T = 800$ & 0.044 & 16.3 & 0.092 & 19.7 & 0.00397 & 0.0274 & 0.0107 & 0.00657 & 0.0274 & 0.0142\\
 & \tiny(0.0085) & \tiny (0.1) & \tiny (0.019) & \tiny (0.019) & \tiny (0.0003) & \tiny (0.00045) & \tiny (0.00038) & \tiny (0.0006) & \tiny (0.0012) & \tiny (0.00088)\\
 $T = 1500$ & 
0.035 & 24.2 & 0.291 & 29.8 & 0.00179 & 0.0182 & 0.00685 & 0.00333 & 0.0182 & 0.00812\\
 & \tiny(0.006) & \tiny (0.14) & \tiny (0.059) & \tiny (0.017) & \tiny (0.00011) & \tiny (0.00027) & \tiny (0.00026) & \tiny (0.0004) & \tiny (0.00091) & \tiny (0.00066)\\
  $T = 3000$ & 
0.023 & 36.3 & 0.129 & 45.8 & 0.000756 & 0.0106 & 0.00301 & 0.0017 & 0.0116 & 0.0043\\
 & \tiny(0.0051) & \tiny (0.19) & \tiny (0.033) & \tiny (0.016) & \tiny (0.000023) & \tiny (0.00015) & \tiny (0.00012) & \tiny (0.00024) & \tiny (0.0007) & \tiny (0.0004)\\
   
			\hline\hline
			
						\multicolumn{11}{c}{Model \ref{mzar:model:5}} \\ \hline
			& \multicolumn{2}{c|}{$E|\hat{q}-q|$}  &  \multicolumn{2}{c|}{$E(D_H)$}  & \multicolumn{3}{c|}{$E\|\hat{\betab} - \betab\|$}  & \multicolumn{3}{c}{$\frac{\mathrm{MSPE(fitted)}}{\mathrm{MSPE(oracle)}}- 1$}    \\
			Method & AMAR & Fused  & AMAR & Fused & AMAR & Fused & AIC & AMAR & Fused & AIC \\\hline
			$T = 400$ & 0.217 & 3.11 & 1.64 & 6.95 & 0.0109 & 0.0106 & 0.0341 & 0.0164 & 0.0151 & 0.0398\\
 & \tiny(0.017) & \tiny (0.085) & \tiny (0.073) & \tiny (0.1) & \tiny (0.00045) & \tiny (0.00038) & \tiny (0.00053) & \tiny (0.0028) & \tiny (0.00099) & \tiny (0.0016)\\
 
$T = 800$ & 0.133 & 4.06 & 0.858 & 12.9 & 0.00414 & 0.00562 & 0.0166 & 0.00517 & 0.00833 & 0.0176\\
 & \tiny(0.013) & \tiny (0.098) & \tiny (0.056) & \tiny (0.17) & \tiny (0.00022) & \tiny (0.00019) & \tiny (0.0003) & \tiny (0.00055) & \tiny (0.00069) & \tiny (0.001)\\
$T = 1500$ & 
0.099 & 5.06 & 0.704 & 22.1 & 0.00167 & 0.00331 & 0.00877 & 0.00237 & 0.00454 & 0.00908\\
 & \tiny(0.012) & \tiny (0.11) & \tiny (0.076) & \tiny (0.25) & \tiny (0.00012) & \tiny (0.000097) & \tiny (0.00017) & \tiny (0.00033) & \tiny (0.00046) & \tiny (0.00065)\\
 $T = 3000$ & 
0.052 & 7.07 & 0.331 & 38.9 & 0.000339 & 0.00171 & 0.00427 & 0.000788 & 0.00278 & 0.00452\\
 & \tiny(0.0086) & \tiny (0.16) & \tiny (0.054) & \tiny (0.27) & \tiny (0.000043) & \tiny (0.000055) & \tiny (0.000087) & \tiny (0.00017) & \tiny (0.00035) & \tiny (0.00044)\\ \hline\hline
			
						\multicolumn{11}{c}{Model \ref{mzar:model:6}} \\ \hline
			& \multicolumn{2}{c|}{$E|\hat{q}-q|$}  &  \multicolumn{2}{c|}{$E(D_H)$}  & \multicolumn{3}{c|}{$E\|\hat{\betab} - \betab\|$}  & \multicolumn{3}{c}{$\frac{\mathrm{MSPE(fitted)}}{\mathrm{MSPE(oracle)}}- 1$}   \\
			Method & AMAR & Fused  & AMAR & Fused & AMAR & Fused & AIC & AMAR & Fused & AIC \\\hline
$T=400$ & 0.407 & 6.87 & 2.3 & 9.26 & 0.0133 & 0.0336 & 0.0372 & 0.023 & 0.0435 & 0.06\\
 & \tiny(0.024) & \tiny (0.097) & \tiny (0.054) & \tiny (0.05) & \tiny (0.00046) & \tiny (0.0018) & \tiny (0.00062) & \tiny (0.0016) & \tiny (0.0027) & \tiny (0.0027)\\
$T=800$ & 0.886 & 9.83 & 3.29 & 13.1 & 0.00902 & 0.0234 & 0.0252 & 0.015 & 0.0311 & 0.0296\\
 & \tiny(0.035) & \tiny (0.13) & \tiny (0.071) & \tiny (0.054) & \tiny (0.00028) & \tiny (0.0017) & \tiny (0.00036) & \tiny (0.00098) & \tiny (0.0025) & \tiny (0.0013)\\
$T=1500$ & 0.455 & 13.8 & 3.08 & 19 & 0.00336 & 0.0174 & 0.0174 & 0.00668 & 0.0241 & 0.0193\\
 & \tiny(0.028) & \tiny (0.17) & \tiny (0.1) & \tiny (0.057) & \tiny (0.00013) & \tiny (0.0016) & \tiny (0.00023) & \tiny (0.00055) & \tiny (0.0023) & \tiny (0.00098)\\
$T=3000$ & 0.642 & 20.9 & 3.52 & 28.8 & 0.00177 & 0.0134 & 0.011 & 0.00395 & 0.0187 & 0.0111\\
 & \tiny(0.037) & \tiny (0.21) & \tiny (0.11) & \tiny (0.063) & \tiny (0.000064) & \tiny (0.0015) & \tiny (0.00012) & \tiny (0.00038) & \tiny (0.0021) & \tiny (0.00068)\\
			\hline\hline
		\end{tabular}
	\end{footnotesize}
	\end{center}
	\caption{\label{Tab:Sim2} Performance of different methods under \ref{mzar:model:4} -- \ref{mzar:model:6}, with estimated errors given in the brackets. Here $\hat{q}$ is the number of the fitted timescales, $D_H$ is the Hausdorff distance between the fitted timescale locations $\{\hat{\tau}_1, \ldots, \hat{\tau}_{\hat{q}}\}$ and the true ones $\{  {\tau}_1, \ldots, {\tau}_{q}\}$, $\|\hat{\betab} - \betab\|$ is the Euclidean distance between the fitted parameter vector and the true one, and MPSE is the mean squared prediction errors of different models.}
\end{table}

\KOMAoptions{fontsize=12pt}
We see that AMAR approach performs consistently better than the fused LASSO for all aspects in all model  settings and with all the sample sizes we consider. In fact, estimates from the fused LASSO do not seem to be consistent in terms of estimating the number and locations of the scales, indicating that the fused LASSO approach (with $L_1$ penalisation) is not appropriate for identifying jumps within the parameter vector. Rather interestingly, the AMAR approach also seems to perform better than the approach based on the AIC in terms of the mean squared prediction errors, illustrating the usefulness and importance of taking into account additional structures in the parameters when they are available.

In Section~\ref{Sec:addnum} of the supplementary materials, we also report results from our sensitivity analysis where we look into the performance of our proposed approach with (i) different choices of $q_\mathrm{max}$ and (ii) a fixed $p$. In addition, We  run  experiments with series simulated from non-stationary AR models with unit roots. To summarise the findings here,  AMAR is generally not sensitive to the choice of $q_\mathrm{max}$ (as long as the truth is no greater). Besides, a fixed $p$ might lead to some very moderate improvement over our current approach of selection via SIC when $T$ is small, but could be problematic when the chosen $p$ is smaller than or close to $\tau_{q}$.  
Finally, even in the setting of non-stationary observations, AMAR still performs much better than its competitors in most settings, though the reported results from all methods are associated with larger standard errors.


\section{Real data examples}
\label{Sec:data}

\subsection{Stock returns}
\begin{figure}[!htbp]
\centering
  \includegraphics[width=1\linewidth]{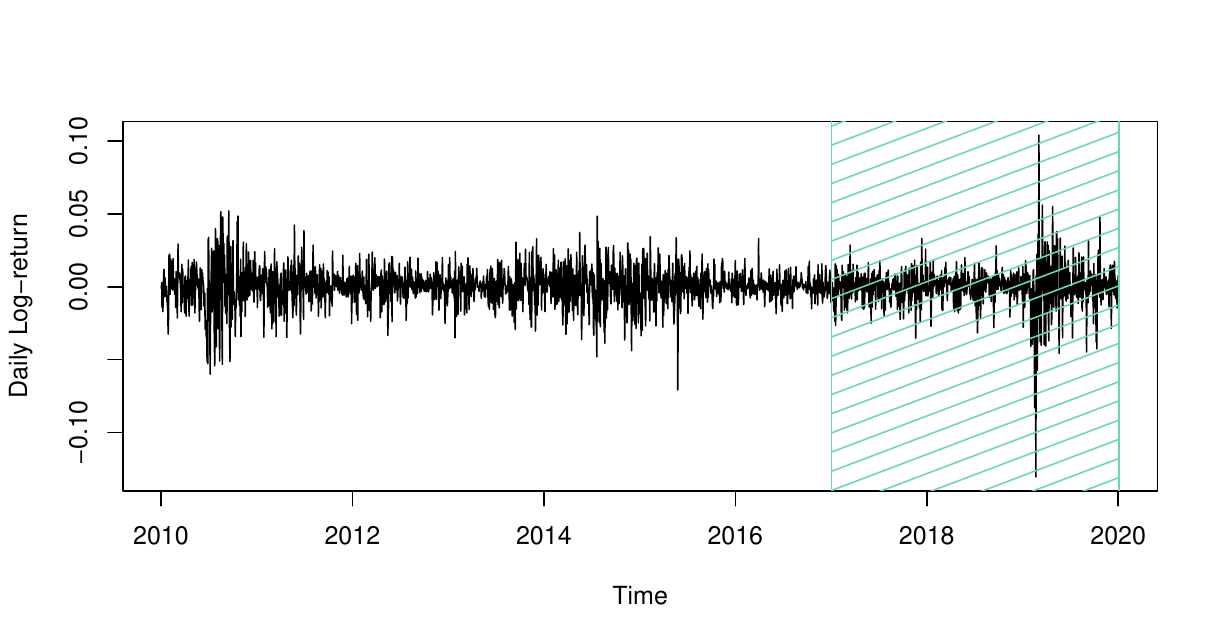}
\caption{\label{fig:DAX}DAX daily log-return from January 2011 to December 2020. The series is divided into two parts for training and testing, with the part for testing highlighted in shade.}
\end{figure}

In this example, we demonstrate the strength of AMAR models for predicting the DAX stock index daily return over the traditional AR. We look at ten years of data from 1 January 2011 to 31 December 2020, with the first seven years of data (70\%) used for training, and the last three years of data (30\%) used for testing. Here we shall work directly on the series of log-return, which we denote as $\{X_t\}$. The visual appearance of the series is illustrated in Figure~\ref{fig:DAX}. 

First, we fit an AMAR model on $\{X_t\}$ using AMAR with the thresholds and $p$ selected automatically via the approach outlined in Section ~\ref{mzar:sec:parameter_choice}. A three-scale AMAR model is selected, with $\tau_1 = 1, \tau_2=5$ and $\tau_3=27$. 

However, for the purpose of interpretability, we note that a two-scale AMAR model might be preferred, with the short scale fixed at $\tau_1=1$ for this particular application. As such, we also fit different AMAR models on only the training data with $\tau_1=1$ and $\tau_2 = \{2,\ldots, 251\}$, and select the corresponding $\tau_2$ that minimises the residual sum of squares. This results in a two-scale AMAR model with  $\tau_2=5$ (and $\tau_1=1$).

Since our focus is on prediction, we also fit the traditional AR model with the order selected by the AIC. This results in an AR(6) model. 

We then examine the performance of these fitted models on the testing data and measure their performance by both the rooted mean squared prediction error (RMSPE) and the hit rate. Here the hit rate is defined as the proportion of time the model predicts the sign of the daily log-return correctly, which is an important performance indicator for financial time series modelling. The results are reported in Table~\ref{Tab:DAX}. In terms of both criteria, AMAR with two scales performs the best. Here AMAR with three scales appears slightly worse, while AR with its order selected by the AIC performs the worst. In addition, we remark that AMAR with $\tau_1=1$ and $\tau_2=5$ can be easily interpreted as having the daily log-return depending on the returns of both the previous trading day and the previous week, a fact that would potentially be appreciated by the practitioners. In summary, we believe that AMAR would be a promising alternative to the traditional AR models in modelling real data of this type. 

\begin{table}[!htpb]
	\begin{center}
			\begin{tabular}{cccc}
				\hline\hline
					Methods & AMAR & AMAR & AR \\
      &  (auto-selection for scales) & (with two scales) & (order via AIC) \\\hline
				RMSPE &  0.014564 & \textbf{0.014521} &  0.014580  \\ 
				 Hit Rate & 0.5013  & \textbf{0.5186} & 0.4775  \\ 
				\hline\hline
			\end{tabular}
	\end{center}
		\caption{\label{Tab:DAX}Performance of different methods in terms of their rooted mean squared prediction error (RMSPE) and hit rate. Results from the better method are highlighted in bold. }
\end{table}

Finally, we note that in this example, more complicated dependence structure, such as heteroscedasticity, has not been taken into account. In principle, the AMAR approach could be extended to the multiscale modelling of both the AR component and the ARCH-type errors.

\subsection{UK and US unemployment data}

\begin{figure}[!htbp]
\centering
  \includegraphics[width=1\linewidth]{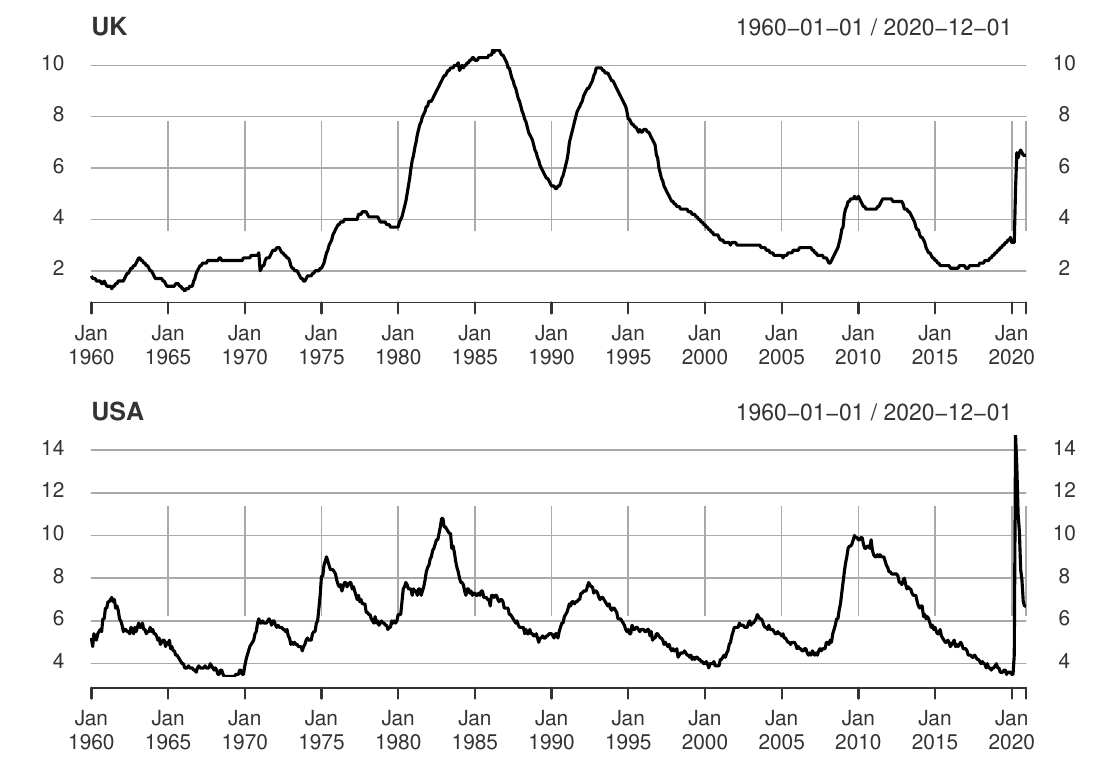}
\caption{\label{fig:unrate}Seasonally adjusted monthly unemployment rates of UK and USA from January 1960 to December 2020.}
\end{figure}

In this example, we first consider the time series of seasonally adjusted UK monthly unemployment rate from 1960 to 2020. The data can be found in \citet{OECD2022}. The series is shown in the top plot of Figure~\ref{fig:unrate}. 
As before, here our aim is not to find models that best fit the data, but to compare AMAR with the AR alternatives normally used in practice, and to demonstrate the potential superiority and practicality of AMAR over other AR approaches. For this analysis, we report our findings on both the original series and the differenced series using AMAR and AR with its order selected via AIC. We also report prediction errors of the different models, where we use the last 5, 10, 20 or 30 years of data for testing and the remaining for training (without specifying the scales or orders a priori). 
We set the maximum AR order to be 48 (i.e. four years) for both methods. For the original series, AMAR fits a model with scales at 1,2 and 3, while AIC selects an AR(13). For the differenced series, AMAR fits a model with scales at 1 and 10, while AIC selects an AR(8). A closer look reveals that the sum of the fitted AR coefficients on the original series is close to one ($>0.99$) for both approaches, reflecting the possibility that the series might not be stationary, while the sum on the differenced series is much smaller (at around 0.7). Looking at the quality of prediction in terms of rooted mean squared errors (at the original series level), we see that AMAR performs better than AR with order selected by AIC in both the original and differenced series in the testing periods of all lengths, though admittedly the difference between these two methods becomes much smaller when considering the differenced series. In fact, for the purpose of prediction, results from Table~\ref{Tab:unrate1}, suggest that it is more appropriate to model the differenced rather than the original series. However, we would like to point out that no matter which series  (i.e. original or differenced) one prefers to work with in this particular example, AMAR always offers better predictive performance than the AR with the order selected using AIC, and possibly also comes with improved interpretability.

\begin{table}
	\begin{center}
			\begin{tabular}{cccccc}
				\hline\hline
				Series & Methods & 5 years & 10 years & 20 years & 30 years \\ \hline
				\multirow{2}{*}{original} & AMAR & \textbf{1.0267} & \textbf{0.7590} & \textbf{0.5849} & \textbf{0.5683}\\
				& AR-AIC & 1.2436 & 0.9452 & 0.7409 & 0.7636\\
				\multirow{2}{*}{differenced} & AMAR & \textbf{0.3359} & \textbf{0.2469} & \textbf{0.1861} & \textbf{0.15295}\\
				& AR-AIC & 0.3753 & 0.2554 & 0.1874 & 0.1557\\
				\hline\hline
			\end{tabular}
	\end{center}
		\caption{\label{Tab:unrate1}Performance of different methods for different testing periods in terms of rooted mean squared prediction errors at the original series level. Here AMAR is the adaptive multiscale autoregression, while AR-AIC is the autoregressive model with order chosen by AIC. Results from the better method are highlighted in bold.}
\end{table}

Next, to demonstrate the potential use of AMAR on multivariate time series, we additionally include the seasonally adjusted US monthly unemployment rate during the same period. The data can be found in \citet{USBLS2022}, with the series also shown in Figure~\ref{fig:unrate}. In view of our previous analysis, we only consider the differenced time series for the purpose of prediction. Let $X_{t,1}$ and  $X_{t,2}$ represent the respective differenced UK and US unemployment rates at time $t$. Then the corresponding Adaptive Multiscale Vector AutoRegressive model (AMVAR) for bivariate observations can be written as 
\begin{align}  
\label{mzar:eq:mzar_model3}
\begin{pmatrix}
X_{t,1}\\X_{t,2}\end{pmatrix}
=\boldsymbol{\alpha}_{1}\begin{pmatrix} \frac{X_{t-1,1}+\ldots+X_{t-\tau_{1},1}}{\tau_{1}} \\ \frac{X_{t-1,2}+\ldots+X_{t-\tau_{1},2}}{\tau_{1}}\end{pmatrix}+\ldots+\boldsymbol{\alpha}_{q}\begin{pmatrix} \frac{X_{t-1,1}+\ldots+X_{t-\tau_{q},1}}{\tau_{q}} \\ \frac{X_{t-1,2}+\ldots+X_{t-\tau_{q},2}}{\tau_{q}}\end{pmatrix}
+\begin{pmatrix}\varepsilon_{t,1}\\\varepsilon_{t,2}\end{pmatrix}
\end{align}
where $\tau_1,\ldots, \tau_q$ are the scales, $\boldsymbol{\alpha}_{1},\ldots,\boldsymbol{\alpha}_{q}$ are $2\times2$ matrices, and where $\boldsymbol{\varepsilon}_{t} = (\varepsilon_{t,1},\varepsilon_{t,2})^T$ are noise vectors. 
The optimal selection of the number of scales and their locations for AMVAR is beyond the scope this paper. One simple approach would be to perform scale selection for each univariate series, and combine all of them to be then used as the scales for AMVAR. 
Though there might be limitations in this approach, we believe it serves as a good starting point for further exploration, as demonstrated below. Recall that AMAR selects scales of 1 and 10 for the differenced UK unemployment series. In addition, AMAR selects a single scale of 11 for the differenced US unemployment series. Consequently, we use scales at 1, 10 and 11 for AMVAR. To facilitate comparison with our previous analysis, we focus on the differenced UK unemployment series, which we now explicit model as
\begin{align*}
X_{t,1} = \alpha_{1,(1,1)} X_{t-1,1} &+ \alpha_{1,(1,2)} X_{t-1,2} 
 + \alpha_{2,(1,1)} \frac{X_{t-1,1}+\ldots+X_{t-10,1}}{10}
+ \alpha_{2,(1,2)} \frac{X_{t-1,2}+\ldots+X_{t-10,2}}{10} \\
& + \alpha_{3,(1,1)} \frac{X_{t-1,1}+\ldots+X_{t-11,1}}{11}
+ \alpha_{3,(1,2)} \frac{X_{t-1,2}+\ldots+X_{t-11,2}}{11} 
+ \varepsilon_{t,1}.
\end{align*}

On the other hand, fitting the data by the Vector Autoregressive (VAR) models and selecting order via AIC leads to a VAR(4). The rooted squared errors for prediction of different models are reported in Table~\ref{Tab:unrate2}, where we use the last five, ten, twenty or thirty years of data for testing and the remaining for training.
\begin{table}
	\begin{center}
			\begin{tabular}{cccccc}
				\hline\hline
					Series & Methods & 5 years & 10 years & 20 years & 30 years \\ \hline
				\multirow{3}{*}{differenced} & AMVAR & \textbf{0.3023} & \textbf{0.2176} & \textbf{0.1682} & \textbf{0.1438}\\
				& VAR-AIC & 0.4383 & 0.3136 &  0.2271 & 0.1912\\ \cline{2-6}
				& AMAR & 0.3359 &  0.2469 &  0.1861 & 0.15295\\ 
				\hline\hline
			\end{tabular}
	\end{center}
		\caption{\label{Tab:unrate2}Performance of different methods for different testing periods in terms of rooted mean squared prediction errors. Here AMVAR is the adaptive multiscale vector autoregression, while VAR-AIC is the vector autoregressive model with order chosen by AIC, both under the bivariate setting. We also include (univariate) AMAR for comparison. Results from the better method are highlighted in bold. }
\end{table}
Our results suggest that for the purpose of prediction, AMVAR performs better than VAR with order selected using AIC for various testing periods in this example under the multivariate setting. Importantly, this appears to be the case even without fine tuning the scale selection procedure. In addition, comparing results in Table~\ref{Tab:unrate1} and Table~\ref{Tab:unrate2}, we note that AMVAR performs better than the univariate AMAR for UK unemployment, indicating that including an extra regressor from a different time series (i.e. US unemployment rate) does indeed improve the predictive power.

\section{Extensions and further discussions}
\label{sec:disc}

The AMAR estimation algorithm can also be used in large-order autoregressions in which the AR coefficients may not necessarily be
piecewise constant, but possess a different type of regularity (e.g. be a piecewise polynomial of a higher degree). As an example, we could consider features that are linearly-weighted averages, instead of the simple averages in \eqref{mzar:eq:mzar_model} for AMAR($q$). To give more details, for some $1 \le \tau_1 < \ldots < \tau_q $, the model is given as
\begin{align*}  
X_{t}=\alpha_{1}\frac{\tau_1 X_{t-1}+ (\tau_1-1) X_{t-2}+\ldots+X_{t-\tau_{1}}}{\tau_1(1+\tau_{1})/2}+\cdots+\alpha_{q}\frac{\tau_q X_{t-1}+ (\tau_q-1) X_{t-2}+\ldots+X_{t-\tau_{q}}}{\tau_q(1+\tau_{q})/2}+\varepsilon_{t}
\end{align*}
Here the influence of the past observations on any given feature decays as the time gap between them and the present widens. The linear decaying form is just one possible way of modelling that results in a more parsimonious parameter structure  of the AR. The other appealing reason for linear decaying is that $\tau_1,\ldots, \tau_q$ can also be estimated using the previous framework, say, by simply changing the contrast function in the NOT algorithm from piecewise-constant contrast to piecewise-linear and continuous contrast.

As briefly illustrated in our second real data example, another interesting venue for applying the AMAR framework is time series data with multivariate, or even high-dimensional observations. In particular, in the high-dimensional setting, instead of applying the same averages (or features) across different components of $(X_{t,1},\ldots,X_{t,p})^T$, some group structures (or even factors) within these components can be introduced to further enhance its interpretability.

Finally, it would also be of interest to investigate how estimation uncertainty could be quantified in AMAR (e.g. via block bootstrap), and whether the AMAR philosophy could be extended to multiscale features of some latent or hidden observations. One such example would be to consider multiscale autoregressive and moving average (ARMA) models, in which there are different scales on both the observed time series and the unobserved innovations, i.e. we observe $\{X_t\}$ following
\begin{align*}  
X_{t}=&\alpha_{1}\frac{X_{t-1}+\ldots+X_{t-\tau_{1}}}{\tau_{1}}+\ldots+\alpha_{q_\mathrm{AR}}\frac{X_{t-1}+\ldots+X_{t-\tau_{q_\mathrm{AR}}}}{\tau_{q_\mathrm{AR}}}\\\notag
&+\varepsilon_{t}
+\beta_{1}\frac{\varepsilon_{t-1}+\ldots+\varepsilon_{t-\rho_{1}}}{\rho_{1}}+\ldots+\beta_{q_\mathrm{MA}}\frac{\varepsilon_{t-1}+\ldots+\varepsilon_{t-\rho_{{q}_\mathrm{MA}}}}{\rho_{{q}_\mathrm{MA}}}
\end{align*}
where ${q}_\mathrm{AR}$ and ${q}_\mathrm{MA}$ are, respectively, the numbers of timescales for AR and  MA, with the AR timescales $1\leq\tau_{1}<\ldots<\tau_{q_\mathrm{AR}}$ and the MA timescales $1\leq \rho_{1}<\ldots<\rho_{q_\mathrm{MA}}$, and with $\alpha_1,\ldots, \alpha_{q_\mathrm{AR}}$, $ \beta_{1},\ldots, \beta_{q_\mathrm{MA}}$ being the scale coefficients.

\section*{Acknowledgements}
We are extremely grateful to the associate editor and three anonymous reviewers for their valuable comments and suggestions, which have helped improve the manuscript.

\newpage
\appendix

\section*{\large{Supplementary materials}}
This part contains (\ref{sec:specialcases}) discussions and illustrations on some special cases of AMAR, (\ref{Sec:addnum}) further simulations, (\ref{Sec:adddata}) an additional real data example, and (\ref{Sec:proof}) proofs of the theoretical results.

\section{Special cases of AMAR}
\label{sec:specialcases}
We now consider some special cases of AMAR, and offer visual insights into their behaviour.
\subsection{Special case I: a single scale}
\label{sec:onescale}

Let $\{X_t\}$ be a series following the AMAR model with a single scale, i.e.
\begin{equation}
	\label{eq:amar1}
	X_t = \alpha_1 \frac{X_{t-1} + \ldots + X_{t-\tau_1}}{\tau_1} + \varepsilon_t, \quad t = 1, \ldots, T.
\end{equation}

Recall that realisations for different values of $\alpha_1$ (from 0.5 to 0.95, the latter corresponds to series that are near unit-root) and $\tau_1$ (from 1 to 10) with standard Gaussian noise are plotted in Figure~\ref{fig:amar_1} of the main paper. It appears that the longer the scale, the noisier the appearance; the overall shape (driven by the low frequencies) is preserved, but the details (driven by the high frequencies) are increasingly obscured by noise. This behaviour can also be understood by considering the spectral properties of the single-scale AMAR model, where
the fact that the corresponding AR coefficients in the single-scale AMAR model (\ref{eq:amar1}) are constant provides a useful simplification in the form of the spectral density. With $\varepsilon_t$ being white noise with unit variance, the spectral density of $X_t$ given by (\ref{eq:amar1}) is
\begin{equation}
	S_X(f) = \left|  1 - \frac{\alpha_1}{\tau_1} e^{-2\pi f i} \frac{1 - e^{-2\pi f \tau_1 i}}{1 - e^{-2\pi f i}}  \right|^{-2},\quad |f| < \frac{1}{2}.
\end{equation}
In view of the boundedness of $e^{-2\pi f i} \frac{1 - e^{-2\pi f \tau_1 i}}{1 - e^{-2\pi f i}}$ as a function of $\tau_1$, we have that $S_X(f) \to 1$ as $\tau_1 \to \infty$, for all $f \in (0, 1/2)$. However, as the sum of the AR coefficients in the single-scale AMAR does not depend on $\tau_1$, we have $S_X(0) = |1 - \alpha_1|^{-2}$. Note that more generally, given $\alpha_1,\ldots,\alpha_q$, the spectral density at zero of any AMAR($q$) process, i.e. its long-run variance, is independent of $\tau_1, \ldots, \tau_q$.

\begin{figure}
	\centering
	\includegraphics[width=0.3\linewidth]{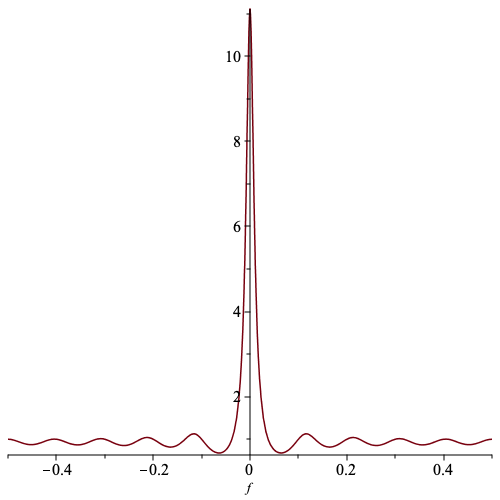}
	\caption{Spectral density of a single-scale AMAR process with $\tau_1 = 10$ and $\alpha_1 = 0.7$. \label{fig:amar_specdens}}
\end{figure}

As a visual illustration, Figure \ref{fig:amar_specdens} shows the spectral density of a single-scale AMAR process with $\tau_1 = 10$ and $\alpha_1 = 0.7$. Due to the limiting behaviour described above, a single-scale
AMAR for a large $\tau_1$ can be approximated as the sum of two independent processes: one band-limited with a sharp peak at zero (and therefore representing a ``slowly-varying'' signal), and the other as white noise. This is in agreement with the appearance of the sample realisations shown in Figure \ref{fig:amar_1}, which begin to resemble a ``signal + white noise'' model for the larger values of $\tau_1$. 

Finally, we note that even though all the series plotted in Figure~\ref{fig:amar_1} are weakly stationary, some of them exhibit behaviour that mimics non-stationarity, at least visually, when $\tau_1$ if large, even for a moderate $\alpha_1$. This hints at the usefulness of AMAR in the modelling of near unit-root or certain non-stationary series. More details can be found in a  simulation study in Section~\ref{Sec:addnum}.


\subsection{Special case II: one short plus one long scale}

We now study the case of the AMAR model in which two timescales are present: one short one, and one long one. We have
\begin{equation}
	\label{eq:amar2}
	X_t = \alpha_1 \frac{X_{t-1} + \ldots + X_{t-\tau_1}}{\tau_1} + \alpha_2 \frac{X_{t-1} + \ldots + X_{t-\tau_2}}{\tau_2} + \varepsilon_t.
\end{equation}

First, if we keep $\alpha_1 + \alpha_2$ constant, and vary both coefficients from $\alpha_1 = 0$ on one extreme to $\alpha_2 = 0$ on the other
extreme, then we obtain a ``smooth transition'' from a single-scale model with scale $\tau_2$ to a single-scale model with scale $\tau_1$.

\begin{figure}[!htbp]
	\centering
	\begin{minipage}{.33\textwidth}
		\centering
		\includegraphics[width=\linewidth]{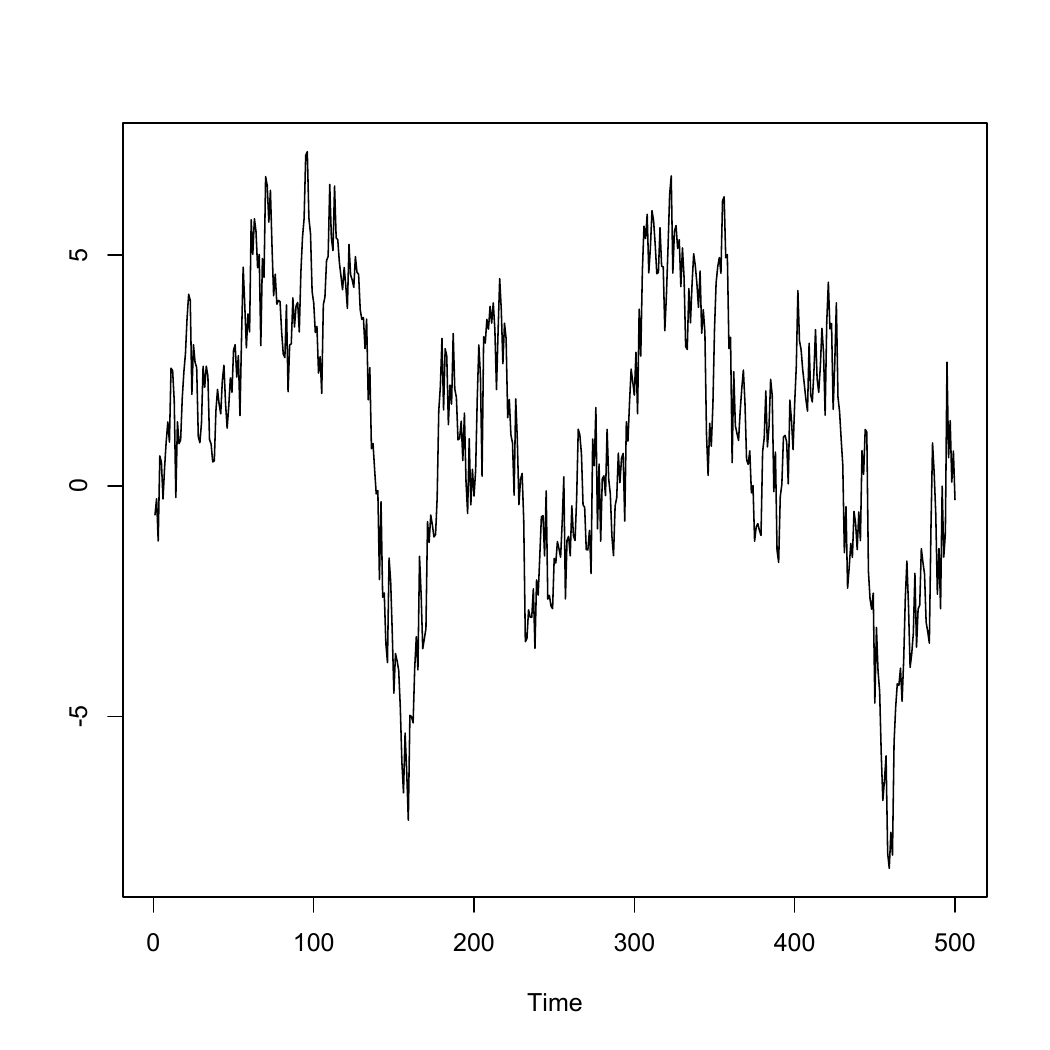}
	\end{minipage}%
	\begin{minipage}{.33\textwidth}
		\centering
		\includegraphics[width=\linewidth]{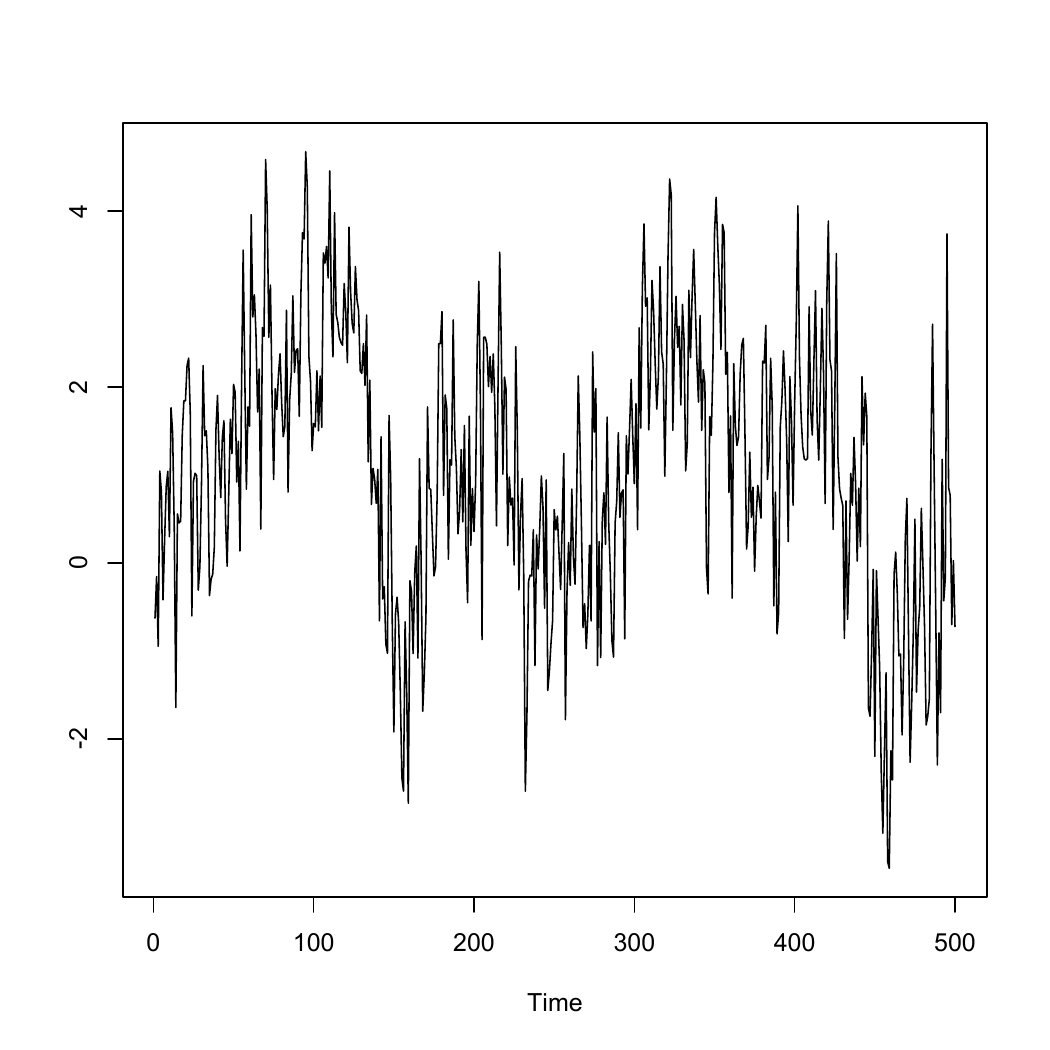}
	\end{minipage}%
	\begin{minipage}{.33\textwidth}
		\centering
		\includegraphics[width=\linewidth]{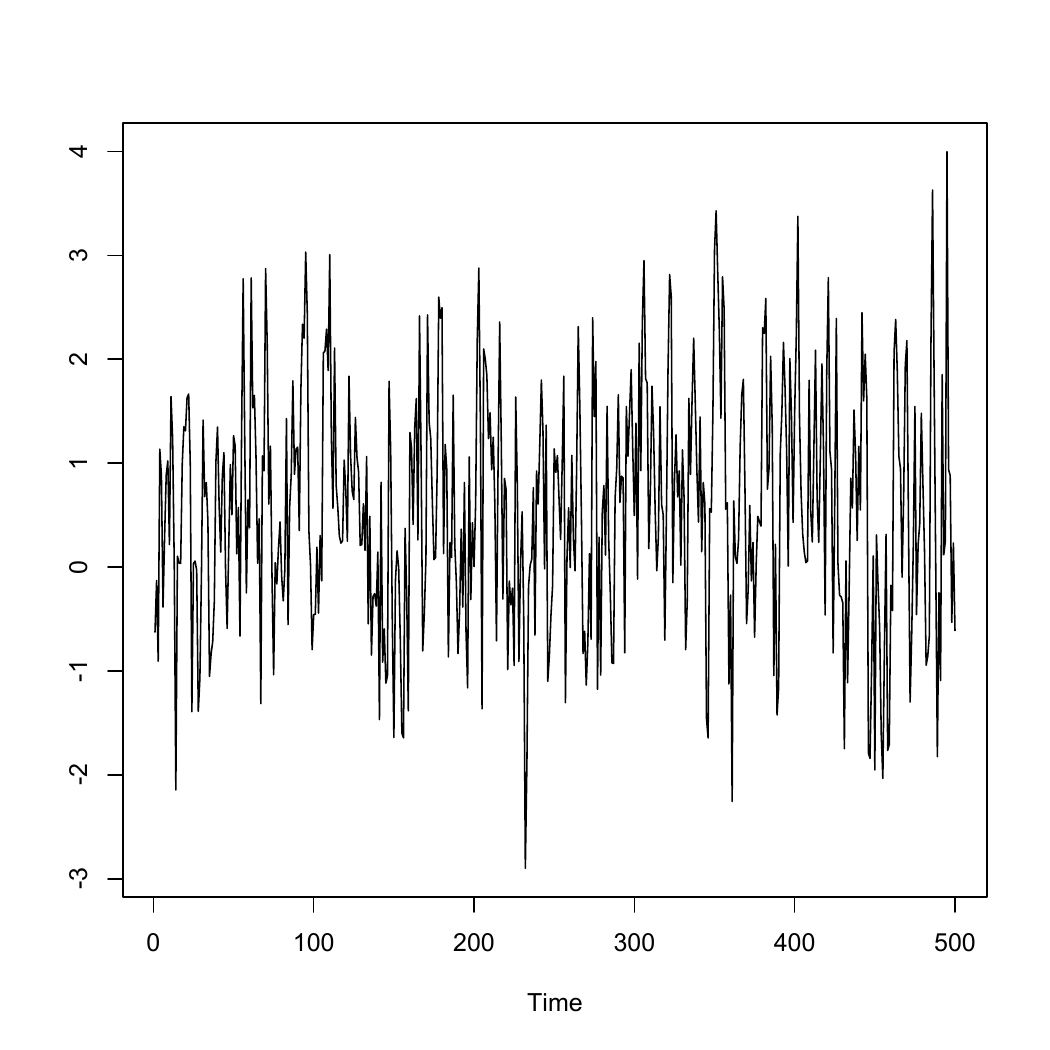}
	\end{minipage}%
	\\
	\centering
	\begin{minipage}{.33\textwidth}
		\centering
		\includegraphics[width=\linewidth]{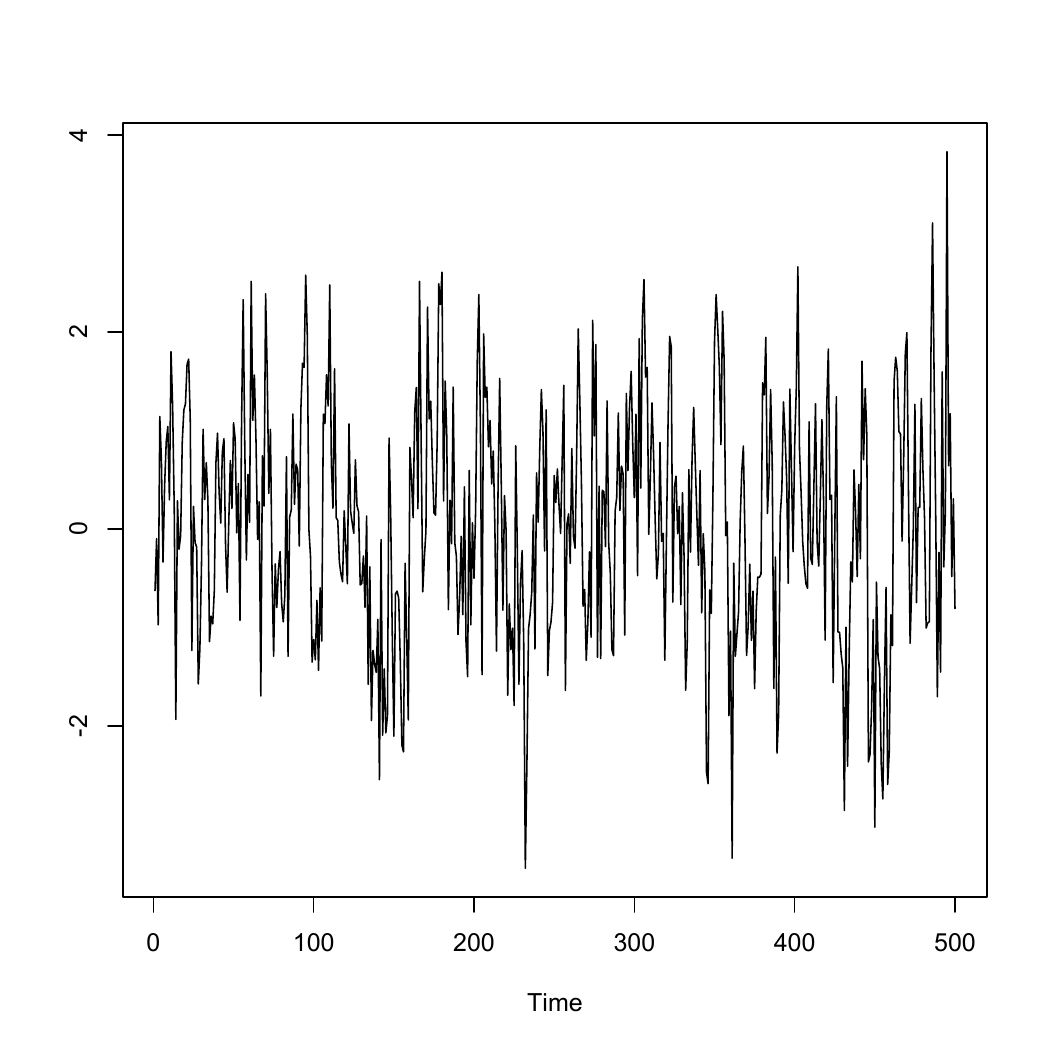}
	\end{minipage}%
	\begin{minipage}{.33\textwidth}
		\centering
		\includegraphics[width=\linewidth]{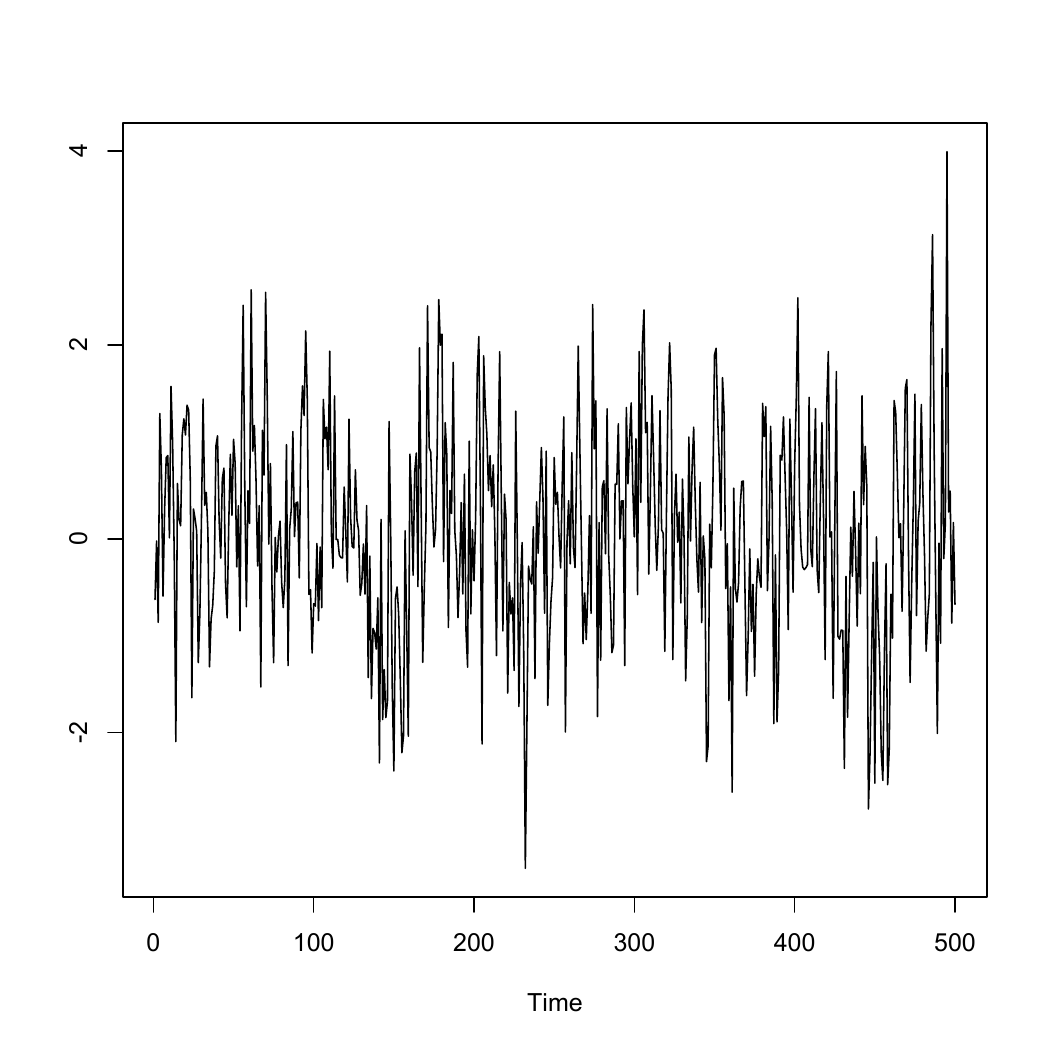}
	\end{minipage}%
	\begin{minipage}{.33\textwidth}
		\centering
		\includegraphics[width=\linewidth]{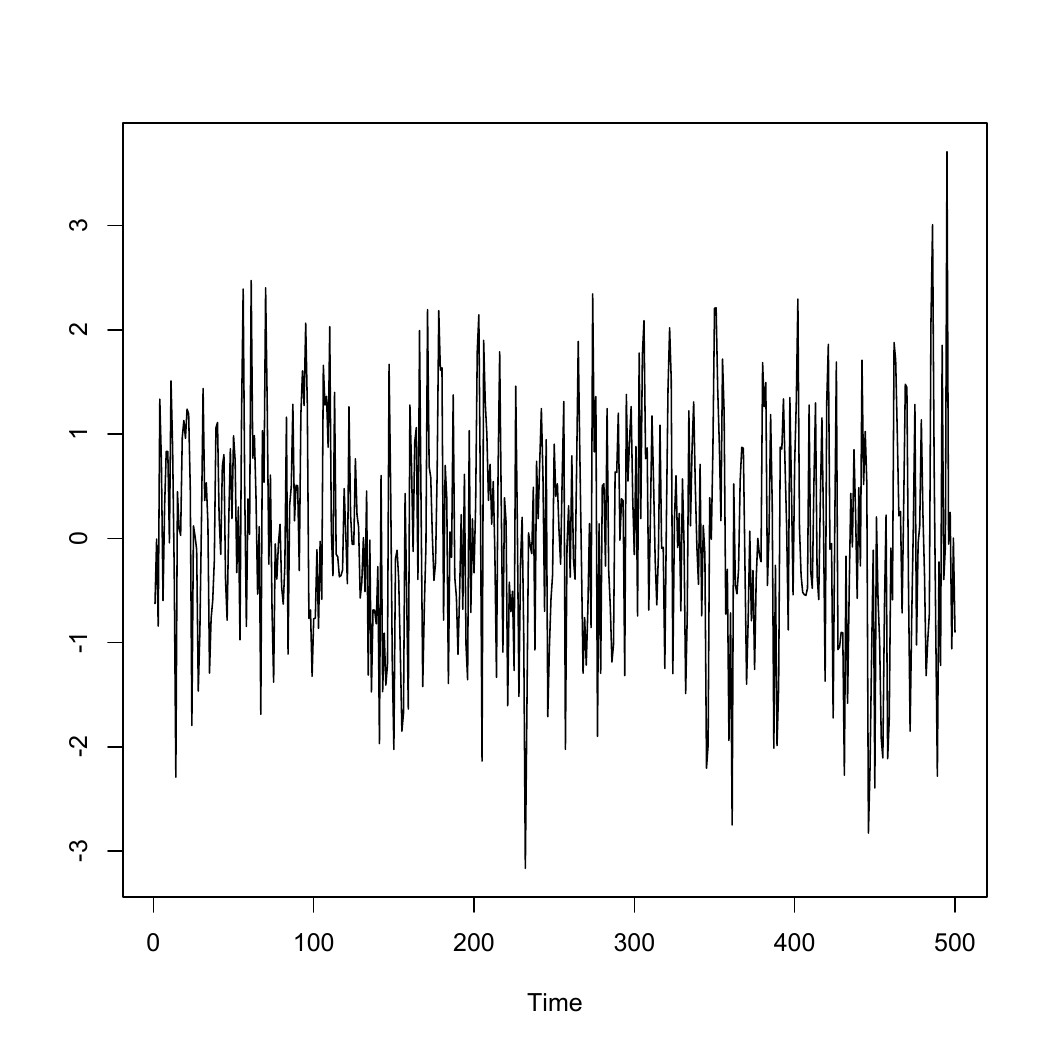}
	\end{minipage}%
	\caption{Simulated sample paths, of length 500, from the two-scale AMAR model (\ref{eq:amar2}), with $\alpha_1 = \alpha_2 = \alpha = 0.49$ in the first row and 
		$\alpha_1 = \alpha_2 = \alpha = 0.3$ in the second. Here  $\tau_1 = 1$ and $\tau_2 = 2, 10, 50$ (respectively from left to right). The same random seed is used to generate path for each row. \label{fig:amar_2}}
\end{figure}

To gain further insight into the behaviour of AMAR with two timescales, now we consider $\alpha_1 = \alpha_2 = \alpha$, take $\tau_1 = 1$ and vary $\tau_2$. Figure \ref{fig:amar_2} illustrates the case in which $\alpha_1 = \alpha_2 = \alpha = 0.49$, $\tau_1 = 1$ and $\tau_2 = 2, 10, 50$. When $\tau_2 = 50$, the longer scale has visually and practically no impact as the coefficients for the individual components (i.e. $\alpha_2/\tau_2$) are small. When $\tau_2 = 2$, we have a simple AR(2) model. On the other hand, when $\tau_2 = 10$, the realisation has the visual appearance of ``a time-varying trend plus a low-order AR model''.  Here, the longer scale is responsible for the changing ``levels'' at low-frequencies, whereas the shorter scale is responsible for instantaneous fluctuations at high-frequencies. This phenomenon is visually not present if $\sum_i \alpha_i$ is small or moderate, e.g. $\alpha = \alpha_1 = \alpha_2 = 0.3$, as demonstrated in Figure \ref{fig:amar_2}, but would show up if $\sum_i \alpha_i$ gets moderately close to 1, e.g. at around 0.8.

Besides, models with a large $\tau_2$ (for different $\alpha_i$'s) also display the same interesting feature (i.e. ``trend + noise'' type, which might visually appear to be non-stationary), but it seems that the longer the scale $\tau_2$, the larger the sample sizes at which we are able to observe this phenomenon. 

From these findings, we would infer that a two-scale AMAR model (with a small $\tau_1$) is perhaps the most useful if (a) the longer scale is not too short or too long (i.e. in the order of 10s in practice), and (b) when the sum of the coefficients  $\alpha_1+\alpha_2$ is moderately close to 1 (say $> 0.8$), with the coefficient $\alpha_2$ of the longer scale not being too small. In this case the two-scale AMAR can imitate a time-varying trend plus a low order AR model, i.e. we are in a situation in which we are able to use a stationary AMAR model to model certain non-stationary-looking phenomena.

\subsection{Special case III: AMAR representation of seasonal models}

In the class of seasonal ARIMA$(p,0,q)\times(P,0,Q)_S$ models, we consider models of the form $\Phi(B^S) \phi(B) X_t = \varepsilon_t$ (i.e. $q = Q = 0$), where
\begin{eqnarray*}
	\Phi(B^S) & = & 1 - \Phi_1 B^S - \ldots - \Phi_P B^{PS}\\
	\phi(B) & = & 1 - \phi_1 B - \ldots - \phi_p B^p,
\end{eqnarray*}
and where $B$ is the lag operator. They belong to the class of AMAR models.
As a simple example, consider ARIMA$(1,0,0)\times(1,0,0)_{12}$, an autoregressive model for monthly time series, with a single non-seasonal lag and yearly seasonality, given by
\begin{equation}
	\label{eq:ss}
	X_t = \phi_1 X_{t-1} + \Phi_1 X_{t-12} - \phi_1 \Phi_1 X_{t-13} + \varepsilon_t.
\end{equation}
A typical characteristic feature of ARIMA$(p,0,q)\times(P,0,Q)_S$ models is its stretches of consecutive zero AR coefficients. For example, in (\ref{eq:ss}), the AR coefficients corresponding to lags 2 to 11 are zero. This means that AMAR models are also able to provide a relatively parsimonious representation of ARIMA$(p,0,q)\times(P,0,Q)_S$ models. As an example, model (\ref{eq:ss}), represented in the AMAR framework, will need four scale parameters (at $\tau_1 = 1$, $\tau_2 = 11$, $\tau_3 = 12$ and $\tau_4 = 13$) and four corresponding AMAR coefficients (i.e. $\alpha_1 = \phi_1$, $\alpha_2 = -11\Phi_1$, $\alpha_3 = 12\Phi_1(1+\phi_1)$ and $\alpha_4 = -13\phi_1\Phi_1$), which is more heavily parameterised than the \emph{optimal} seasonal representation (\ref{eq:ss}) (with  $\phi_1$ and $\Phi_1$) but much less than the full AR representation of (\ref{eq:ss}). 

This (relative) parsimony of representation of seasonal models in the AMAR framework, plus the fact that the AMAR estimation framework is able to estimate the number of timescales and their spans automatically, makes AMAR a viable exploratory tool for identifying time series seasonality in data. 
In fact, we have demonstrated in the simulation study in Section~\ref{mzar:sec:simulation_study} of the main paper that the AMAR estimation procedure is capable of identifying the right timescales rather effectively even with relatively small number of observations, confirming good potential of AMAR for the identification and exploratory analysis of seasonal models.

\section{Additional numerical experiments}
\label{Sec:addnum}

\subsection{Sensitivity analysis}

Several tuning parameters are required in the algorithm of our approach. The notable ones are the maximum number of scales $q_{\mathrm{max}}$, and the autoregressive order $p$ used in the initial step. Besides, the choice of number of intervals $M$ would also be required, but it should be apparent from our algorithm that it only plays a minor role under a large $p$ (which, in the setup of our current algorithm, would imply $T > 250000$). 

Based on our experiments, we find that the proposed approach is not too sensitive to the choice of all the aforementioned tuning parameters. Detailed results are given below.

\subsubsection{Maximum number of timescales -- \texorpdfstring{$q_{\mathrm{max}}$}{qmax}}

Here we run the same experiments listed in the main manuscript, but set $q_{\mathrm{max}} = 5, 20$. The same evaluation metrics are used. Results are given in Table~\ref{Tab:Sim3} and Table~\ref{Tab:Sim4}.

By comparing the results with those from Table~\ref{Tab:Sim1} and Table~\ref{Tab:Sim2} in the main manuscript (where by default $q_{\mathrm{max}} = 10$), it becomes evident that our approach does not appear to be sensitive to the choice of $q_{\mathrm{max}}$. In particular, for different choices of $q_{\mathrm{max}}$, every corresponding AMAR performs better than the competitors.

\begin{table}[!htbp]
	
	\begin{center}
		\begin{footnotesize}
			\begin{tabular}{c|cc|cc|cc|cc}
				\hline\hline
				\multicolumn{9}{c}{Model \ref{mzar:model:1}} \\ \hline
				& \multicolumn{2}{c|}{$E|\hat{q}-q|$}  &  \multicolumn{2}{c|}{$E(D_H)$}  & \multicolumn{2}{c|}{$E\|\hat{\betab} - \betab\|$}  & \multicolumn{2}{c}{$\frac{\mathrm{MSPE(fitted)}}{\mathrm{MSPE(oracle)}}- 1$}   \\
				$q_\mathrm{max}$ & 5 & 20  & 5 & 20 & 5 & 20 & 5 & 20 \\\hline
				$T=400$ & 0.164 & 0.17 & 0.539 & 0.589 & 0.016 & 0.0162 & 0.018 & 0.0134\\
				& \tiny(0.013) & \tiny (0.013) & \tiny (0.043) & \tiny (0.046) & \tiny (0.00086) & \tiny (0.00082) & \tiny (0.0052) & \tiny (0.00092)\\
				$T=800$ & 0.051 & 0.051 & 0.187 & 0.206 & 0.00351 & 0.00385 & 0.00446 & 0.00469\\
				& \tiny(0.0072) & \tiny (0.0074) & \tiny (0.032) & \tiny (0.034) & \tiny (0.00026) & \tiny (0.00036) & \tiny (0.00049) & \tiny (0.00054)\\
				$T=1500$ & 0.022 & 0.021 & 0.143 & 0.117 & 0.00116 & 0.00117 & 0.00138 & 0.00145\\
				& \tiny(0.0046) & \tiny (0.0045) & \tiny (0.045) & \tiny (0.043) & \tiny (0.000088) & \tiny (0.000088) & \tiny (0.00024) & \tiny (0.00024)\\
				$T=3000$ & 0.01 & 0.011 & 0.021 & 0.049 & 0.000546 & 0.000549 & 0.000671 & 0.000685\\
				& \tiny(0.0031) & \tiny (0.0033) & \tiny (0.0083) & \tiny (0.029) & \tiny (0.000027) & \tiny (0.000027) & \tiny (0.00017) & \tiny (0.00017)\\
				\hline\hline
				
				\multicolumn{9}{c}{Model \ref{mzar:model:2}} \\ \hline
				& \multicolumn{2}{c|}{$E|\hat{q}-q|$}  &  \multicolumn{2}{c|}{$E(D_H)$}  & \multicolumn{2}{c|}{$E\|\hat{\betab} - \betab\|$}  & \multicolumn{2}{c}{$\frac{\mathrm{MSPE(fitted)}}{\mathrm{MSPE(oracle)}}- 1$}   \\
				$q_\mathrm{max}$ & 5 & 20  & 5 & 20 & 5 & 20 & 5 & 20 \\\hline
				$T=400$ & 0.251 & 0.289 & 1.07 & 1.27 & 0.0207 & 0.0206 & 0.0187 & 0.0243\\
				& \tiny(0.017) & \tiny (0.017) & \tiny (0.064) & \tiny (0.071) & \tiny (0.0015) & \tiny (0.0014) & \tiny (0.002) & \tiny (0.0049)\\
				$T=800$ & 0.134 & 0.149 & 0.463 & 0.538 & 0.00551 & 0.00574 & 0.00649 & 0.00911\\
				& \tiny(0.011) & \tiny (0.012) & \tiny (0.044) & \tiny (0.049) & \tiny (0.0006) & \tiny (0.00059) & \tiny (0.0011) & \tiny (0.0013)\\
				$T=1500$ & 0.125 & 0.136 & 1.18 & 1.26 & 0.00152 & 0.00142 & 0.00234 & 0.0025\\
				& \tiny(0.011) & \tiny (0.011) & \tiny (0.13) & \tiny (0.13) & \tiny (0.00029) & \tiny (0.00026) & \tiny (0.00036) & \tiny (0.00039)\\
				$T=3000$ & 0.064 & 0.069 & 0.673 & 0.663 & 0.000159 & 0.000181 & 0.000776 & 0.00118\\
				& \tiny(0.008) & \tiny (0.008) & \tiny (0.1) & \tiny (0.1) & \tiny (0.00011) & \tiny (0.000074) & \tiny (0.0002) & \tiny (0.00034)\\
				\hline\hline
				\multicolumn{9}{c}{Model \ref{mzar:model:3}} \\ \hline
				& \multicolumn{2}{c|}{$E|\hat{q}-q|$}  &  \multicolumn{2}{c|}{$E(D_H)$}  & \multicolumn{2}{c|}{$E\|\hat{\betab} - \betab\|$}  & \multicolumn{2}{c}{$\frac{\mathrm{MSPE(fitted)}}{\mathrm{MSPE(oracle)}}- 1$}   \\
				$q_\mathrm{max}$ & 5 & 20  & 5 & 20 & 5 & 20 & 5 & 20 \\\hline
				$T=400$ & 0.511 & 0.706 & 1.46 & 1.35 & 0.0238 & 0.0209 & 0.0355 & 0.0289\\
				& \tiny(0.023) & \tiny (0.035) & \tiny (0.054) & \tiny (0.046) & \tiny (0.00094) & \tiny (0.00075) & \tiny (0.0021) & \tiny (0.0016)\\
				$T=800$ & 0.262 & 0.344 & 0.631 & 0.64 & 0.00756 & 0.00701 & 0.0103 & 0.00918\\
				& \tiny(0.018) & \tiny (0.026) & \tiny (0.034) & \tiny (0.034) & \tiny (0.00039) & \tiny (0.00031) & \tiny (0.00088) & \tiny (0.00075)\\
				$T=1500$ & 0.068 & 0.078 & 0.285 & 0.297 & 0.00197 & 0.00201 & 0.00341 & 0.00343\\
				& \tiny(0.0089) & \tiny (0.011) & \tiny (0.04) & \tiny (0.042) & \tiny (0.0001) & \tiny (0.00011) & \tiny (0.00039) & \tiny (0.0004)\\
				$T=3000$ & 0.052 & 0.054 & 0.192 & 0.196 & 0.000677 & 0.000671 & 0.00152 & 0.00151\\
				& \tiny(0.0078) & \tiny (0.0082) & \tiny (0.04) & \tiny (0.04) & \tiny (0.000042) & \tiny (0.000041) & \tiny (0.00023) & \tiny (0.00023)\\
				\hline\hline
			\end{tabular}
		\end{footnotesize}
	\end{center}
	\caption{\label{Tab:Sim3} Performance of AMAR using different $q_\mathrm{max}$ under \ref{mzar:model:1} -- \ref{mzar:model:3}, with estimated errors given in the brackets. Here $\hat{q}$ is the number of the fitted timescales, $D_H$ is the Hausdorff distance between the fitted timescale locations $\{\hat{\tau}_1, \ldots, \hat{\tau}_{\hat{q}}\}$ and the true ones $\{  {\tau}_1, \ldots, {\tau}_{q}\}$, $\|\hat{\betab} - \betab\|$ is the Euclidean distance between the fitted parameter vector and the true one, and MPSE is the mean squared prediction errors of different models.}
\end{table}

\begin{table}[!htbp]
	
	\begin{center}
		\begin{footnotesize}
			\begin{tabular}{c|cc|cc|cc|cc}
				\hline\hline
				\multicolumn{9}{c}{Model \ref{mzar:model:4}} \\ \hline
				& \multicolumn{2}{c|}{$E|\hat{q}-q|$}  &  \multicolumn{2}{c|}{$E(D_H)$}  & \multicolumn{2}{c|}{$E\|\hat{\betab} - \betab\|$}  & \multicolumn{2}{c}{$\frac{\mathrm{MSPE(fitted)}}{\mathrm{MSPE(oracle)}}- 1$}   \\
				$q_\mathrm{max}$ & 5 & 20  & 5 & 20 & 5 & 20 & 5 & 20 \\\hline
				$T=400$ & 0.065 & 0.106 & 0.154 & 0.252 & 0.0104 & 0.00932 & 0.015 & 0.0145\\
				& \tiny(0.0079) & \tiny (0.013) & \tiny (0.022) & \tiny (0.031) & \tiny (0.00085) & \tiny (0.00061) & \tiny (0.0011) & \tiny (0.001)\\
				$T=800$ & 0.041 & 0.061 & 0.129 & 0.131 & 0.00399 & 0.00489 & 0.00627 & 0.00722\\
				& \tiny(0.0063) & \tiny (0.0098) & \tiny (0.024) & \tiny (0.022) & \tiny (0.0004) & \tiny (0.00042) & \tiny (0.00057) & \tiny (0.00063)\\
				$T=1500$ & 0.041 & 0.035 & 0.274 & 0.202 & 0.0018 & 0.00193 & 0.00313 & 0.00352\\
				& \tiny(0.0063) & \tiny (0.006) & \tiny (0.053) & \tiny (0.046) & \tiny (0.0001) & \tiny (0.00025) & \tiny (0.0004) & \tiny (0.00056)\\
				$T=3000$ & 0.015 & 0.023 & 0.112 & 0.128 & 0.000753 & 0.000752 & 0.0017 & 0.00173\\
				& \tiny(0.0038) & \tiny (0.0049) & \tiny (0.037) & \tiny (0.036) & \tiny (0.000023) & \tiny (0.000023) & \tiny (0.00024) & \tiny (0.00024)\\
				\hline\hline
				
				\multicolumn{9}{c}{Model \ref{mzar:model:5}} \\ \hline
				& \multicolumn{2}{c|}{$E|\hat{q}-q|$}  &  \multicolumn{2}{c|}{$E(D_H)$}  & \multicolumn{2}{c|}{$E\|\hat{\betab} - \betab\|$}  & \multicolumn{2}{c}{$\frac{\mathrm{MSPE(fitted)}}{\mathrm{MSPE(oracle)}}- 1$}   \\
				$q_\mathrm{max}$ & 5 & 20  & 5 & 20 & 5 & 20 & 5 & 20 \\\hline
				$T=400$ & 0.211 & 0.222 & 1.63 & 1.66 & 0.0107 & 0.011 & 0.0129 & 0.0143\\
				& \tiny(0.017) & \tiny (0.018) & \tiny (0.072) & \tiny (0.073) & \tiny (0.00043) & \tiny (0.00045) & \tiny (0.00092) & \tiny (0.0016)\\
				$T=800$ & 0.137 & 0.137 & 0.86 & 0.859 & 0.0042 & 0.00421 & 0.00535 & 0.00533\\
				& \tiny(0.013) & \tiny (0.013) & \tiny (0.055) & \tiny (0.055) & \tiny (0.00023) & \tiny (0.00022) & \tiny (0.00056) & \tiny (0.00056)\\
				$T=1500$ & 0.101 & 0.104 & 0.729 & 0.736 & 0.00167 & 0.00171 & 0.0023 & 0.00223\\
				& \tiny(0.012) & \tiny (0.013) & \tiny (0.078) & \tiny (0.078) & \tiny (0.00011) & \tiny (0.00012) & \tiny (0.00034) & \tiny (0.00034)\\
				$T=3000$ & 0.052 & 0.052 & 0.327 & 0.336 & 0.000343 & 0.00034 & 0.000757 & 0.000761\\
				& \tiny(0.0086) & \tiny (0.0086) & \tiny (0.054) & \tiny (0.056) & \tiny (0.000044) & \tiny (0.000043) & \tiny (0.00018) & \tiny (0.00018)\\
				\hline\hline
				\multicolumn{9}{c}{Model \ref{mzar:model:6}} \\ \hline
				& \multicolumn{2}{c|}{$E|\hat{q}-q|$}  &  \multicolumn{2}{c|}{$E(D_H)$}  & \multicolumn{2}{c|}{$E\|\hat{\betab} - \betab\|$}  & \multicolumn{2}{c}{$\frac{\mathrm{MSPE(fitted)}}{\mathrm{MSPE(oracle)}}- 1$}   \\
				$q_\mathrm{max}$ & 5 & 20  & 5 & 20 & 5 & 20 & 5 & 20 \\\hline
				$T=400$ & 0.378 & 0.4 & 2.27 & 2.29 & 0.0133 & 0.013 & 0.0229 & 0.0221\\
				& \tiny(0.022) & \tiny (0.023) & \tiny (0.054) & \tiny (0.053) & \tiny (0.00048) & \tiny (0.00045) & \tiny (0.0016) & \tiny (0.0013)\\
				$T=800$ & 0.823 & 0.881 & 3.31 & 3.3 & 0.00909 & 0.009 & 0.0158 & 0.0152\\
				& \tiny(0.031) & \tiny (0.035) & \tiny (0.072) & \tiny (0.071) & \tiny (0.00029) & \tiny (0.00027) & \tiny (0.00099) & \tiny (0.00097)\\
				$T=1500$ & 0.428 & 0.462 & 3.09 & 3.1 & 0.00342 & 0.00341 & 0.00676 & 0.00667\\
				& \tiny(0.025) & \tiny (0.028) & \tiny (0.1) & \tiny (0.1) & \tiny (0.00014) & \tiny (0.00013) & \tiny (0.00055) & \tiny (0.00055)\\
				$T=3000$ & 0.533 & 0.644 & 3.57 & 3.52 & 0.00192 & 0.00178 & 0.00444 & 0.00394\\
				& \tiny(0.029) & \tiny (0.038) & \tiny (0.12) & \tiny (0.11) & \tiny (0.000084) & \tiny (0.000064) & \tiny (0.00045) & \tiny (0.00038)\\
				\hline\hline
			\end{tabular}
		\end{footnotesize}
	\end{center}
	\caption{\label{Tab:Sim4} Performance of AMAR using different $q_\mathrm{max}$ under \ref{mzar:model:4} -- \ref{mzar:model:6}, with estimated errors given in the brackets. Here $\hat{q}$ is the number of the fitted timescales, $D_H$ is the Hausdorff distance between the fitted timescale locations $\{\hat{\tau}_1, \ldots, \hat{\tau}_{\hat{q}}\}$ and the true ones $\{  {\tau}_1, \ldots, {\tau}_{q}\}$, $\|\hat{\betab} - \betab\|$ is the Euclidean distance between the fitted parameter vector and the true one, and MPSE is the mean squared prediction errors of different models.}
\end{table}

\subsubsection{The initial order of AR -- \texorpdfstring{$p$}{p}}
We run the same experiments listed in the main manuscript, but use a fixed $p = 25$. 
The same evaluation metrics are used. Results are given in Table~\ref{Tab:Sim5} and Table~\ref{Tab:Sim6}. For the ease of comparison, here we also recall the performance results of the default AMAR that uses $p$ selected via SIC, for which details can be founded in Section~\ref{mzar:sec:parameter_choice} of the main manuscript.  

Here we carefully fixed $p$ at $25$, so that it is larger than the timescales among all cases. Here the largest timescale is equal to $\lfloor 3000^{0.4} \rfloor = 24$, from Model~\ref{mzar:model:6} with $T=3000$. It can be seen that for most cases, both approaches perform similarly. Indeed, AMAR with a fixed $p$ might lead to some very moderate improvement over our current approach of selection via SIC in a few settings. Still, not surprisingly, using a fixed $p$ could be quite problematic when the chosen $p$ is close to or bigger than $\tau_{q_{\mathrm{max}}}$, as is evident in the setting of Model~\ref{mzar:model:6} with $T=3000$, where its performance is more than 100\% worse in every evaluation metric.

\begin{table}[!htbp]
	
	\begin{center}
		\begin{footnotesize}
			\begin{tabular}{c|cc|cc|cc|cc}
				\hline\hline
				\multicolumn{9}{c}{Model \ref{mzar:model:1}} \\ \hline
				& \multicolumn{2}{c|}{$E|\hat{q}-q|$}  &  \multicolumn{2}{c|}{$E(D_H)$}  & \multicolumn{2}{c|}{$E\|\hat{\betab} - \betab\|$}  & \multicolumn{2}{c}{$\frac{\mathrm{MSPE(fitted)}}{\mathrm{MSPE(oracle)}}- 1$}   \\
				$p$ & SIC & fixed  & SIC & fixed & SIC & fixed & SIC & fixed \\\hline
				$T=400$ & 0.172 & 0.196 & 0.593 & 0.738 & 0.0159 & 0.0178 & 0.0133 & 0.0147\\
				& \tiny(0.014) & \tiny (0.016) & \tiny (0.047) & \tiny (0.07) & \tiny (0.0008) & \tiny (0.00091) & \tiny (0.00093) & \tiny (0.001)\\
				$T=800$ & 0.051 & 0.047 & 0.181 & 0.252 & 0.0035 & 0.00401 & 0.0046 & 0.00483\\
				& \tiny(0.0072) & \tiny (0.0071) & \tiny (0.03) & \tiny (0.046) & \tiny (0.00026) & \tiny (0.00032) & \tiny (0.00048) & \tiny (0.00052)\\
				$T=1500$ & 0.018 & 0.02 & 0.085 & 0.073 & 0.00116 & 0.00115 & 0.00138 & 0.0014\\
				& \tiny(0.0042) & \tiny (0.0046) & \tiny (0.03) & \tiny (0.027) & \tiny (0.000088) & \tiny (0.000084) & \tiny (0.00024) & \tiny (0.00025)\\
				$T=3000$ & 0.012 & 0.009 & 0.072 & 0.016 & 0.000546 & 0.000546 & 0.000662 & 0.000681\\
				& \tiny(0.0034) & \tiny (0.003) & \tiny (0.035) & \tiny (0.0063) & \tiny (0.000027) & \tiny (0.000027) & \tiny (0.00017) & \tiny (0.00017)\\
				\hline\hline
				
				\multicolumn{9}{c}{Model \ref{mzar:model:2}} \\ \hline
				& \multicolumn{2}{c|}{$E|\hat{q}-q|$}  &  \multicolumn{2}{c|}{$E(D_H)$}  & \multicolumn{2}{c|}{$E\|\hat{\betab} - \betab\|$}  & \multicolumn{2}{c}{$\frac{\mathrm{MSPE(fitted)}}{\mathrm{MSPE(oracle)}}- 1$}   \\
				$p$ & SIC & fixed  & SIC & fixed & SIC & fixed & SIC & fixed \\\hline
				$T=400$ & 0.303 & 0.235 & 1.33 & 1.67 & 0.02 & 0.0233 & 0.0281 & 0.0259\\
				& \tiny(0.018) & \tiny (0.017) & \tiny (0.072) & \tiny (0.11) & \tiny (0.0013) & \tiny (0.0018) & \tiny (0.01) & \tiny (0.007)\\
				$T=800$ & 0.194 & 0.154 & 0.764 & 1.13 & 0.00635 & 0.00638 & 0.00852 & 0.00815\\
				& \tiny(0.014) & \tiny (0.012) & \tiny (0.06) & \tiny (0.1) & \tiny (0.00071) & \tiny (0.00065) & \tiny (0.0013) & \tiny (0.0011)\\
				$T=1500$ & 0.108 & 0.122 & 0.921 & 0.821 & 0.00171 & 0.000986 & 0.00666 & 0.00386\\
				& \tiny(0.01) & \tiny (0.011) & \tiny (0.11) & \tiny (0.092) & \tiny (0.00038) & \tiny (0.00019) & \tiny (0.0038) & \tiny (0.0019)\\
				$T=3000$ & 0.07 & 0.056 & 0.646 & 0.446 & 0.0000979 & 0.0000896 & 0.000793 & 0.000687\\
				& \tiny(0.0081) & \tiny (0.0073) & \tiny (0.099) & \tiny (0.072) & \tiny (0.000021) & \tiny (0.000021) & \tiny (0.0002) & \tiny (0.00017)\\
				\hline\hline
				\multicolumn{9}{c}{Model \ref{mzar:model:3}} \\ \hline
				& \multicolumn{2}{c|}{$E|\hat{q}-q|$}  &  \multicolumn{2}{c|}{$E(D_H)$}  & \multicolumn{2}{c|}{$E\|\hat{\betab} - \betab\|$}  & \multicolumn{2}{c}{$\frac{\mathrm{MSPE(fitted)}}{\mathrm{MSPE(oracle)}}- 1$}   \\
				$p$ & SIC & fixed  & SIC & fixed & SIC & fixed & SIC & fixed  \\\hline
				$T=400$ & 0.711 & 0.314 & 1.37 & 1.17 & 0.0211 & 0.0187 & 0.0296 & 0.0297\\
				& \tiny(0.035) & \tiny (0.024) & \tiny (0.046) & \tiny (0.057) & \tiny (0.00076) & \tiny (0.00072) & \tiny (0.0016) & \tiny (0.0017)\\
				$T=800$ & 0.344 & 0.146 & 0.643 & 0.481 & 0.00699 & 0.00571 & 0.00922 & 0.00825\\
				& \tiny(0.026) & \tiny (0.015) & \tiny (0.034) & \tiny (0.036) & \tiny (0.00031) & \tiny (0.00029) & \tiny (0.00075) & \tiny (0.00069)\\
				$T=1500$ & 0.083 & 0.087 & 0.31 & 0.254 & 0.00203 & 0.0022 & 0.0034 & 0.00359\\
				& \tiny(0.011) & \tiny (0.011) & \tiny (0.043) & \tiny (0.03) & \tiny (0.00011) & \tiny (0.00018) & \tiny (0.0004) & \tiny (0.00044)\\
				$T=3000$ & 0.054 & 0.055 & 0.219 & 0.126 & 0.000673 & 0.000685 & 0.0015 & 0.00147\\
				& \tiny(0.0082) & \tiny (0.0091) & \tiny (0.045) & \tiny (0.023) & \tiny (0.000041) & \tiny (0.000041) & \tiny (0.00023) & \tiny (0.00023)\\
				\hline\hline
			\end{tabular}
		\end{footnotesize}
	\end{center}
	\caption{\label{Tab:Sim5} Performance of AMAR under \ref{mzar:model:1} -- \ref{mzar:model:3} with the initial AR order $p$ either selected via SIC, or fixed at $p=25$. The estimated errors given in the brackets. Here $\hat{q}$ is the number of the fitted timescales, $D_H$ is the Hausdorff distance between the fitted timescale locations $\{\hat{\tau}_1, \ldots, \hat{\tau}_{\hat{q}}\}$ and the true ones $\{  {\tau}_1, \ldots, {\tau}_{q}\}$, $\|\hat{\betab} - \betab\|$ is the Euclidean distance between the fitted parameter vector and the true one, and MPSE is the mean squared prediction errors of different models.}
\end{table}

\begin{table}[!htbp]
	
	\begin{center}
		\begin{footnotesize}
			\begin{tabular}{c|cc|cc|cc|cc}
				\hline\hline
				\multicolumn{9}{c}{Model \ref{mzar:model:4}} \\ \hline
				& \multicolumn{2}{c|}{$E|\hat{q}-q|$}  &  \multicolumn{2}{c|}{$E(D_H)$}  & \multicolumn{2}{c|}{$E\|\hat{\betab} - \betab\|$}  & \multicolumn{2}{c}{$\frac{\mathrm{MSPE(fitted)}}{\mathrm{MSPE(oracle)}}- 1$}   \\
				$p$ & SIC & fixed  & SIC & fixed & SIC & fixed & SIC & fixed  \\\hline
				$T=400$ & 0.098 & 0.078 & 0.2 & 0.27 & 0.00892 & 0.0105 & 0.0145 & 0.0157\\
				& \tiny(0.012) & \tiny (0.012) & \tiny (0.027) & \tiny (0.041) & \tiny (0.00065) & \tiny (0.00074) & \tiny (0.0011) & \tiny (0.0012)\\
				$T=800$ & 0.044 & 0.039 & 0.092 & 0.172 & 0.00397 & 0.00446 & 0.00657 & 0.00653\\
				& \tiny(0.0085) & \tiny (0.008) & \tiny (0.019) & \tiny (0.035) & \tiny (0.0003) & \tiny (0.0004) & \tiny (0.0006) & \tiny (0.00057)\\
				$T=1500$ & 0.035 & 0.039 & 0.291 & 0.158 & 0.00179 & 0.00194 & 0.00333 & 0.00324\\
				& \tiny(0.006) & \tiny (0.0063) & \tiny (0.059) & \tiny (0.03) & \tiny (0.00011) & \tiny (0.00011) & \tiny (0.0004) & \tiny (0.00039)\\
				$T=3000$ & 0.023 & 0.024 & 0.129 & 0.077 & 0.000756 & 0.000753 & 0.0017 & 0.00162\\
				& \tiny(0.0051) & \tiny (0.005) & \tiny (0.033) & \tiny (0.019) & \tiny (0.000023) & \tiny (0.000023) & \tiny (0.00024) & \tiny (0.00024)\\
				\hline\hline
				
				\multicolumn{9}{c}{Model \ref{mzar:model:5}} \\ \hline
				& \multicolumn{2}{c|}{$E|\hat{q}-q|$}  &  \multicolumn{2}{c|}{$E(D_H)$}  & \multicolumn{2}{c|}{$E\|\hat{\betab} - \betab\|$}  & \multicolumn{2}{c}{$\frac{\mathrm{MSPE(fitted)}}{\mathrm{MSPE(oracle)}}- 1$}   \\
				$p$ & SIC & fixed  & SIC & fixed & SIC & fixed & SIC & fixed \\\hline
				
				$T=400$ & 0.217 & 0.265 & 1.64 & 2.29 & 0.0109 & 0.0123 & 0.0164 & 0.0159\\
				& \tiny(0.017) & \tiny (0.02) & \tiny (0.073) & \tiny (0.1) & \tiny (0.00045) & \tiny (0.00051) & \tiny (0.0028) & \tiny (0.0014)\\
				$T=800$ & 0.133 & 0.144 & 0.858 & 1.05 & 0.00414 & 0.00447 & 0.00517 & 0.00588\\
				& \tiny(0.013) & \tiny (0.014) & \tiny (0.056) & \tiny (0.073) & \tiny (0.00022) & \tiny (0.00026) & \tiny (0.00055) & \tiny (0.00065)\\
				$T=1500$ & 0.099 & 0.092 & 0.704 & 0.574 & 0.00167 & 0.00168 & 0.00237 & 0.0023\\
				& \tiny(0.012) & \tiny (0.011) & \tiny (0.076) & \tiny (0.058) & \tiny (0.00012) & \tiny (0.00011) & \tiny (0.00033) & \tiny (0.00034)\\
				$T=3000$ & 0.052 & 0.049 & 0.331 & 0.271 & 0.000339 & 0.000346 & 0.000788 & 0.000746\\
				& \tiny(0.0086) & \tiny (0.0082) & \tiny (0.054) & \tiny (0.042) & \tiny (0.000043) & \tiny (0.000043) & \tiny (0.00017) & \tiny (0.00017)\\
				\hline\hline
				\multicolumn{9}{c}{Model \ref{mzar:model:6}} \\ \hline
				& \multicolumn{2}{c|}{$E|\hat{q}-q|$}  &  \multicolumn{2}{c|}{$E(D_H)$}  & \multicolumn{2}{c|}{$E\|\hat{\betab} - \betab\|$}  & \multicolumn{2}{c}{$\frac{\mathrm{MSPE(fitted)}}{\mathrm{MSPE(oracle)}}- 1$}   \\
				$p$ & SIC & fixed  & SIC & fixed & SIC & fixed & SIC & fixed \\\hline
				$T=400$ & 0.407 & 0.43 & 2.3 & 3.9 & 0.0133 & 0.015 & 0.023 & 0.0279\\
				& \tiny(0.024) & \tiny (0.024) & \tiny (0.054) & \tiny (0.12) & \tiny (0.00046) & \tiny (0.00058) & \tiny (0.0016) & \tiny (0.0016)\\
				$T=800$ & 0.886 & 0.486 & 3.29 & 3.18 & 0.00902 & 0.00718 & 0.015 & 0.0129\\
				& \tiny(0.035) & \tiny (0.026) & \tiny (0.071) & \tiny (0.083) & \tiny (0.00028) & \tiny (0.00023) & \tiny (0.00098) & \tiny (0.00086)\\
				$T=1500$ & 0.455 & 0.639 & 3.08 & 3.04 & 0.00336 & 0.0038 & 0.00668 & 0.00712\\
				& \tiny(0.028) & \tiny (0.035) & \tiny (0.1) & \tiny (0.085) & \tiny (0.00013) & \tiny (0.00014) & \tiny (0.00055) & \tiny (0.00056)\\
				$T=3000$ & 0.642 & 2.04 & 3.52 & 6.8 & 0.00177 & 0.00392 & 0.00395 & 0.0071\\
				& \tiny(0.037) & \tiny (0.063) & \tiny (0.11) & \tiny (0.12) & \tiny (0.000064) & \tiny (0.000096) & \tiny (0.00038) & \tiny (0.00055)\\
				\hline\hline
			\end{tabular}
		\end{footnotesize}
	\end{center}
	\caption{\label{Tab:Sim6} Performance of AMAR under \ref{mzar:model:4} -- \ref{mzar:model:6}, with the initial AR order $p$ either selected via SIC, or fixed at $p=25$. Here $\hat{q}$ is the number of the fitted timescales, $D_H$ is the Hausdorff distance between the fitted timescale locations $\{\hat{\tau}_1, \ldots, \hat{\tau}_{\hat{q}}\}$ and the true ones $\{  {\tau}_1, \ldots, {\tau}_{q}\}$, $\|\hat{\betab} - \betab\|$ is the Euclidean distance between the fitted parameter vector and the true one, and MPSE is the mean squared prediction errors of different models.}
\end{table}

\subsection{(More conventional) higher-order AR}
In this part, we compare AMAR and conventional
AR models (selected both by AIC and BIC) over the data that are generated from more conventional high-order stationary AR models. In particular, we consider the following settings, with $\tau_q =16$ and $q = 16,12,8$.

\begin{enumerate}[label=(M\arabic*)] 
	\setcounter{enumi}{6}
	\item \label{mzar:model:P1} $q=16$ and $\tau_{i}=i$ for $i = 1,\ldots,16$,  with the corresponding AR coefficients 
	\[\betab = (0.2,-0.2,0.2,-0.2,\ldots, 0.2,-0.2)^T.\] 
	\item  \label{mzar:model:P2} $q=12$ and $\{\tau_1,\ldots,\tau_{12}\}=\{1,\ldots,16\}\backslash\{2,6,10,14\}$,  with the corresponding AR coefficients 
	\[\betab = (0.2,0, 0, -0.2,0.2,0,0,-0.2,\ldots,0.2,0,0,-0.2)^T.\]
	\item  \label{mzar:model:P3}  $q=8$ and $\tau_{i}=2i$ for $i = 1,\ldots,8$,  with the corresponding AR coefficients 
	\[\betab = (0.2,0.2,-0.2,-0.2,\ldots, 0.2,0.2,-0.2,-0.2)^T.\] 
\end{enumerate}
Here Model~\ref{mzar:model:P1} is a conventional high-order AR. Models~\ref{mzar:model:P2} and \ref{mzar:model:P3} are also high-order, but their AR coefficients are more structured (though $\tau_q =16$ and $q$ are still at the same order). In particular, all three models are stationary. 

For these models, we run the experiments using the same settings as listed in the main manuscript, but set $q_{\mathrm{max}} = 20$ (as here $q$ can be as high as 16). For the evaluation metrics, we look at the accuracy of the estimated order of AR, denoted by $|\hat{\tau}_{\hat{q}} - \tau_q|$, the Euclidean distance between the fitted parameter vector and the true one, denoted by $\|\hat{\betab} - \betab\|$, and the mean squared prediction errors (MSPE) of different models. Results are given in Table~\ref{Tab:Sim9}.

We see that with in Model~\ref{mzar:model:P1}, unsurprisingly AR with order selected via BIC performs the best among all the evaluation measures. However, the performance of AMAR is only slightly worse (and better than AR with order selected via AIC). In particular, it tends to estimates the number of scales (which is the same as the AR order) correctly when $T$ is reasonably large, implying little efficiency loss for using AMAR even when there is no meaningful AMAR-type structure in the parameter vector of AR coefficients. On the other hand, as we move to Model~\ref{mzar:model:P2} and Model~\ref{mzar:model:P3} where the AR parameter vectors have more structures embedded (though here $\tau_q$ and $q$ are still at the same order), AMAR tends to perform better than its competitors in terms of both the parameter estimation and prediction accuracy. The improvement is more visible in the setting of Model~\ref{mzar:model:P3}, as it has less scales than Model~\ref{mzar:model:P2}, so is intuitively more favourable to AMAR.

\begin{table}[!htbp]
	
	\begin{center}
		\begin{footnotesize}
			\begin{tabular}{c|ccc|ccc|ccc}
				\hline\hline
				\multicolumn{10}{c}{Model \ref{mzar:model:P1}} \\ \hline
				& \multicolumn{3}{c|}{$E|\hat{\tau}_{\hat{q}} - \tau_q|$}   & \multicolumn{3}{c|}{$E\|\hat{\betab} - \betab\|$}  & \multicolumn{3}{c}{$\frac{\mathrm{MSPE(fitted)}}{\mathrm{MSPE(oracle)}}- 1$}   \\
				Method & AMAR & AIC  & BIC & AMAR & AIC & BIC & AMAR & AIC & BIC \\\hline
				$T=400$ & 6.73 & 0.877 & 12.6 & 0.577 & 0.051 & 0.543 & 0.204 & 0.0534 & 0.177\\
				& \tiny(0.14) & \tiny (0.042) & \tiny (0.17) & \tiny (0.0038) & \tiny (0.00084) & \tiny (0.0068) & \tiny (0.0031) & \tiny (0.0018) & \tiny (0.003)\\
				$T=800$ & 2.05 & 1.68 & 0.23 & 0.139 & 0.027 & 0.0293 & 0.0826 & 0.0274 & 0.0265\\
				& \tiny(0.1) & \tiny (0.09) & \tiny (0.055) & \tiny (0.0053) & \tiny (0.00053) & \tiny (0.0023) & \tiny (0.0026) & \tiny (0.0012) & \tiny (0.0016)\\
				$T=1500$ & 0.916 & 2 & 0.014 & 0.012 & 0.0148 & 0.0105 & 0.0124 & 0.0141 & 0.0114\\
				& \tiny(0.09) & \tiny (0.11) & \tiny (0.0037) & \tiny (0.00026) & \tiny (0.0004) & \tiny (0.00014) & \tiny (0.00074) & \tiny (0.00079) & \tiny (0.00071)\\
				$T=3000$ & 0.662 & 2.14 & 0.015 & 0.00582 & 0.00733 & 0.00532 & 0.00597 & 0.00701 & 0.00555\\
				& \tiny(0.077) & \tiny (0.13) & \tiny (0.0038) & \tiny (0.00015) & \tiny (0.00017) & \tiny (0.000074) & \tiny (0.00052) & \tiny (0.00057) & \tiny (0.00049)\\
				\hline\hline

				\multicolumn{10}{c}{Model \ref{mzar:model:P2}} \\ \hline
				& \multicolumn{3}{c|}{$E|\hat{\tau}_{\hat{q}} - \tau_q|$}   & \multicolumn{3}{c|}{$E\|\hat{\betab} - \betab\|$}  & \multicolumn{3}{c}{$\frac{\mathrm{MSPE(fitted)}}{\mathrm{MSPE(oracle)}}- 1$}   \\
				Method & AMAR & AIC  & BIC & AMAR & AIC & BIC & AMAR & AIC & BIC \\\hline
				$T=400$ & 2.93 & 0.895 & 7.45 & 0.151 & 0.0503 & 0.175 & 0.132 & 0.051 & 0.14\\
				& \tiny(0.085) & \tiny (0.04) & \tiny (0.21) & \tiny (0.0019) & \tiny (0.00075) & \tiny (0.0038) & \tiny (0.003) & \tiny (0.0017) & \tiny (0.0039)\\
				$T=800$ & 1.62 & 1.69 & 0.307 & 0.08 & 0.0263 & 0.025 & 0.0767 & 0.0281 & 0.0273\\
				& \tiny(0.034) & \tiny (0.091) & \tiny (0.041) & \tiny (0.001) & \tiny (0.00042) & \tiny (0.00077) & \tiny (0.002) & \tiny (0.0011) & \tiny (0.0014)\\
				$T=1500$ & 0.669 & 2.02 & 0.019 & 0.00969 & 0.014 & 0.0109 & 0.00957 & 0.0122 & 0.0102\\
				& \tiny(0.077) & \tiny (0.12) & \tiny (0.0052) & \tiny (0.00029) & \tiny (0.00025) & \tiny (0.00015) & \tiny (0.0007) & \tiny (0.00075) & \tiny (0.00068)\\
				$T=3000$ & 0.597 & 1.95 & 0.014 & 0.00401 & 0.00687 & 0.00533 & 0.00452 & 0.0062 & 0.00516\\
				& \tiny(0.077) & \tiny (0.14) & \tiny (0.004) & \tiny (0.00011) & \tiny (0.00013) & \tiny (0.000068) & \tiny (0.00041) & \tiny (0.0005) & \tiny (0.00045)\\
				
				\hline\hline
				\multicolumn{10}{c}{Model \ref{mzar:model:P3}} \\ \hline
				& \multicolumn{3}{c|}{$E|\hat{\tau}_{\hat{q}} - \tau_q|$}   & \multicolumn{3}{c|}{$E\|\hat{\betab} - \betab\|$}  & \multicolumn{3}{c}{$\frac{\mathrm{MSPE(fitted)}}{\mathrm{MSPE(oracle)}}- 1$}   \\
				Method & AMAR & AIC  & BIC & AMAR & AIC & BIC & AMAR & AIC & BIC \\\hline
				$T=400$ & 1.37 & 0.849 & 0.161 & 0.104 & 0.0499 & 0.0485 & 0.117 & 0.0493 & 0.0487\\
				& \tiny(0.015) & \tiny (0.039) & \tiny (0.022) & \tiny (0.0011) & \tiny (0.00082) & \tiny (0.0012) & \tiny (0.003) & \tiny (0.0017) & \tiny (0.0017)\\
				$T=800$ & 1.04 & 1.77 & 0.021 & 0.0738 & 0.0269 & 0.0199 & 0.0695 & 0.0253 & 0.0212\\
				& \tiny(0.0061) & \tiny (0.093) & \tiny (0.0048) & \tiny (0.00069) & \tiny (0.00051) & \tiny (0.00029) & \tiny (0.002) & \tiny (0.0011) & \tiny (0.001)\\
				$T=1500$ & 0.214 & 1.93 & 0.017 & 0.00469 & 0.0147 & 0.0107 & 0.00691 & 0.0131 & 0.011\\
				& \tiny(0.042) & \tiny (0.12) & \tiny (0.0041) & \tiny (0.00039) & \tiny (0.0003) & \tiny (0.00016) & \tiny (0.00067) & \tiny (0.0008) & \tiny (0.00073)\\
				$T=3000$ & 0.174 & 1.75 & 0.018 & 0.00215 & 0.00703 & 0.00528 & 0.00384 & 0.00646 & 0.00551\\
				& \tiny(0.041) & \tiny (0.12) & \tiny (0.0047) & \tiny (0.00022) & \tiny (0.00015) & \tiny (0.000079) & \tiny (0.00044) & \tiny (0.00051) & \tiny (0.00047)\\
				
				\hline\hline
			\end{tabular}
		\end{footnotesize}
	\end{center}
	\caption{\label{Tab:Sim9} Performance of different methods under \ref{mzar:model:P1} -- \ref{mzar:model:P3}, with estimated errors given in the brackets. Here $|\hat{\tau}_{\hat{q}} - \tau_q|$ is the difference between the estimated and true order of AR, $\|\hat{\betab} - \betab\|$ is the Euclidean distance between the fitted parameter vector and the true one, and MPSE is the mean squared prediction errors of different models.}
\end{table}

\subsection{Non-stationary AR}

Here we report the results from experiments with series simulated from non-stationary AR models with unit roots. The scenarios we consider are similar to \ref{mzar:model:1} -- \ref{mzar:model:6} listed in the main manuscript, with their details outlined below.

\begin{enumerate}[label=(M\arabic*')] 
	\item \label{mzar:model:1UR} Same as \ref{mzar:model:1}  but with $\alpha_{1}= 0.4$, $\alpha_{2}=0.6$ (i.e. $\betab = (0.6,0.2,0.2)^T$).
	\item  \label{mzar:model:2UR}  Same as \ref{mzar:model:2} but with $\alpha_{1}= 1.5$, $\alpha_{2}=-0.5$ (i.e. $\betab = (0.65,0.65,-0.1,-0.1,-0.1)^T$).
	\item  \label{mzar:model:3UR}  Same as \ref{mzar:model:3} but with  $\alpha_{1}= 0.5$, $\alpha_{2}=-1$, $\alpha_{3}=1.4$ 
	\\(i.e. $\betab = (0.5, -0.1,-0.1,-0.1,-0.1,0.1,\ldots,0.1)^T$).
	\item  \label{mzar:model:4UR}  Same as \ref{mzar:model:4} but with $\alpha_{1}= 1$, $\alpha_{2}=-4.8$, $\alpha_{3}=10.2$, $\alpha_{4}=-6.4$ (i.e. $\betab = (1,0,\ldots,0,0.8,-0.8)^T$, so $\varepsilon_{t} = (1-0.8B^7)(1-B)X_t$).
	\item  \label{mzar:model:5UR}  Same as \ref{mzar:model:5} but with $\alpha_{1}= 1$ (i.e. $\betab = (0.1,\ldots,0.1)^T$).
	\item  \label{mzar:model:6UR}  Same as \ref{mzar:model:6} but with $\alpha_{1}= \alpha_{2}=0.5$ (i.e. $\betab = (0.5+0.5/\lfloor T^{0.4} \rfloor, 0.5/\lfloor T^{0.4} \rfloor \ldots,0.5/\lfloor T^{0.4} \rfloor)^T$).
\end{enumerate}

Here we use AMAR with default choice of its tuning parameters outlined in Section~\ref{mzar:sec:parameter_choice}. The corresponding results are summarised in Table~\ref{Tab:Sim7} and Table~\ref{Tab:Sim8}, where as before, we report the estimates for $|q-\hat{q}|$, with $\hat{q}$ being the number of the fitted timescales,  the Hausdorff distance $D_H$ between the fitted timescale locations $\{\hat{\tau}_1, \ldots, \hat{\tau}_{\hat{q}}\}$ and the true ones $\{  {\tau}_1, \ldots, {\tau}_{q}\}$, the Euclidean distance between the fitted parameter vector and the true one, denoted by $\|\hat{\betab} - \betab\|$, and the ratio between the mean squared prediction error (MPSE) using the fitted model and that with the oracle over the next $T^*=100$ unseen observations.

\begin{table}[!htbp]
	
	\begin{center}
		\begin{footnotesize}
			\begin{tabular}{c|cc|cc|ccc|ccc}
				\hline\hline
				\multicolumn{11}{c}{Model \ref{mzar:model:1UR}} \\ \hline
				& \multicolumn{2}{c|}{$E|\hat{q}-q|$}  &  \multicolumn{2}{c|}{$E(D_H)$}  & \multicolumn{3}{c|}{$E\|\hat{\betab} - \betab\|$}  & \multicolumn{3}{c}{$\frac{\mathrm{MSPE(fitted)}}{\mathrm{MSPE(oracle)}}- 1$}   \\
				Method & AMAR & Fused  & AMAR & Fused & AMAR & Fused & AIC & AMAR & Fused & AIC \\\hline
				$T=400$ & 0.469 & 1.21 & 2.31 & 18.6 & 0.0449 & 0.381 & 0.0268 & 38.6 & 28.7 & 32.5\\
				& \tiny(0.019) & \tiny (0.049) & \tiny (0.11) & \tiny (0.042) & \tiny (0.0022) & \tiny (0.0014) & \tiny (0.0008) & \tiny (23) & \tiny (1.2) & \tiny (3.1)\\
				$T=800$ & 0.33 & 1.43 & 1.97 & 26.4 & 0.0347 & 0.398 & 0.0201 & 1.45 & 29 & 7.79\\
				& \tiny(0.016) & \tiny (0.081) & \tiny (0.11) & \tiny (0.063) & \tiny (0.0022) & \tiny (0.001) & \tiny (0.00068) & \tiny (1.4) & \tiny (1.3) & \tiny (0.6)\\
				$T=1500$ & 0.367 & 2.13 & 4.47 & 36.1 & 0.0302 & 0.409 & 0.017 & 0.0231 & 26.9 & 1.95\\
				& \tiny(0.016) & \tiny (0.14) & \tiny (0.25) & \tiny (0.088) & \tiny (0.0022) & \tiny (0.00073) & \tiny (0.00059) & \tiny (0.0018) & \tiny (1.1) & \tiny (0.16)\\
				$T=3000$ & 0.249 & 2.61 & 3.44 & 51.8 & 0.0192 & 0.42 & 0.0161 & 0.0148 & 29.8 & 0.536\\
				& \tiny(0.014) & \tiny (0.18) & \tiny (0.23) & \tiny (0.094) & \tiny (0.0019) & \tiny (0.00026) & \tiny (0.00063) & \tiny (0.0016) & \tiny (1.3) & \tiny (0.043)\\
				\hline\hline
				
				\multicolumn{11}{c}{Model \ref{mzar:model:2UR}} \\ \hline
				& \multicolumn{2}{c|}{$E|\hat{q}-q|$}  &  \multicolumn{2}{c|}{$E(D_H)$}  & \multicolumn{3}{c|}{$E\|\hat{\betab} - \betab\|$}  & \multicolumn{3}{c}{$\frac{\mathrm{MSPE(fitted)}}{\mathrm{MSPE(oracle)}}- 1$}    \\
				Method & AMAR & Fused  & AMAR & Fused & AMAR & Fused & AIC & AMAR & Fused & AIC \\\hline
				$T=400$ & 0.399 & 3.85 & 2.09 & 13.5 & 0.0254 & 0.609 & 0.116 & 0.0374 & 4.41 & 0.226\\
				& \tiny(0.019) & \tiny (0.1) & \tiny (0.075) & \tiny (0.12) & \tiny (0.00093) & \tiny (0.0092) & \tiny (0.0031) & \tiny (0.0024) & \tiny (0.14) & \tiny (0.014)\\
				$T=800$ & 0.269 & 6.75 & 1.52 & 18.5 & 0.0134 & 0.69 & 0.0892 & 0.0175 & 6 & 0.0971\\
				& \tiny(0.017) & \tiny (0.16) & \tiny (0.073) & \tiny (0.21) & \tiny (0.00061) & \tiny (0.0073) & \tiny (0.0025) & \tiny (0.0015) & \tiny (0.2) & \tiny (0.0055)\\
				$T=1500$ & 0.187 & 10 & 2.11 & 24.5 & 0.00572 & 0.733 & 0.0681 & 0.00601 & 7.33 & 0.0506\\
				& \tiny(0.013) & \tiny (0.22) & \tiny (0.16) & \tiny (0.33) & \tiny (0.00036) & \tiny (0.0059) & \tiny (0.0021) & \tiny (0.00074) & \tiny (0.25) & \tiny (0.0024)\\
				$T=3000$ & 0.086 & 14.2 & 0.985 & 35.3 & 0.0018 & 0.774 & 0.0399 & 0.0023 & 8.7 & 0.0246\\
				& \tiny(0.009) & \tiny (0.31) & \tiny (0.12) & \tiny (0.5) & \tiny (0.0002) & \tiny (0.0039) & \tiny (0.0014) & \tiny (0.00051) & \tiny (0.25) & \tiny (0.0013)\\
				
				\hline\hline
				
				\multicolumn{11}{c}{Model \ref{mzar:model:3UR}} \\ \hline
				& \multicolumn{2}{c|}{$E|\hat{q}-q|$}  &  \multicolumn{2}{c|}{$E(D_H)$}  & \multicolumn{3}{c|}{$E\|\hat{\betab} - \betab\|$}  & \multicolumn{3}{c}{$\frac{\mathrm{MSPE(fitted)}}{\mathrm{MSPE(oracle)}}- 1$}   \\
				Method & AMAR & Fused  & AMAR & Fused & AMAR & Fused & AIC & AMAR & Fused & AIC \\\hline
				$T=400$ & 0.747 & 2.47 & 1.37 & 17.4 & 0.0246 & 0.297 & 0.0606 & 0.0351 & 0.766 & 17.4\\
				& \tiny(0.034) & \tiny (0.058) & \tiny (0.043) & \tiny (0.13) & \tiny (0.0011) & \tiny (0.003) & \tiny (0.00089) & \tiny (0.0021) & \tiny (0.022) & \tiny (2.1)\\
				$T=800$ & 0.499 & 1.72 & 0.897 & 24.7 & 0.0201 & 0.318 & 0.0342 & 0.0365 & 0.767 & 3.76\\
				& \tiny(0.029) & \tiny (0.089) & \tiny (0.039) & \tiny (0.11) & \tiny (0.0015) & \tiny (0.0027) & \tiny (0.00059) & \tiny (0.0036) & \tiny (0.018) & \tiny (0.34)\\
				$T=1500$ & 0.177 & 2.4 & 0.744 & 34.1 & 0.0047 & 0.323 & 0.0216 & 0.0104 & 0.788 & 0.954\\
				& \tiny(0.015) & \tiny (0.16) & \tiny (0.073) & \tiny (0.13) & \tiny (0.00076) & \tiny (0.0031) & \tiny (0.00043) & \tiny (0.0019) & \tiny (0.019) & \tiny (0.09)\\
				$T=3000$ & 0.104 & 1.93 & 0.506 & 49.2 & 0.00336 & 0.339 & 0.0141 & 0.00609 & 0.854 & 0.288\\
				& \tiny(0.011) & \tiny (0.16) & \tiny (0.069) & \tiny (0.15) & \tiny (0.00073) & \tiny (0.0027) & \tiny (0.00035) & \tiny (0.0015) & \tiny (0.02) & \tiny (0.028)\\
				\hline\hline
			\end{tabular}
		\end{footnotesize}
	\end{center}
	\caption{\label{Tab:Sim7} Performance of different methods under \ref{mzar:model:1UR} -- \ref{mzar:model:3UR}, with estimated errors given in the brackets. Here $\hat{q}$ is the number of the fitted timescales, $D_H$ is the Hausdorff distance between the fitted timescale locations $\{\hat{\tau}_1, \ldots, \hat{\tau}_{\hat{q}}\}$ and the true ones $\{  {\tau}_1, \ldots, {\tau}_{q}\}$, $\|\hat{\betab} - \betab\|$ is the Euclidean distance between the fitted parameter vector and the true one, and MPSE is the mean squared prediction errors of different models.}
\end{table}

\begin{table}[!htbp]
	
	\begin{center}
		\begin{footnotesize}
			\begin{tabular}{c|cc|cc|ccc|ccc}
				\hline\hline
				\multicolumn{11}{c}{Model \ref{mzar:model:4UR}} \\ \hline
				& \multicolumn{2}{c|}{$E|\hat{q}-q|$}  &  \multicolumn{2}{c|}{$E(D_H)$}  & \multicolumn{3}{c|}{$E\|\hat{\betab} - \betab\|$}  & \multicolumn{3}{c}{$\frac{\mathrm{MSPE(fitted)}}{\mathrm{MSPE(oracle)}}- 1$}   \\
				Method & AMAR & Fused  & AMAR & Fused & AMAR & Fused & AIC & AMAR & Fused & AIC \\\hline
				
				$T=400$ & 0.527 & 2.64 & 1.45 & 18.1 & 0.161 & 2.22 & 0.888 & 0.254 & 216 & 28.1\\
				& \tiny(0.021) & \tiny (0.024) & \tiny (0.07) & \tiny (0.074) & \tiny (0.013) & \tiny (0.0019) & \tiny (0.013) & \tiny (0.021) & \tiny (9.3) & \tiny (2)\\
				$T=800$ & 0.27 & 2.55 & 0.702 & 25.4 & 0.21 & 2.24 & 0.84 & 0.295 & 227 & 7.28\\
				& \tiny(0.017) & \tiny (0.026) & \tiny (0.052) & \tiny (0.11) & \tiny (0.014) & \tiny (0.00068) & \tiny (0.012) & \tiny (0.022) & \tiny (10) & \tiny (0.5)\\
				$T=1500$ & 0.225 & 2.4 & 2.02 & 34.2 & 0.124 & 2.25 & 0.847 & 0.189 & 267 & 2.79\\
				& \tiny(0.014) & \tiny (0.041) & \tiny (0.15) & \tiny (0.17) & \tiny (0.011) & \tiny (0.00035) & \tiny (0.011) & \tiny (0.018) & \tiny (13) & \tiny (0.13)\\
				$T=3000$ & 0.175 & 2.97 & 1.69 & 48.2 & 0.0678 & 2.26 & 0.841 & 0.0921 & 293 & 1.4\\
				& \tiny(0.012) & \tiny (0.092) & \tiny (0.14) & \tiny (0.22) & \tiny (0.0085) & \tiny (0.00026) & \tiny (0.0096) & \tiny (0.012) & \tiny (14) & \tiny (0.052)\\
				\hline\hline
				\multicolumn{11}{c}{Model \ref{mzar:model:5UR}} \\ \hline
				& \multicolumn{2}{c|}{$E|\hat{q}-q|$}  &  \multicolumn{2}{c|}{$E(D_H)$}  & \multicolumn{3}{c|}{$E\|\hat{\betab} - \betab\|$}  & \multicolumn{3}{c}{$\frac{\mathrm{MSPE(fitted)}}{\mathrm{MSPE(oracle)}}- 1$}    \\
				Method & AMAR & Fused  & AMAR & Fused & AMAR & Fused & AIC & AMAR & Fused & AIC \\\hline
				$T=400$ & 0.354 & 0.575 & 1.94 & 9.74 & 0.0134 & 0.0461 & 0.0464 & 5.81 & 0.439 & 10\\
				& \tiny(0.022) & \tiny (0.037) & \tiny (0.077) & \tiny (0.039) & \tiny (0.00061) & \tiny (0.00035) & \tiny (0.00082) & \tiny (4.5) & \tiny (0.032) & \tiny (0.89)\\
				$T=800$ & 0.322 & 1.33 & 1.44 & 17.5 & 0.00602 & 0.0591 & 0.0276 & 0.254 & 0.467 & 2.47\\
				& \tiny(0.02) & \tiny (0.065) & \tiny (0.074) & \tiny (0.07) & \tiny (0.00052) & \tiny (0.00046) & \tiny (0.00062) & \tiny (0.19) & \tiny (0.027) & \tiny (0.19)\\
				$T=1500$ & 0.154 & 2.3 & 1.2 & 27.6 & 0.00212 & 0.069 & 0.0203 & 2.32 & 0.569 & 0.692\\
				& \tiny(0.013) & \tiny (0.12) & \tiny (0.11) & \tiny (0.062) & \tiny (0.0003) & \tiny (0.00044) & \tiny (0.0006) & \tiny (2.2) & \tiny (0.034) & \tiny (0.064)\\
				$T=3000$ & 0.114 & 4.19 & 0.818 & 43.4 & 0.000744 & 0.0776 & 0.014 & 0.72 & 0.622 & 0.197\\
				& \tiny(0.01) & \tiny (0.19) & \tiny (0.088) & \tiny (0.1) & \tiny (0.00024) & \tiny (0.00036) & \tiny (0.00048) & \tiny (0.54) & \tiny (0.03) & \tiny (0.022)\\
				\hline\hline
				
				\multicolumn{11}{c}{Model \ref{mzar:model:6UR}} \\ \hline
				& \multicolumn{2}{c|}{$E|\hat{q}-q|$}  &  \multicolumn{2}{c|}{$E(D_H)$}  & \multicolumn{3}{c|}{$E\|\hat{\betab} - \betab\|$}  & \multicolumn{3}{c}{$\frac{\mathrm{MSPE(fitted)}}{\mathrm{MSPE(oracle)}}- 1$}   \\
				Method & AMAR & Fused  & AMAR & Fused & AMAR & Fused & AIC & AMAR & Fused & AIC \\\hline
				$T=400$ & 0.889 & 1.32 & 3.67 & 17.5 & 0.0177 & 0.234 & 0.0441 & 0.0251 & 0.762 & 10.4\\
				& \tiny(0.034) & \tiny (0.064) & \tiny (0.076) & \tiny (0.11) & \tiny (0.0007) & \tiny (0.002) & \tiny (0.00068) & \tiny (0.0013) & \tiny (0.038) & \tiny (0.87)\\
				$T=800$ & 1.54 & 1.52 & 7.63 & 24.2 & 0.0184 & 0.228 & 0.0329 & 0.0336 & 0.631 & 3.37\\
				& \tiny(0.049) & \tiny (0.11) & \tiny (0.072) & \tiny (0.17) & \tiny (0.00073) & \tiny (0.0022) & \tiny (0.00041) & \tiny (0.0015) & \tiny (0.031) & \tiny (0.3)\\
				$T=1500$ & 0.931 & 2.01 & 5.6 & 33.4 & 0.00501 & 0.23 & 0.0229 & 0.01 & 0.499 & 1.07\\
				& \tiny(0.04) & \tiny (0.15) & \tiny (0.13) & \tiny (0.22) & \tiny (0.00025) & \tiny (0.0022) & \tiny (0.00025) & \tiny (0.00075) & \tiny (0.018) & \tiny (0.094)\\
				$T=3000$ & 2.09 & 2.95 & 12.4 & 47.9 & 0.00515 & 0.236 & 0.0159 & 0.0122 & 0.427 & 0.328\\
				& \tiny(0.066) & \tiny (0.18) & \tiny (0.13) & \tiny (0.28) & \tiny (0.00011) & \tiny (0.0019) & \tiny (0.00015) & \tiny (0.00076) & \tiny (0.012) & \tiny (0.036)\\
				\hline\hline
			\end{tabular}
		\end{footnotesize}
	\end{center}
	\caption{\label{Tab:Sim8} Performance of different methods under \ref{mzar:model:4UR} -- \ref{mzar:model:6UR}, with estimated errors given in the brackets. Here $\hat{q}$ is the number of the fitted timescales, $D_H$ is the Hausdorff distance between the fitted timescale locations $\{\hat{\tau}_1, \ldots, \hat{\tau}_{\hat{q}}\}$ and the true ones $\{  {\tau}_1, \ldots, {\tau}_{q}\}$, $\|\hat{\betab} - \betab\|$ is the Euclidean distance between the fitted parameter vector and the true one, and MPSE is the mean squared prediction errors of different models.}
\end{table}

We see that even in the setting of non-stationary observations, AMAR still performs much better than its competitors in most settings, even though all methods seem to perform worse as compared to the stationary settings. Unsurprisingly,  here the reported results are associated with larger estimation errors. 

In addition, we note that the fused LASSO approach performs much worse than its competitors in terms of MSPE, especially in \ref{mzar:model:1UR} and \ref{mzar:model:4UR}. This is because the fused LASSO approach tends to over-estimate the number of scales, resulting in less accurate $\hat{\betab}$, which could greatly affect the corresponding MSPE when the series is non-stationary.

\section{Additional real data example: well-log}
\label{Sec:adddata}

\begin{figure}[!htbp]
	\centering
	\begin{minipage}{.45\textwidth}
		\centering
		\includegraphics[width=\linewidth]{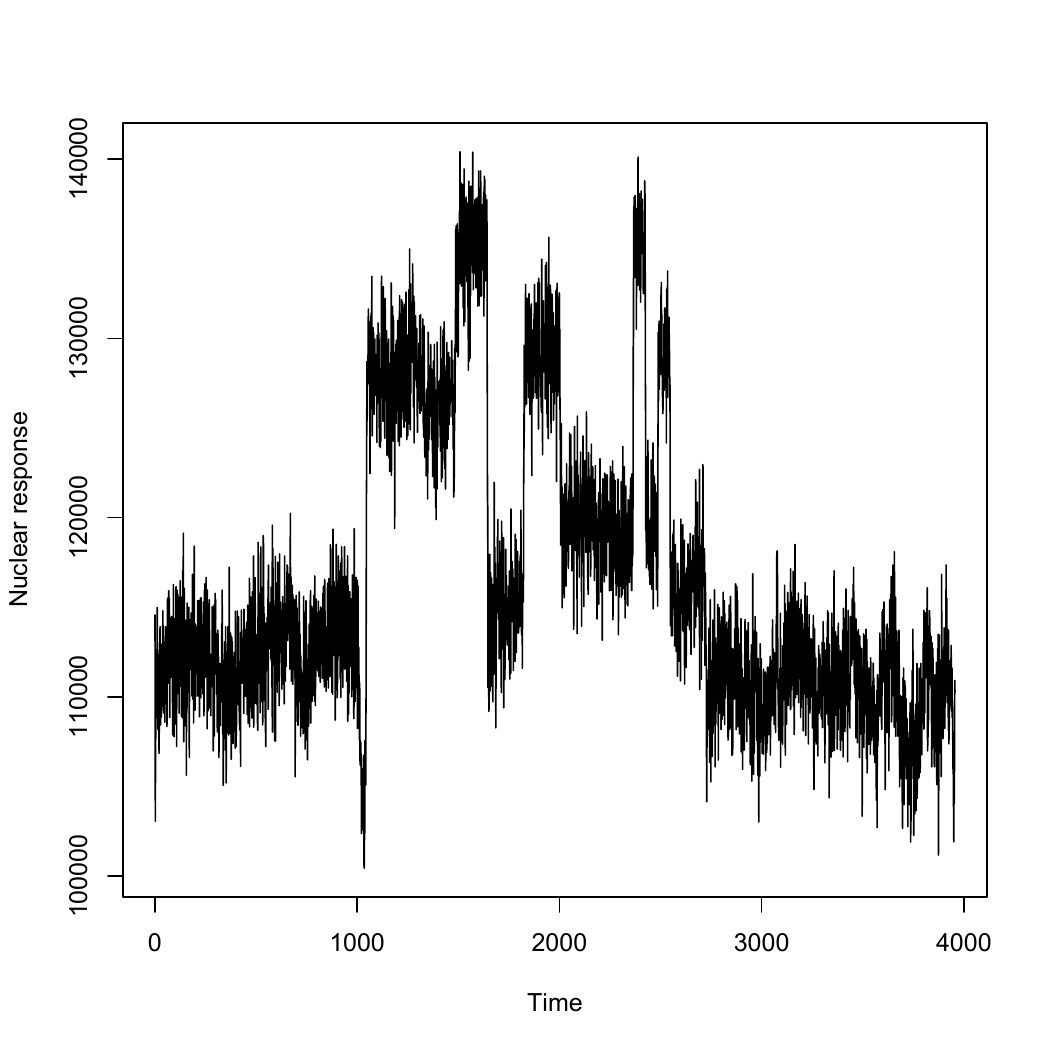}
	\end{minipage}%
	\begin{minipage}{.45\textwidth}
		\centering
		\includegraphics[width=\linewidth]{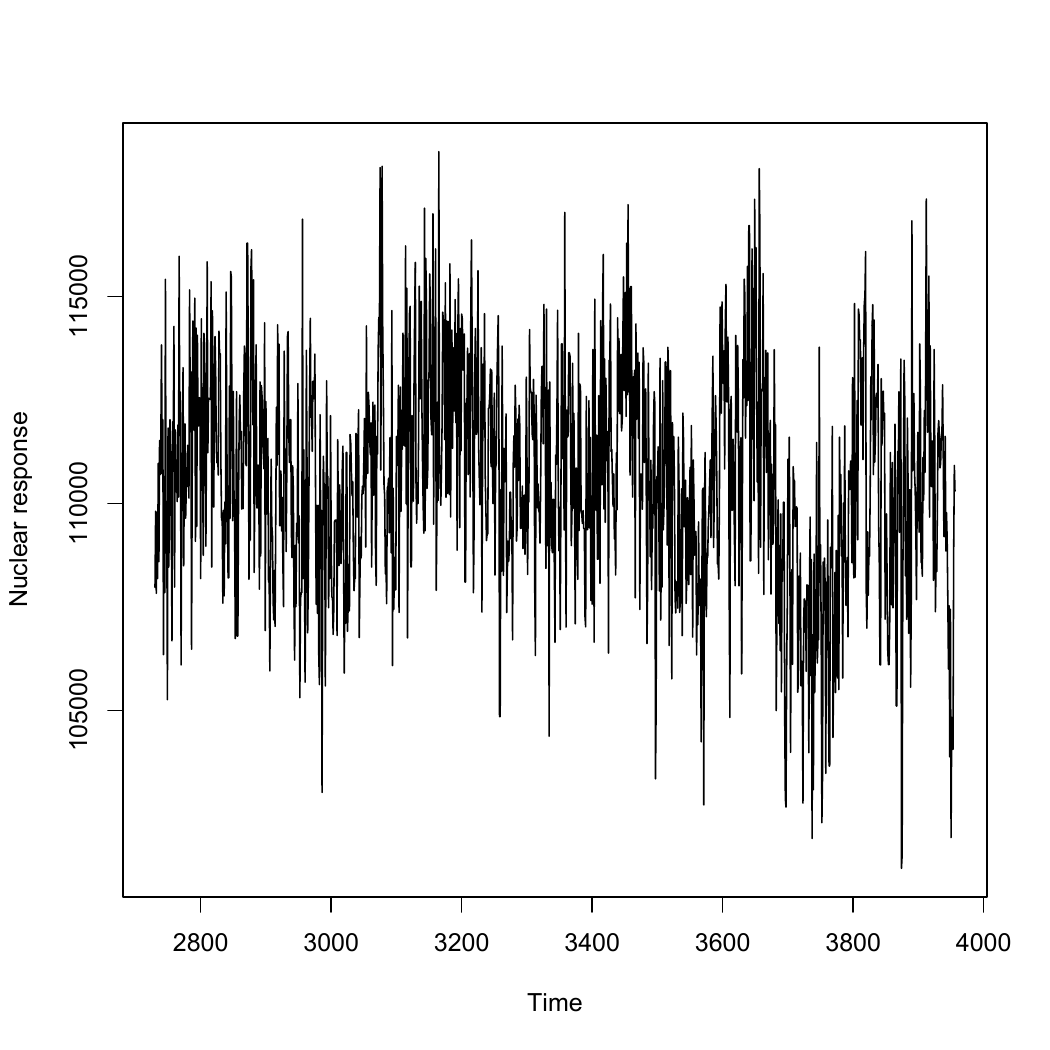}
	\end{minipage}%
	\caption{Left: the well-log data from \cite{of96}, cleaned as described in the text. Right: the end part of the data, from time location 2730. \label{fig:wl}}
\end{figure}

We consider the well-log data from \cite{of96}. Prior to use, the data is cleaned 
by removing outliers, taken here to be the observations that differ from the median-fliter fit to the data (with span 25) by at least 7500. This retains 97.7\% of the data points. The cleaned data, denoted as $\{X_t\}_{t=1}^{3956}$, is shown in the left plot of Figure \ref{fig:wl}.

As summarised in \cite{fc03}, the data represents measurements of the nuclear magnetic response of underground rocks. The underlying (unobserved) signal is assumed to be piecewise constant, with each constant segment representing a stratum of rock. The jumps occur when a new rock stratum is met. The problem of detecting these change-points in the underlying signal is of practical importance in oil drilling.

It is known (for instance, see \cite{cf21} and the references therein) 
that the problem of multiple change-point detection in a piecewise-constant signal observed in noise is much more challenging if the noise displays autocorrelation, as the natural fluctuations of the autocorrelated process can be mistaken for change-points, and vice versa. This appears to be the case in the well-log data: the right-hand plot of Figure \ref{fig:wl} shows the end portion of the data, from the observation after the last visually obvious change-point (at location 2729) to the end. As discussed earlier, the visual appearance of the data fluctuations in this region of the dataset suggests that the AMAR model may be appropriate. Our aim is therefore to: (a) estimate the appropriate AMAR model on $\{X_{2730},\ldots, X_{3956}\}$, (b) fit the estimated model from the previous step on the entire dataset (i.e. $\{X_1,\ldots, X_{3956}\}$) to remove the autocorrelations in the data, and (c) estimate change-point locations in the thus-decorrelated dataset using a method suitable for multiple change-point detection in uncorrelated (Gaussian) noise.

\begin{figure}
	\centering
	\begin{minipage}{.45\textwidth}
		\centering
		\includegraphics[width=\linewidth]{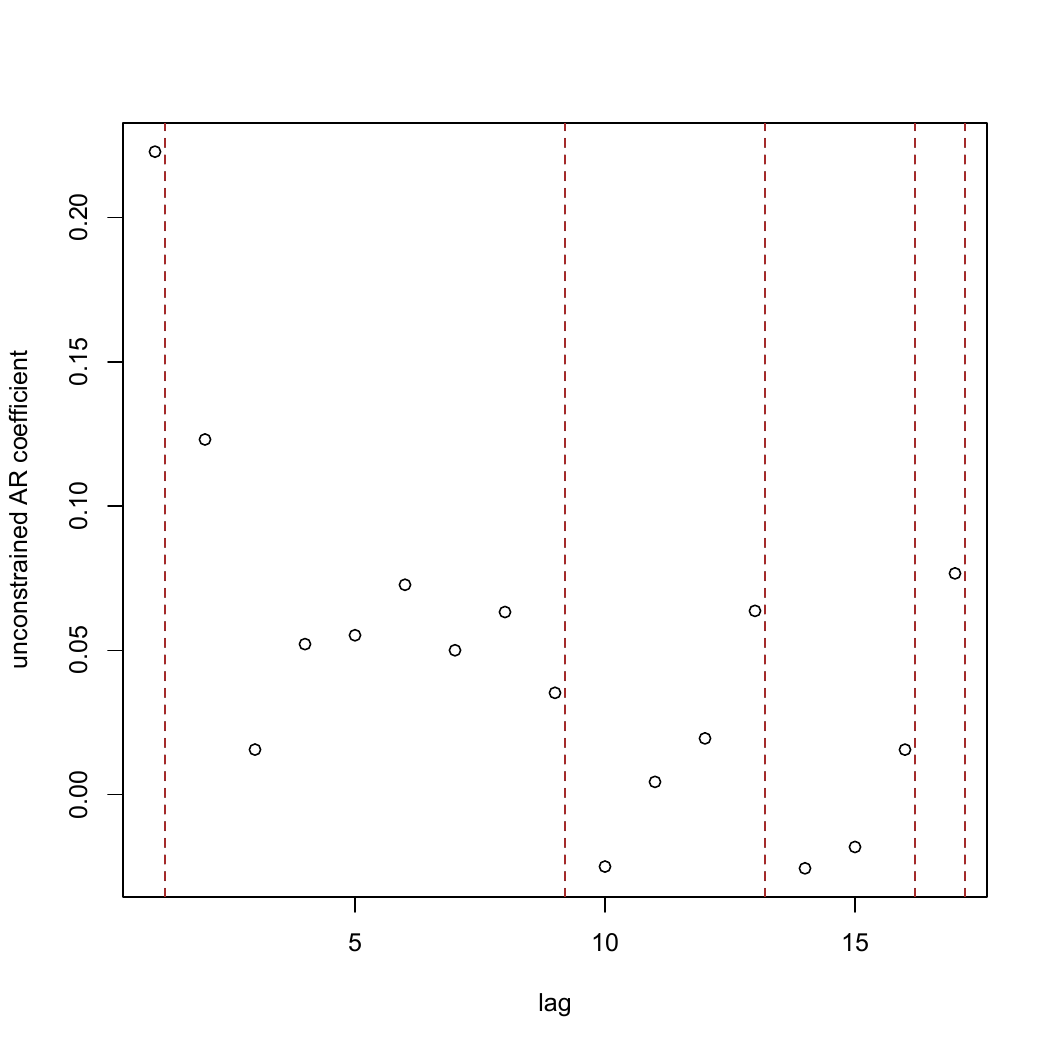}
	\end{minipage}%
	\begin{minipage}{.45\textwidth}
		\centering
		\includegraphics[width=\linewidth]{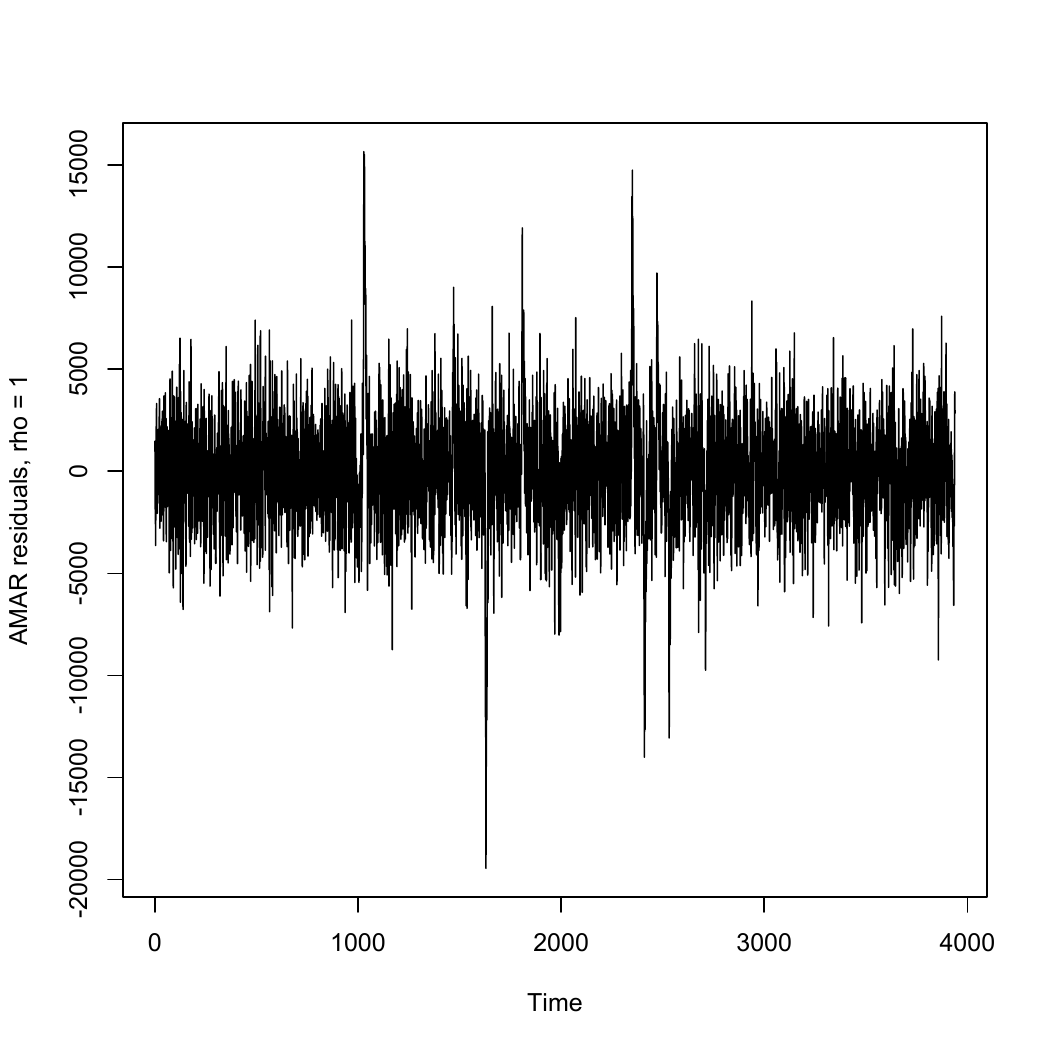}
	\end{minipage}\\
	\begin{minipage}{.45\textwidth}
		\centering
		\includegraphics[width=\linewidth]{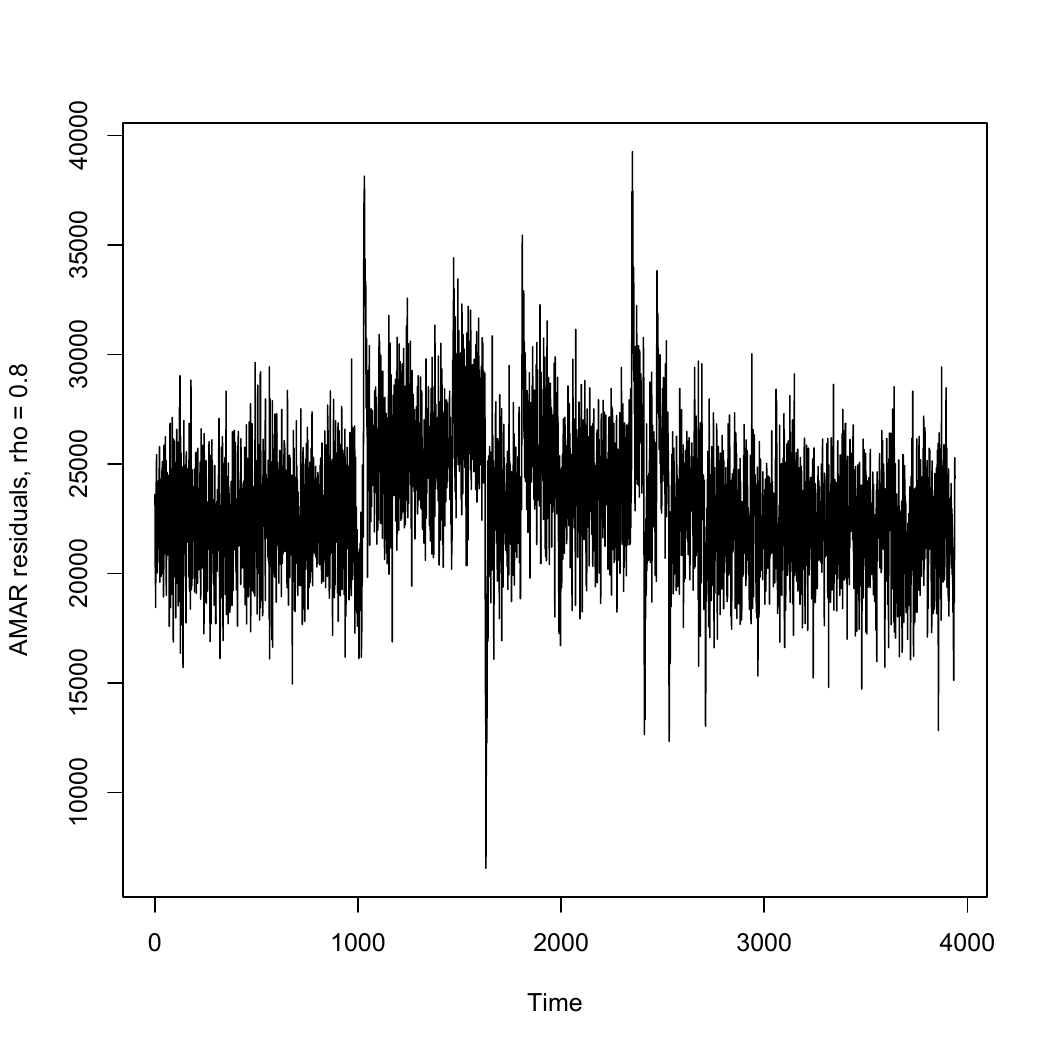}
	\end{minipage}%
	\begin{minipage}{.45\textwidth}
		\centering
		\includegraphics[width=\linewidth]{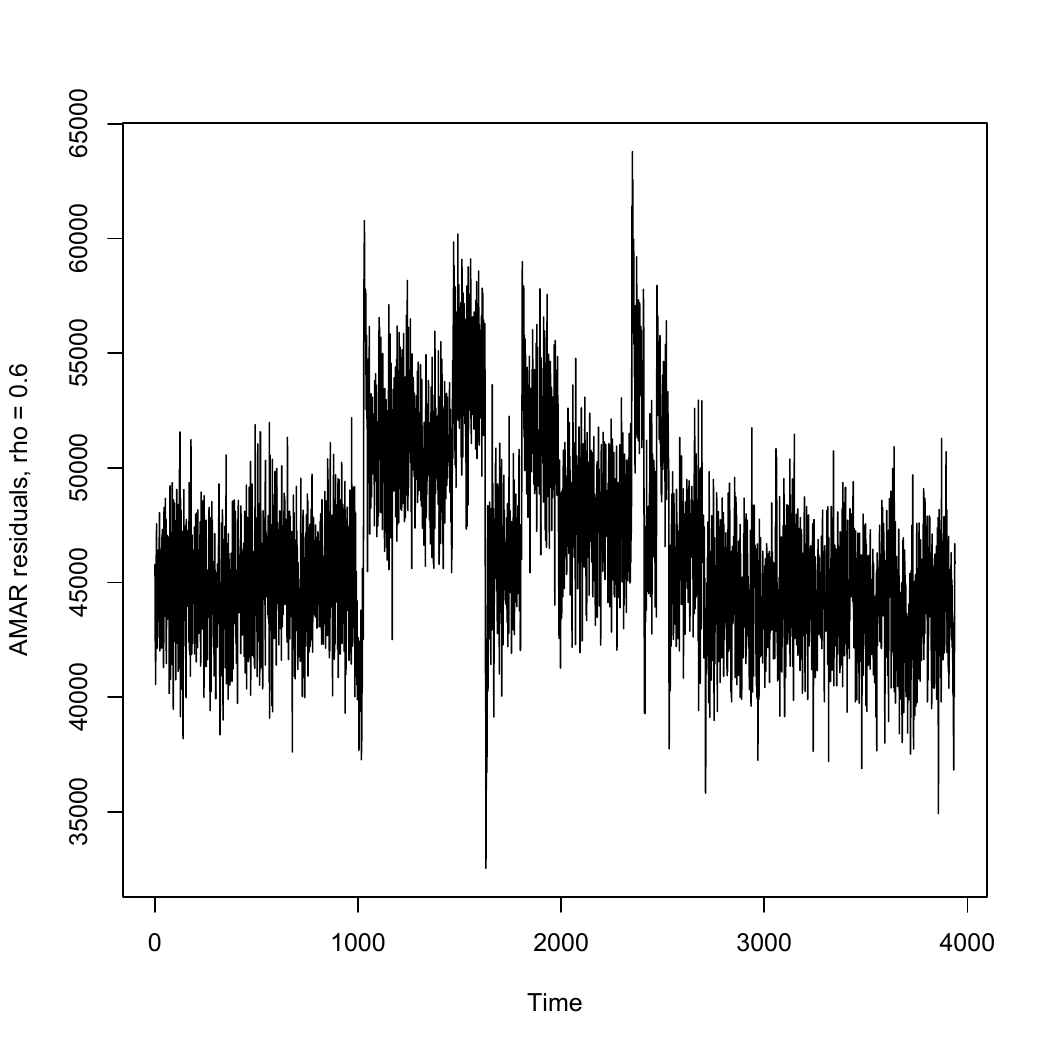}
	\end{minipage}
	\caption{Top left: the unconstrained estimated AR coefficients for $X_{2730:3956}$ (circles); extents of estimated AMAR scales (dashed lines). Top right: unshrunk AMAR residuals.
		Bottom left: AMAR residuals with $\rho = 0.8$. Bottom right: AMAR residuals with $\rho = 0.6$.\label{fig:wl2}}
\end{figure}

We start with a preliminary time series analysis of $\{X_{2730},\ldots, X_{3956}\}$. The unconstrained AR fit to this subset of the data, with the AR order chosen via AIC yields order 17, and the estimated coefficients are shown in the left panel of Figure \ref{fig:wl2}. The appearance of the vector of the estimated coefficients suggests that a piecewise-constant model (as dictated by AMAR) may be suitable here.
The  fitted   AMAR model returns estimated scales $1,  9, 13, 16, 17$ (see Figure \ref{fig:wl2}).

Prior to fitting the estimated AMAR model to the entire dataset, however, we shrink the estimated AMAR coefficients by a factor of $\rho \in (0, 1)$, i.e. we replace each estimated AMAR coefficient $\hat{\alpha}_r$ by $\rho \hat{\alpha}_r$. This is done because the original estimated AMAR coefficients sum up to practically 1 (0.9998), and therefore fitting such a ``near-unit-root" AMAR model has a strong differencing effect, which as well as successfully removing the autocorrelations, could also potentially remove too much of the structure of the signal for successful detection of change-points in the levels.

\begin{figure}
	\centering
	\includegraphics[width=0.65\linewidth, height = 0.45\linewidth]{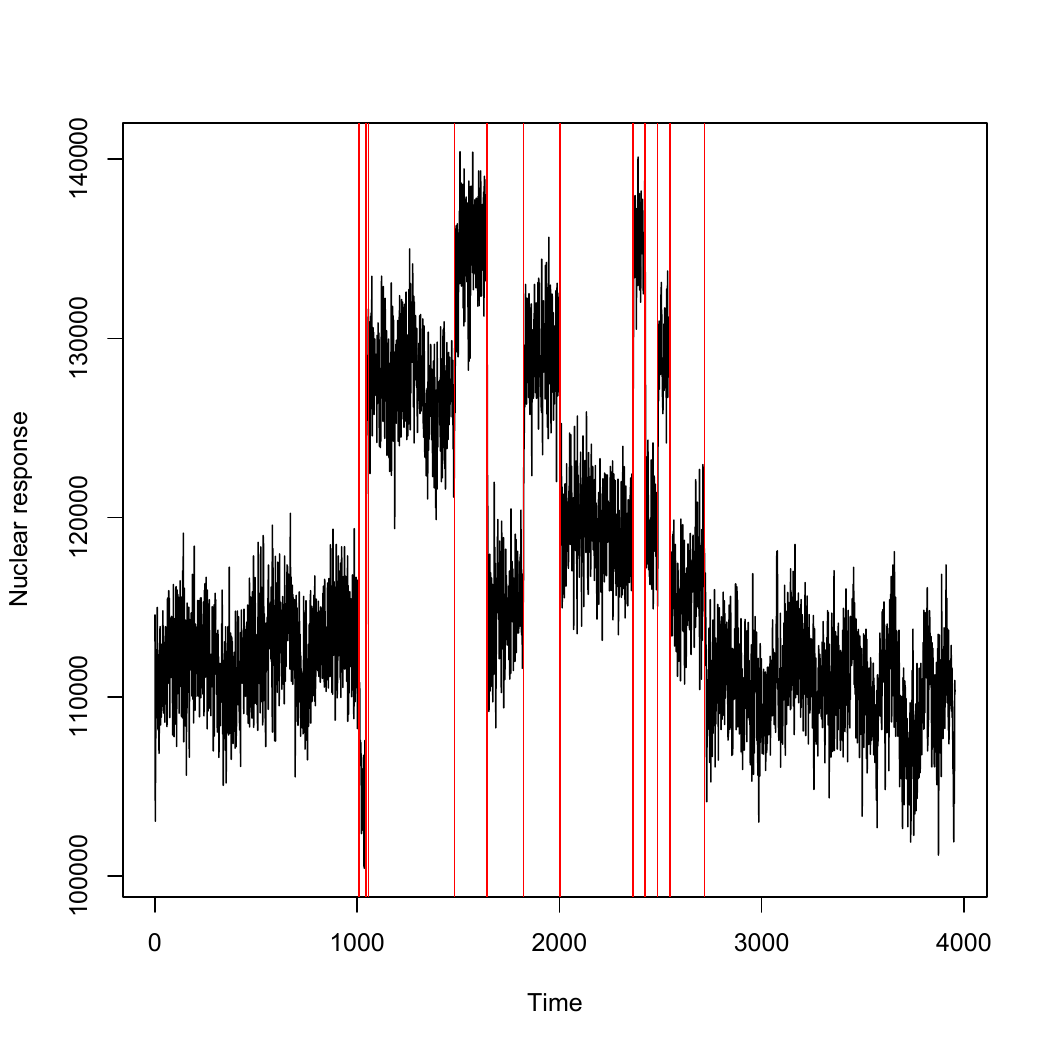}
	\caption{The well-log data with the change-point locations estimated via the shrunk AMAR fit with $\rho = 0.78$ and using WBS+sSIC on the residuals.\label{fig:wlwbs}}
\end{figure}

We choose $\rho$ as follows. Starting with $\rho = 0$, we increase $\rho$ in steps of $0.01$ until our selected procedure(s) for change-point detection under lack of serial correlation do not indicate any change-points after time $t = 2730$ (since we initially fitted an AMAR model on this portion of the data under the assumption of stationarity there). This is first achieved for $\rho = 
0.78$, for both Wild Binary Segmentation \citep{f14} and Narrowest-Over-Threshold \citep{bcf19}, both with model selection via the strengthened Schwarz Information Criterion, and using the implementation from the R package \pkg{breakfast} \citep{breakfastpackage} with otherwise default parameters.
These two procedures indicate, respectively, 12 and 10 change-points in the signal.
The change-point locations estimated via Wild Binary Segmentation are shown in Figure \ref{fig:wlwbs}. With the exception of the possible double detection at times $t = 1043, 1056$, the estimated change-point locations visually align with the signal very well.

\section{Proofs of the theoretical results}
\label{Sec:proof}

\subsection{Proof of Proposition \ref{Prop:stationary}}
For AR($p$) processes, it has a stationary and causal solution if and only if all the roots of $b(z) = 0$ lie outside $\mathbb{T}$. 

For any AMAR($q$) with $\alpha_1,\ldots,\alpha_q$ (and the corresponding AR parameters $\beta_1,\ldots,\beta_p $), $\sum_{j=1}^q |\alpha_j| < 1$ under the AMAR framework is equivalent to
\[
\sum_{j=1}^p |\beta_j| = \sum_{j=1}^p \Big(\sum_{k: \tau_{k} \geq j} \frac{|\alpha_{k}|}{\tau_{k}}\Big) = \frac{|\alpha_1|}{\tau_1}\tau_1 + \cdots + \frac{|\alpha_q|}{\tau_q}\tau_q < 1 
\]
in view of Equations~\eqref{mzar:eq:mzar_model} and \eqref{mzar:eq:ar_p} and \eqref{mzar:eq:mzar_ar_beta}. Now since $\sum_{j=1}^p |\beta_j|<1$, $b(z): = 1-\beta_1z - \cdots - \beta_p z^p \ge 1 - |\beta_1|\|z\| - \cdots - |\beta_p|\|z\|^p \ge 1 - \sum_{j=1}^p |\beta_j| > 0$ for any $z \in \mathbb{T}$. As such,  all the roots of $b(z) = 0$ lie outside $\mathbb{T}$, which implies the existence of a causal stationary solution. 

Next, given $\alpha_1,\ldots,\alpha_q \ge 0$, we have that $\beta_1,\ldots,\beta_p \ge 0$. The existence of a causal stationary solution implies that all the roots of $b(z) = 0$ lie outside $\mathbb{T}$. Since $b(0)=1$ and $b(\cdot)$ is continuous, one would necessarily require $b(1)>0$. i.e. $\beta_1+\cdots+\beta_p < 1$.  This condition under the AMAR framework is equivalent to
\[
\sum_{j=1}^p \Big(\sum_{k: \tau_{k} \geq j} \frac{\alpha_{k}}{\tau_{k}}\Big) = \frac{\alpha_1}{\tau_1}\tau_1 + \cdots + \frac{\alpha_q}{\tau_q}\tau_q < 1,
\]
which is the same as $\sum_{j=1}^q |\alpha_j| < 1$ under non-negativity.
\hfill$\square$

\subsection{Proof of Theorem \ref{mzar:theorem:lse_in_ar_conistency_bound}}
\label{mzar:sec:large_deviations}

We write the \ar{p} model as
\begin{align}	
	\label{mzar:eq:ar_p_as_var_1}	\Yb_{t} = \Bb\Yb_{t-1} + \varepsilon_{t} \ub,\quad t = 1,\ldots, T,
\end{align}
where $\Yb_{t}=(X_{t},X_{t-1},...,X_{t-p+1})'$, the matrix of the coefficients  
\begin{align}
	\label{mzar:eq:B_matrix}
	\Bb=\begin{pmatrix}
		\beta_{1} & \beta_{2} & \cdots & \beta_{p} \\
		&\Ib_{p-1}&  & \mathbf{0}
	\end{pmatrix}
\end{align}
and $\ub=(1,0,\ldots,0)' \in \mathbb{R}^{p}$. We start with a few auxiliary results.

\begin{Lemma}[Parseval's identity, Theorem 1.9 in \cite{duoandikoetxea2000fourier}]
	\label{mzar:theorem:parsevals_identity}
	For any complex-valued sequence $\left\{f_{k}\right\}_{{k\in\Z}}$ such that $\sum_{{k\in\Z}}|f_{k}|^{2} < \infty$, the following identity holds
	\begin{align}\label{mzar:eq:parsevals_identity}
		\sum_{k\in\Z} |f_{k}|^{2} = \int_{\T} |f(z)|^{2}dm(z),
	\end{align}
	where $f(z) = \sum_{k\in\Z} f_{k}z^{k}$, $\T=\{z\in\mathbb{C}:\;|z|=1\}$,  $dm(z)= \frac{d|z|}{2\pi}$.
\end{Lemma}

\begin{Lemma}[Cauchy's integral formula]
	\label{mzar:lemma:cauchy_integral_formula} 
	Let $\Mb\in\R^{p\times p}$ be a real- or complex- valued matrix. Then for any curve $\Gamma$ enclosing all eigenvalues of $\Mb$ and any $j\in\N$ the following holds 
	\begin{align}
		\Mb^{j} = \frac{1}{2\pi i}\int_{\Gamma} z^{j} (z\Ib_{p}-\Mb)^{-1} dz= \frac{1}{2\pi i}\int_{\Gamma} z^{j-1} (\Ib_{p}-z^{-1}\Mb)^{-1} dz.\label{mzar:eq:cauchy_integral_formula}
	\end{align}	
\end{Lemma}

\begin{Lemma}
	\label{mzar:lemma:taylor_coef_x_over_b}
	Let $\Bb$ given by \eqref{mzar:eq:B_matrix} be the matrix of coefficients of a stationary $AR(p)$ process and let $\vb = (v_1,\ldots, v_p)'\in\R^p$. For all $z\in\C$ such that $\sum_{i=0}^{\infty} |\ip{\vb}{\Bb^{i}\ub}||z^{i}| < \infty$, we have
	\begin{align}
		\label{mzar:eq:taylor_coef_x_over_b}
		b(z)\sum_{i=0}^{\infty} \ip{\vb}{\Bb^{i}\ub} z^{i} = b(z)\ip{\vb}{(\Ib_{p}-z\Bb)^{-1}\ub}  = v(z),
	\end{align}
	where $v(z)=v_{1} + v_{2}z +\ldots+v_{p}z^{p-1}$, and where $b(z)$ is the AR polynomial.    
\end{Lemma}
{\bf Proof.} As $\sum_{i=0}^{\infty} |\ip{\vb}{\Bb^{i}\ub}||z^{i}|<0$,  we can change the order of summation in the left-hand side of \eqref{mzar:eq:taylor_coef_x_over_b}
\begin{align*}		(1-\beta_{1}z-\ldots-\beta_{p}z^{p})\sum_{i=0}^{\infty} \ip{\vb}{\Bb^{i}\ub} z^{i} 	= \ip{\vb}{ \left(\sum_{i=0}^{\infty}(1-\beta_{1}z-\ldots-\beta_{p}z^{p})z^{i}\Bb^{i}\right)\ub}.
\end{align*}
Define $\beta_{0} = -1$, $\beta_{k}=0$ for $k>p$. By direct algebraic manipulation, 
\begin{align*}
	\sum_{i=0}^{\infty}(1-\beta_{1}z-\ldots-\beta_{p}z^{p})z^{i}\Bb^{i} 			=
	- \sum_{i=0}^{\infty} \left(\sum_{k=0}^{i}\beta_{k}\Bb^{i-k}\right)z^{i} := - \sum_{i=0}^{\infty} \Db_{i}z^{i}.
\end{align*}
The characteristic polynomial of $\Bb$ is given by $\phi(z)= \sum_{k=0}^{p}\beta_{k}z^{p-k}$. From the Cayley--Hamilton theorem, 
$\Bb$ is a root of $\phi$, and, consequently for $i\geq p$,
\begin{align*}
	\Db_{i} = \Bb^{i-p}\sum_{k=0}^{i}\beta_{k}\Bb^{p-k} = \Bb^{i-p}\sum_{k=0}^{p}\beta_{k}\Bb^{p-k} =0.
\end{align*}
It remains to demonstrate that $\ip{\vb}{\Db_{i}\ub} = -v_{i+1}$ for $i=0,\ldots,p-1$, which we show by induction. For $i=0$, $\ip{\vb}{\Db_{i}\ub}= \beta_{0} \ip{\vb}{\ub} = -v_{1}$. When $i \geq 1$, matrices $\Db_{i}$ satisfy
$
\Db_{i} = \Bb\Db_{i-1} + \beta_{i}\Ib_{p} ,
$
therefore
\begin{align*}
	\ip{\vb}{\Db_{i}\ub} & = 
	\ip{\vb}{\Bb\Db_{i-1}\ub} + \beta_{i}\ip{\vb}{\ub}  =
	\ip{\Bb'\vb}{\Db_{i-1}\ub} + \beta_{i}\ip{\vb}{\ub} \\
	& =
	\ip{v_{1}(\beta_{1},\ldots,\beta_{p})'+(0,v_{2},\ldots,v_{p})'}{\Db_{i-1}\ub} + \beta_{i} \ip{\vb}{\ub}  
	=-v_{1}\beta_{i} -v_{i+1} +v_{1}\beta_{i} \\
	& = -v_{i+1},
\end{align*}
which completes the proof.\hfill$\square$

\begin{Lemma}	
	\label{mzar:lemma:exp_ineq_gauss_prod}
	Let $Z_{1},Z_{2},\ldots$ be a sequence of i.i.d. $\mathcal{N}(0,1)$ random variables. Then for any integers $l\neq 0$ and $k>0$, the following exponential probability bound holds  for any $x>0$:
	\begin{align}
		\label{mzar:eq:exp_ineq_gauss_prod}
		\Pb{\left|\sum_{t=1}^{k}Z_{t}Z_{t+l}\right|  >  kx} &\leq 2\Exp{-\frac{1}{8}\frac{kx^2}{4+x}}.
	\end{align}
\end{Lemma}
{\bf Proof.} 
We will show that $\Pb{\sum_{t=1}^{k}Z_{t}Z_{t+l}  >  kx} \leq \Exp{-\frac{1}{8}\frac{kx^2}{4+x}}$, which would then imply  \eqref{mzar:eq:exp_ineq_gauss_prod} by symmetry.  
By Markov's inequality, for any $x>0$ and $\lambda>0$, it holds that 
\begin{align*}
	\Pb{\sum_{t=1}^{k}Z_{t}Z_{t+l}  >  kx} \leq \Exp{-kx\lambda} \E\Exp{\lambda \sum_{t=1}^{k}Z_{t}Z_{t+l}}.
\end{align*}
By the convexity of $y\mapsto \Exp{\lambda y}$ for any $\lambda>0$, Theorem~1 in \cite{vershynin2011simple} implies
\begin{align*}
	\E\Exp{\lambda \sum_{t=1}^{k}Z_{t}Z_{t+l}} \leq \E\Exp{4\lambda \sum_{t=1}^{k}Z_{t}\tilde{Z}_{t}},
\end{align*}
where $\tilde{Z}_{1},\ldots, \tilde{Z}_{k}$ are independent copies of $Z_{1},\ldots,Z_{k}$. Using the independence and by direct computation  (see also \cite{Craig1936}), we get
\begin{align*}
	\E\Exp{4\lambda \sum_{t=1}^{k}Z_{t}\tilde{Z}_{t}} =\left(\E\Exp{4\lambda Z_{1}\tilde{Z}_{1}}\right)^k =\left(\E\Exp{8\lambda^2\tilde{Z}_{1}^2}\right)^k
	= \left(1 - 16\lambda^{2}\right)^{-\frac{1}{2}k}  
\end{align*}
provided that $0<\lambda<\frac{1}{4}$, therefore $\Pb{\sum_{t=1}^{k}Z_{t}Z_{t+l}  >  kx} \leq   \Exp{-kx\lambda - \frac{k}{2}\log \left(1 - 16\lambda^{2}\right)}$.
Taking $\lambda=\frac{-2 + \sqrt{4 + x^2}}{4 x}$ minimises the right-hand side of this inequality. With this value of $\lambda$ and using $\log(x) \leq x-1$, we have
\begin{align*}
	\Pb{\sum_{t=1}^{k}Z_{t}Z_{t+l}  >  kx} &\leq \Exp{\frac{k}{4} \left(2-\sqrt{x^2+4} + 2\log \left(\frac{1}{4} \left(\sqrt{x^2+4}+2\right)\right)\right)} \\
	&\leq \Exp{\frac{k}{4}\left(2-\sqrt{x^2+4} +  \frac{1}{2} \left(\sqrt{x^2+4}+2\right) -2\right)}\\
	&= \Exp{\frac{k}{8}\left(2-\sqrt{x^2+4}\right)} =  \Exp{-\frac{1}{8}\frac{kx^2}{2+\sqrt{x^2+4}}}\\
	&\leq \Exp{-\frac{1}{8}\frac{kx^2}{4+x}},
\end{align*}
which completes the proof.\hfill$\square$
\begin{Lemma}[Lemma~1 in \cite{laurent2000adaptive}]
	\label{mzar:lemma:exp_ineq_gauss_squares}
	Let $Z_{1},Z_{2},\ldots$ be a sequence of i.i.d. $\mathcal{N}(0,1)$ random variables. For any integer $k>0$ and $x>0$, the following exponential probability bounds hold
	\begin{align} 
		\label{mzar:eq:exp_ineq_gauss_squares_upper}\Pb{\sum_{t=1}^{k}Z_{t}^2 \geq k+2\sqrt{kx}+2x} &\leq \Exp{-x}, \\\
		\label{mzar:eq:exp_ineq_gauss_squares_lower}\Pb{\sum_{t=1}^{k}Z_{t}^2 \leq k-2\sqrt{kx}} &\leq \Exp{-x}.
	\end{align}
\end{Lemma}

\vspace{10pt}

{\bf Proof of Theorem~\ref{mzar:theorem:lse_in_ar_conistency_bound}.}
For $\Cb_{T}= \sum_{t=1}^{T-1}\Yb_{t}\Yb_{t}'$ and $\Ab_{T}= \sum_{t=1}^{T-1} \varepsilon_{t+1}\Yb_{t} $, we have $\betahb-\betab=\Cb_{T}^{-1}\Ab_{T}$. Here the distribution of $\betahb-\betab$ is invariant to the value of $\sigma$. As such, in the following, we assume $\sigma=1$ for notational convenience. Consequently, 
\begin{align} 
	\label{mzar:eq:simple_bound_on_l2_norm_of_lse_error}
	\norm{\betahb-\betab}\leq \lambda_{max}(\Cb_{T}^{-1})\norm{\Ab_{T}}=\lambda_{min}^{-1}(\Cb_{T})\norm{\Ab_{T}},
\end{align}
where $\lambda_{min}(\Mb)$ and $\lambda_{max}(\Mb)$ denote, respectively, the smallest and the largest eigenvalues of a symmetric matrix $\Mb$. To provide an upper bound on $\norm{\betahb-\betab}$ given in Theorem~\ref{mzar:theorem:lse_in_ar_conistency_bound}, we will bound $\lambda_{min}(\Cb_{T})$ from below and $\norm{\Ab_{T}}$ from above, working on a set whose probability is large.

In the calculations below,  we will repeatedly use the following representation of $\Yb_{t}$, which follows from applying \eqref{mzar:eq:ar_p_as_var_1} recursively:
\begin{align}
	\label{mzar:eq:vma_1_representation}
	\Yb_{t} = \Bb^{t} \Yb_{0} +  \sum_{j=1}^{t} \varepsilon_{t-j+1} \Bb^{j-1}\ub,\quad t=1,\ldots,T.
\end{align}
In addition, to improve the presentational aspect of the proof, here we shall take $\Yb_{0} = \textbf{0}$. All the results would go through (with minor modifications to handle the extra terms) if one instead assumes that $\Yb_{0}$ is a realization from a stationary solution.

In the arguments below, we will show result more specific than \eqref{mzar:eq:bound_on_lse_norm}, i.e. 
\begin{align}
	\label{mzar:eq:lse_in_ar_consit_proof_ub}
	\norm{\Ab_{T}} & \leq 	 \left(32 \bl^{-2} \sqrt{1+\norm{\betab}^{2}}\right) p \log(T)\sqrt{(1+\log(T+p))T},\\
	\label{mzar:eq:lse_in_ar_consit_proof_lb}\lambda_{min}(\Cb_{T}) &\geq \bu^{-2} \left(T-p(1+ 32 \log(T) \sqrt{T})\right),
\end{align}
on the event
\begin{align}
	\label{mzar:eq:lse_in_ar_consit_proof_ev_n}
	\Ec_{T} = \Ec_{T}^{(1)} \cap \Ec_{T}^{(2)} \cap \Ec_{T}^{(3)},
\end{align}
where 
\begin{align*}
	\Ec_{T}^{(1)}  &= \bigcap_{1\leq i < j \leq p}\left\{\left|\sum_{t=1}^{T-\max(i,j)}\varepsilon_{t}\varepsilon_{t+|i-j|}\right|< 32 \log(T) \sqrt{T-\max(i,j)}\right\}, \\
	\Ec_{T}^{(2)}  &= \bigcap_{j=1}^{T}\left\{\left|\sum_{t=1}^{T-j}\varepsilon_{t}\varepsilon_{t+j}\right|< 32 \log(T) \sqrt{T-j}\right\},\\
	\Ec_{T}^{(3)}  &= \left\{\sum_{t=1}^{T-p}\varepsilon_{t}^2 > T-p-2\sqrt{\log(T)(T-p)} \right\}.
\end{align*} 
Finally, we will demonstrate that $\Ec_{T}$ satisfies
\begin{align}
	\label{mzar:eq:lse_in_ar_consit_proof_ev_n_prob}
	\Pb{\Ec_{T}} \geq 1-\frac{5}{T}.
\end{align}
Thus, \eqref{mzar:eq:simple_bound_on_l2_norm_of_lse_error}, \eqref{mzar:eq:lse_in_ar_consit_proof_ub}, \eqref{mzar:eq:lse_in_ar_consit_proof_lb} and \eqref{mzar:eq:lse_in_ar_consit_proof_ev_n_prob} combined together imply the statement of Theorem~\ref{mzar:theorem:lse_in_ar_conistency_bound}.
The remaining part of the proof is split into three parts, in which we show  \eqref{mzar:eq:lse_in_ar_consit_proof_ub}, \eqref{mzar:eq:lse_in_ar_consit_proof_lb} and \eqref{mzar:eq:lse_in_ar_consit_proof_ev_n_prob} in turn. 

\vspace{10pt}

{\bf Upper bound for \texorpdfstring{$\norm{\Ab_{T}}$}{the norm of A matrix)}.} 
The Euclidean norm satisfies $\norm{\Ab_{T}}=\sup_{\vb:\in\R^{p},\norm{\vb}=1}|\ip{\vb}{\Ab_{T}}|$, therefore we consider inner products $\ip{\vb}{\Ab_{T}}$ where $\vb\in\R^{p}$ is any unit vector. By \eqref{mzar:eq:vma_1_representation},
\begin{align*}
	\ip{\vb}{\Ab_{T}} & = \sum_{t=1}^{T-1} \ip{\vb}{\Yb_{t}}\varepsilon_{t+1} = \sum_{t=1}^{T-1}\sum_{j=1}^{t} \ip{\vb}{\Bb^{j-1}\ub}\varepsilon_{t-j+1}\varepsilon_{t+1} 
	\\
	& =\sum_{j=1}^{T-1} \ip{\vb}{\Bb^{j-1}\ub} a_{j}, 
\end{align*}
where $a_{j}=  \sum_{t=j}^{T-1}\varepsilon_{t-j+1}\varepsilon_{t+1}= \sum_{t=1}^{T-j}\varepsilon_{t}\varepsilon_{t+j}$.

Lemma~\ref{mzar:lemma:cauchy_integral_formula} and Lemma~\ref{mzar:lemma:taylor_coef_x_over_b} applied to 
the right-hand side of the above equation yield
\begin{align*}
	\sum_{j=1}^{T-1} \ip{\vb}{\Bb^{j-1}\ub} a_{j} &=  \frac{1}{2\pi i} \int_{\T} \left(\sum_{j=1}^{T-1} z^{j-1} a_{j}\right)\ip{\vb}{(z\Ib_{p}-\Bb)^{-1}\ub}dz \\
	&= \frac{1}{2\pi i} \int_{\T} \left(\sum_{j=1}^{T-1} z^{j-1} a_{j}\right)\left(\sum_{j=1}^{p} z^{p-j} v_{j}\right) q(z) dz \\
	&= \frac{1}{2\pi i} \int_{\T} \left(\sum_{j=0}^{T+p-1} z^{j} c_{j}\right) q(z) dz,
\end{align*} 
where $q(z)=(z^{p}b(z^{-1}))^{-1}$ and $c_{j}=\sum_{i=0}^{j} a_{i+1}v_{p-j+i} $. Integrating by parts, we get
\begin{align*}
	\frac{1}{2\pi i} \int_{\T} \left(\sum_{j=0}^{T+p-1} z^{j} c_{j}\right) q(z) dz &= -\frac{1}{2\pi i} \int_{\T} \left(\sum_{j=0}^{T+p-1} z^{j+1} \frac{c_{j}}{j+1}\right) q'(z) dz,
\end{align*}
where $q'(\cdot)$ is the derivative of $q(\cdot)$.
Combining the calculations above and using the fact that $\T=\{z\in\mathbb{C}:\;|z|=1\}$, Cauchy's inequality and Lemma~\ref{mzar:theorem:parsevals_identity}, we obtain
\begin{align}
	\label{mzar:eq:lse_in_ar_consit_ub_proof_1}
	\left |\sum_{j=1}^{T-1} \ip{\vb}{\Bb^{j-1}\ub} a_{j} \right|  \leq  \sqrt{\sum_{j=0}^{T+p-1}\left(\frac{c_{j}}{j+1}\right)^2} \sqrt{\int_{\T}|q'(z)|^2 dm(z)},
\end{align}
where we recall that $dm(z)= \frac{d|z|}{2\pi}$. 
To further bound the first term on the right-hand side of \eqref{mzar:eq:lse_in_ar_consit_ub_proof_1}, we recall that on the event $\Ec_{T}$ coefficients $|a_{j}| \leq 32 \log(T) \sqrt{T}$, hence
\begin{align*}
	\sqrt{\sum_{j=0}^{T+p-1}\left(\frac{c_{j}}{j+1}\right)^2} & =
	\sqrt{\sum_{j=0}^{T+p-1}\frac{1}{(j+1)^2}\left(\sum_{i=0}^{j} a_{i+1}v_{p-j+i}\right)^2} \\
	& \leq \max_{j=0,\ldots,T+p-1}|a_{j}|\sqrt{\sum_{j=0}^{T+p-1}\frac{1}{(j+1)^2}\left(\sum_{i=0}^{j} |v_{p-j+i}|\right)^2}\\
	& \leq 32 \log(T) \sqrt{T} \sqrt{\sum_{j=0}^{T+p-1}\frac{j+1}{(j+1)^2}}  \\
	&\leq 32 \log(T) \sqrt{(1+\log(T+p))T}.
\end{align*}	
For the second term in \eqref{mzar:eq:lse_in_ar_consit_ub_proof_1}, we calculate the derivative $$q'(z) = -\frac{pz^{p-1} - \sum_{j=1}^{p}(p-j)\beta_{j}z^{p-j-1}}{(z^{p}b(z^{-p}))^2}$$ and use Lemma~\ref{mzar:theorem:parsevals_identity} to bound
\begin{align*}
	\sqrt{\int_{\T} |q'(z)|^2 dm(z)} &= 
	\sqrt{\int_{\T} \left|\frac{pz^{p-1} - \sum_{j=1}^{p}(p-j)\beta_{j}z^{p-j}}{(z^{p}b(z^{-p}))^2}\right|^2 dm(z)}	\\
	&\leq \frac{\sqrt{\int_{\T} \left|pz^{p-1} - \sum_{j=1}^{p}(p-j)\beta_{j}z^{p-j}\right|^2 dm(z)}}{\min_{|z|=1}|(z^{p}b(z^{-p}))|^2} \\
	&=	\bl^{-2} \sqrt{\left(p^{2}+ \sum_{j=1}^{p} (p-j)^2 \beta_{j}^{2}\right)} \leq  \bl^{-2} p \sqrt{1+\norm{\betab}^{2}}.
\end{align*}
Combining the bounds on the two terms, we obtain
\begin{align*}
	\sum_{j=1}^{T-1} \ip{\vb}{\Bb^{j-1}\ub} a_{j} \leq \left(32 \bl^{-2} \sqrt{1+\norm{\betab}^{2}}\right) p \log(T)\sqrt{(1+\log(T+p))T}.
\end{align*}
Taking supremum over $\vb\in\R^p$ such that $\norm{\vb}=1$ proves \eqref{mzar:eq:lse_in_ar_consit_proof_ub}.

\vspace{10pt}

{\bf Lower bound for $\lambda_{min}(\Cb_{T})$.}
Let $\vb=(v_1,\ldots,v_p)'$ be a unit vector in $\R^{p}$. We begin the proof by establishing the following inequality
\begin{align}
	\label{mzar:eq:lse_in_ar_consit_lb_proof_1}
	\ip{\vb}{\Cb_{T}\vb} \geq \bu^{-2} \sum_{i,j=1}^{p} v_{i}v_{j} \sum_{t=1}^{T-1}\varepsilon_{t-j+1}\varepsilon_{t-i+1},
\end{align} 
where $\varepsilon_{t}=0$ for $t\leq0$ and $\bu = \max_{z\in\T}|b(z)|$. By Lemma~\ref{mzar:theorem:parsevals_identity} and \eqref{mzar:eq:vma_1_representation}, we rewrite the quadratic form on the left-hand side of \eqref{mzar:eq:lse_in_ar_consit_lb_proof_1} to 
\begin{align}
	\label{mzar:eq:lse_in_ar_consit_lb_proof_2}
	\ip{\vb}{\Cb_{T}\vb} &=  \sum_{t=1}^{T-1}\ip{\vb}{\Yb_{t}}^2 \\
	&= \int_{\T} \left|\sum_{t=1}^{T-1} \ip{\vb}{\sum_{j=1}^{t} \varepsilon_{j} \Bb^{t-j}\ub}z^{t}\right|^2 dm(z) \\
	&=  \int_{\T} \left|\sum_{t=1}^{T-1}\sum_{j=1}^{T-1}  \varepsilon_{j} \omega_{t-j} z^t\right|^2 dm(z) 
\end{align}
where $\omega_{j} = \ip{\vb}{\Bb^{j}\ub}$ for $j\geq0$, $\omega_{j}=0$ for $j<0$. Changing the order of summation and by a simple substitution we get
\begin{align}
	\label{mzar:eq:lse_in_ar_consit_lb_proof_3}
	\sum_{t=1}^{T-1}\sum_{j=1}^{T-1}  \varepsilon_{j} \omega_{t-j} z^t = \sum_{j=1}^{T-1}  \varepsilon_{j} z^{j}  \sum_{t=1}^{T-1} \omega_{t-j} z^{t-j} =
	\sum_{j=1}^{T-1}  \varepsilon_{j} z^{j} \sum_{t=0}^{T-j-1} \omega_{t} z^{t}.
\end{align}
Using the  definition of $\omega_j$, the fact that all eigenvalues of $\Bb$ have modulus strictly lower than one and Lemma~\ref{mzar:lemma:taylor_coef_x_over_b}, \eqref{mzar:eq:lse_in_ar_consit_lb_proof_3} simplifies to 
\begin{align*}
	\sum_{j=1}^{T-1}  \varepsilon_{j} z^{j} \sum_{t=0}^{T-j-1} \omega_{t} z^{t} &= 
	\sum_{j=1}^{T-1}  \varepsilon_{j} z^{j} \ip{\vb}{(\Ib_{p}-(\Bb z)^{T-j})(\Ib_{p}-\Bb z)^{-1}\ub} \\
	&=\sum_{j=1}^{T-1}  \varepsilon_{j} \left(z^{j} \ip{\vb}{(\Ib_{p}-\Bb z)^{-1}\ub} - z^{T} \ip{\Bb^{T-j}\vb}{(\Ib_{p}-\Bb z)^{-1}\ub}\right) \\
	&= b(z)^{-1} \sum_{j=1}^{T-1}  \varepsilon_{j} \left(z^{j} v(z) - z^{T} w_{j}(z)\right),
\end{align*}
where $v(z)=\sum_{k=1}^{p}v_{k}z_{k-1}$ and $w_{j}(z)=\sum_{k=1}^{p}(\Bb^{T-j}v)_{k}z^{k-1}$ for $j=0,\ldots, T-1$. The equation above, \eqref{mzar:eq:lse_in_ar_consit_lb_proof_2} and \eqref{mzar:eq:lse_in_ar_consit_lb_proof_3} combined together imply the following inequality
\begin{align*}
	\ip{\vb}{\Cb_{T}\vb} &= \int_{\T} \left|b(z)^{-1} \sum_{j=1}^{T-1}  \varepsilon_{j} \left(z^{j} v(z) - z^{T} w_{j}(z)\right)\right|^{2} dm(z) \\
	&\geq \bu^{-2} \int_{\T} \left|\sum_{j=1}^{T-1}  \varepsilon_{j} \left(z^{j} v(z) - z^{T} w_{j}(z)\right)\right|^{2} dm(z).
\end{align*}
Observe that $\sum_{j=1}^{T-1}  \varepsilon_{j} \left(z^{j} v(z) - z^{T} w_{j}(z)\right) = \sum_{j=1}^{T-1}  \varepsilon_{j} \left(z^{j} v(z) - z^{T} w_{j}(z)\right) = \sum_{t=1}^{T+p-1} c_{t} z^{t}$  is a trigonometric polynomial, therefore by Lemma~\ref{mzar:theorem:parsevals_identity} and simple algebra
\begin{align*}
	\int_{\T} \left|\sum_{j=1}^{T-1}  \varepsilon_{j} \left(z^{j} v(z) - z^{T} w_{j}(z)\right)\right| dm(z) &=  \sum_{t=1}^{T+p-1} |c_{t}|^2 \geq 
	\sum_{t=1}^{T-1} |c_{t}|^2 = \sum_{t=1}^{T-1} \left(\sum_{j=1}^{p} v_{j}\varepsilon_{t-j+1}\right)^2 = \\
	&= \sum_{i,j=1}^{p} v_{j}v_{i} \sum_{t=1}^{T-1}\varepsilon_{t-j+1}\varepsilon_{t-i+1},
\end{align*} 
which proves \eqref{mzar:eq:lse_in_ar_consit_lb_proof_1}.

We are now in a position to bound $\ip{\vb}{\Cb_{T}\vb}$ from below. Rearranging terms in \eqref{mzar:eq:lse_in_ar_consit_lb_proof_1} yields
\begin{align*}
	\ip{\vb}{\Cb_{T}\vb} &\geq \bu^{-2} \left(\sum_{i=1}^{p} v_{i}^2 \sum_{t=1}^{n-i}\varepsilon_{t}^2 + \sum_{1\leq i<j\leq p} v_{i}v_{j} \sum_{t=1}^{T-\max(i,j)}\varepsilon_{t}\varepsilon_{t+|j-i|}\right)  \\
	&\geq  \bu^{-2} \left(\sum_{t=1}^{T-p}\varepsilon_{t}^2 \sum_{i=1}^{p} v_{i}^2 - \max_{1\leq i<j\leq p}\left|\sum_{t=1}^{T-\max(i,j)}\varepsilon_{t}\varepsilon_{t+|j-i|}\right|\left(\left(\sum_{i=1}^{p} |v_{i}|\right)^2 - \sum_{i=1}^{p} v_{i}^2 \right) \right) \\
	& \geq \bu^{-2} \left(\sum_{t=1}^{T-p}\varepsilon_{t}^2  - \left(p-1\right)\max_{1\leq i<j\leq p}\left|\sum_{t=1}^{T-\max(i,j)}\varepsilon_{t}\varepsilon_{t+|j-i|}\right| \right).
\end{align*}
Recalling the definition of $\Ec_{T}$, we conclude that on this event
\begin{align*}
	\ip{\vb}{\Cb_{T}\vb} &\geq \bu^{-2} \left(T-p - 2 \sqrt{\log(T) (T-p)} -(p-1) 32 \log(T) \sqrt{T}\right) \\
	&\geq \bu^{-2} \left(T-p(1+ 32 \log(T) \sqrt{T})\right).
\end{align*}
Taking infimum over $\vb\in\R^p$ such that $\norm{\vb}=1$ in the inequality above proves \eqref{mzar:eq:lse_in_ar_consit_proof_lb}.

\vspace{10pt}

{\bf Lower bound for $\Pb{\Ec_{T}}$.} Recalling \eqref{mzar:eq:lse_in_ar_consit_proof_ev_n} and using a simple Bonferroni bound, we get
\begin{align*}
	\Pb{\Ec_T^c} &\leq p^{2} \max_{1\leq i < j\leq p} \Pb{\left|\sum_{t=1}^{T-\max(i,j)}\varepsilon_{t}\varepsilon_{t+|i-j|}\right| \geq 32 \log(T) \sqrt{T-\max(i,j)}} \\	
	&+ T 	\max_{1\leq j\leq T} \Pb{\left|\sum_{t=1}^{T-j}\varepsilon_{t}\varepsilon_{t+j}\right|< 32 \log(T) \sqrt{T-j}}\\
	&+  \Pb{\sum_{t=1}^{T-p}\varepsilon_{t}^2 > T-p-2\sqrt{\log(T)(T-p)}}\\
	&:= p^{2} \max_{1\leq i < j\leq p} P_{i,j}^{(1)} + T \max_{1\leq j\leq T}  P_{j}^{(2)} + P^{(3)}.
\end{align*}
Lemma~\ref{mzar:lemma:exp_ineq_gauss_prod} implies that
\begin{align*}
	P_{i,j}^{(1)} &\leq 2\Exp{-\frac{1}{8} \frac{(32\log(T))^2}{4+(\sqrt{T-\max(i,j)})^{-1}32\log(T)}} \leq 2\Exp{-2\log(T)} = \frac{2}{T^2},\\
	P_{j}^{(2)} &\leq 2\Exp{-\frac{1}{8} \frac{(32\log(T))^2}{4+(\sqrt{T-j})^{-1}32\log(T)}} \leq 2\Exp{-2\log(T)} = \frac{2}{T^2}.
\end{align*}
Moreover, by Lemma~\ref{mzar:lemma:exp_ineq_gauss_squares}, $P^{(3)} \leq \Exp{-\log(T)} = \frac{1}{T}$,
hence, given that $p^2 < T$, we have $\Pb{\Ec_T^c} \leq \frac{5}{T}$, which completes the proof.\hfill$\square$


\subsection{Proof of Theorem~\ref{mzar:theorem:consistency_result}}
\label{mzar:sec:proof_consistency}
In the proof below, we shall focus on the case where $F_T^M$ consists of randomly drawn intervals (which is what Algorithm~\ref{mzar:alg:mzar_algorithm} does when $p$ is large). For the case where all sub-intervals of $[1,p]$ are used, the same arguments would go through, because Algorithm~\ref{mzar:alg:mzar_algorithm} then produces a larger set $F_T^M$ compared to the approach of random drawing. 

We now split the proof into four steps.

\vspace{10pt}

{\bf Step 1.} Consider the event	$\left\{\norm{\betahb-\betab} \leq \kappa_{1}(\bl/\bu)^2 \norm{\betab}\frac{p\log(T)\sqrt{\log(T+p)}}{\sqrt{T}-\kappa_{2}p\log(T) } \right\}$
where $\kappa_{1}, \kappa_{2}$ are as in Theorem~\ref{mzar:theorem:lse_in_ar_conistency_bound}. Assumption~\ref{mzar:as:strict_stationarity} implies that $\bl/\bu$ and $\norm{\betab}$ are bounded from above by constants. Furthermore, by Assumption~\ref{mzar:as:maximum_changepoint}, $p \leq c_{1} T^{\theta}$, which implies that 
\begin{align}
	\label{eq:c_3}
	\kappa_{1}(\bl/\bu)^2 \norm{\betab}\frac{p\log(T)\sqrt{\log(T+p)}}{\sqrt{T}-\kappa_{2}p\log(T) } \leq c_3 T^{\theta-1/2} (\log(T))^{3/2} = c_3 \underline{\lambda}_{T} =: \lambda_{T}
\end{align}
for some constant $c_3>0$ and a sufficiently large $T$. Define now
\begin{align}
	A_{T} = \left\{\norm{\betahb-\betab} \leq \lambda_{T}\right\}
\end{align}
By Theorem~\ref{mzar:theorem:lse_in_ar_conistency_bound},
\begin{align}
	\Pb{A_{T}} \geq \Pb{\norm{\betahb-\betab} \leq \kappa_{1}(\bl/\bu)^2 \norm{\betab}\frac{p\log(T)\sqrt{\log(T+p)}}{\sqrt{T}-\kappa_{2}p\log(T) }} \geq 1 - \kappa_{3}T^{-1},
\end{align}
for some constant $\kappa_{3}>0$.

\vspace{10pt}

{\bf Step 2.}
For $j=1,\ldots,q$, define the intervals 
\begin{align}
	\Ic_{j}^{L} &= (\tau_{j} -\delta_T/3, \tau_{j} - \delta_T/6) \label{mzar:eq:intervals_arnd_cpts_left}\\
	\Ic_{j}^{R} &= (\tau_{j} +\delta_T/6, \tau_{j}+\delta_T/3)\label{mzar:eq:intervals_arnd_cpts_right}
\end{align}
Recall that $F_{T}^{M}$ is the set of $M$ randomly drawn intervals with endpoints in $\{1,\ldots, p\}$.  Denote by $[s_{1},e_{1}], \ldots, [s_{M},e_{M}]$ the elements of $F_{T}^{M}$ and let 
\begin{align}
	D^{M}_{T}  = \Big\{\forall j=1,\ldots,q, \exists k \in \{1,\ldots,M\}, \; \mbox{s.t.} \; s_k \times e_k \in \Ic_{j}^{L}\times\Ic_{j}^{R}\Big\} \label{mzar:eq:event_D_M_T}.
\end{align}
We have that 
\begin{align*}
	\Pb{(D^{M}_{T})^{c}} &\leq \sum_{j=1}^{q}\Pi_{m=1}^{M}\Big(1-\Pb{s_m \times e_m \in \Ic_{j}^{L}\times\Ic_{j}^{R}}\Big) \\
	&\le q  \left(1-\frac{\delta_{T}^{2}}{6^2 p^{2}}\right)^{M} \leq \frac{p}{\delta_{T}} \left(1 - \frac{\delta_{T}^{2}}{36 p^{2}}\right)^{M}.
\end{align*}
Therefore, $\Pb{A_{T}\cap D^{M}_T} \geq 1-  \kappa_{3}T^{-1} - p \delta_{T}^{-1}(1-\delta_T^{2}p^{-2}/36)^M \rightarrow 1$. 
Note that the same conclusion still holds if $F_{T}^M$ contains all the intervals with endpoints in $\{1,\ldots, p\}$. 
In the remainder of the proof, assume that  $A_{T}$ and $D^{M}_{T}$ all hold. 

Note that Assumption~\ref{mzar:as:changepoints_spacing} implies that there exists $\cl >0$ such that $\delta_{T}^{1/2} \all_{T} > \cl \lambda_T$ for all sufficiently large $T$. We are now in the position to specify the constants explicitly as
\[C_1 = 2\sqrt{C_3} + c_3, \quad C_2 = \frac{1}{\sqrt{6}} - \frac{1}{\cl}, \quad C_3 = (4\sqrt{2}+6)c_3^2,\]
where $c_3$ is in Equation~\eqref{eq:c_3}.

\vspace{10pt}

{\bf Step 3.} We focus on a generic interval $[s,e]$ such that
\begin{align}
	\exists j \in \{1,\ldots,q\}, \; \exists k \in \{1,\ldots,M\}, \; \mbox{s.t.}\; [s_{k},e_{k}]\subset [s,e] \mbox{ and } s_k \times e_k \in\Ic_{j}^{L}\times\Ic_{j}^{R}. \label{mzar:eq:main_th_proof_there exists interval}
\end{align}
Fix such an interval $[s,e]$ and let $j \in \{1,\ldots,q\}$ and $k \in \{1,\ldots,M\}$ be such that \eqref{mzar:eq:main_th_proof_there exists interval} is satisfied. 
Let $b_k^{*}=\argmax_{s_{k}\leq b\le  e_{k}} \cont{s_k}{e_k}{b}{\betahb}$. By construction, $[s_k,e_k]$ satisfies $\tau_j-s_{k}+1 \ge \delta_{T}/6$ and $e_{k}-\tau_j > \delta_{T}/6$.  
Let
\begin{align*}
	\Mc_{s,e}&=\left\{m:[s_{m},e_{m}]\in F_{T}^{M}, [s_{m},e_{m}]\subset[s,e] \right\},\\
	\Oc_{s,e}&=\{m\in\Mc_{s,e}:\max_{s_m \leq b < e_m} \cont{s_m}{e_m}{b}{\betahb} > \zeta_{T}\}.
\end{align*}
Our first aim is to show that $\Oc_{s,e}$ is non-empty. This follows from Lemma 2 in \cite{bcf19}, the Cauchy--Schwarz inequality, and the calculation below, as
\begin{align*}
	\cont{s_k}{e_k}{b^{*}_k}{\betahb} &\geq \cont{s_k}{e_k}{\tau_{j}}{\betahb} \\
	&\geq \cont{s_k}{e_k}{b^{*}_k}{\betab}-\lambda_{T} \geq \left(\frac{\delta_{T}}{6}\right)^{1/2} |\alpha_{j}\tau_{j}^{-1}|- \lambda_{T}\geq \left(\frac{\delta_{T}}{6}\right)^{1/2} \all_{T} - \lambda_{T} \\
	& = \left( \frac{1}{\sqrt{6}} - \frac{\lambda_{T}}{\delta_{T}^{1/2}\all_{T}}\right)\delta_{T}^{1/2}\all_{T}\geq \left(\frac{1}{\sqrt{6}} - \frac{1}{\cl} \right)\delta_{T}^{1/2}\all_{T}  =  C_2 \delta_{T}^{1/2}\all_{T} > \zeta_T.
\end{align*}

Let $m^{*}=\argmin_{m\in\Oc_{s,e}} (e_{m}-s_{m}+1)$ and $b^{*} = \argmax_{s_{m^*} \leq b < e_{m^*}} \cont{s_{m^*}}{e_{m^*}}{b}{\betahb}$. Observe that $[s_{m^*},e_{m^*})$ must contain at least one change in $\betahb$. Indeed, if this were not the case, we would have $\cont{s_{m^*}}{e_{m^*}}{b}{\betab}=0$ and
\begin{align*}
	\cont{s_{m^*}}{e_{m^*}}{b^{*}}{\betahb}=	|\cont{s_{m^*}}{e_{m^*}}{b^{*}}{\betahb} - 	\cont{s_{m^*}}{e_{m^*}}{b^{*}}{\betab}| \le \lambda_{T} < \frac{C_{1}}{c_3}\lambda_{T} =  C_{1}\underline{\lambda}_{T}  \leq\zeta_{T},
\end{align*}
which contradicted $\cont{s_{m^*}}{e_{m^*}}{b^{*}}{\betahb} >  \zeta_{T}$. On the other hand, $[s_{m^*},e_{m^*})$ cannot contain more than one change-points, because $e_{m^*}- s_{m^*} + 1 \le  e_{k}- s_{k} + 1 \le \delta_{T}$.

Without loss of generality, assume $\tau_{j} \in [s_{m^*}, e_{m^*})$. Let  $\eta_L=\tau_{j}-s_{m^*}+1$,
$\eta_R = e_{m^*}-\tau_{j}$ and $\eta_{T}= (C_1/c_3 - 1)^2 \alpha_{j}^{2}\tau_{j}^{-2}\lambda_{T}^{2}$.
We claim that $\min(\eta_L, \eta_R) > \eta_{T}$, because otherwise $\min(\eta_L, \eta_R) \le  \eta_{T}$ and Lemma 2 in \cite{bcf19}
would have implied
\begin{align*}	
	\cont{s_{m^*}}{e_{m^*}}{b^{*}}{\betahb} &\leq \cont{s_{m^*}}{e_{m^*}}{b^{*}}{\betab} +\lambda_{T} \leq  	\cont{s_{m^*}}{e_{m^*}}{\tau_{j}}{\betab} +\lambda_{T} \leq \eta_{T}^{1/2}|\alpha_{j}\tau_{j}^{-1}|+ \lambda_{T} \\
	&= (C_1/c_3 - 1+ 1) \lambda_{T} = C_1\underline{\lambda}_{T} < \zeta_{T},
\end{align*}
which contradicted  $\cont{s_{m^*}}{e_{m^*}}{b^{*}}{\betahb} > \zeta_{T}$.  

We are now in the position to prove $|b^* - \tau_j| \le  C_{3}\underline{\lambda}_{T}\all_{T}^{-2}$. Our aim is to find $\epsilon_T$ such that for any $b \in \{s_{m^*}, s_{m^*}+1,\ldots,e_{m^*}-1\}$ with $|b-\tau_j| > \epsilon_{T}$, we always have
\begin{align}
	\left\{\cont{s_{m^*}}{e_{m^*}}{\tau_j}{\betahb}\right\}^2 -  	\left\{\cont{s_{m^*}}{e_{m^*}}{b}{\betahb}\right\}^2 > 0. 
	\label{mzar:eq:main_th_proof_interval_rate1}
\end{align}
This would then imply that $|b^* - \tau_j| \le \epsilon_T$. By expansion and rearranging the terms, we see that (\ref{mzar:eq:main_th_proof_interval_rate1}) is equivalent to 
\begin{align}
	\notag
	& \langle\betab, \psib_{s_{m^*},e_{m^*}}^{\tau_j} \rangle^2 - \langle\betab, \psib_{s_{m^*},e_{m^*}}^{b} \rangle^2  >  \langle \betahb -\betab, \psib_{s_{m^*},e_{m^*}}^{b} \rangle^2  -  \langle \betahb -\betab, \psib_{s_{m^*},e_{m^*}}^{\tau_j} \rangle^2 \\
	& + 2 \Big\langle \betahb -\betab, \psib_{s_{m^*},e_{m^*}}^b  \langle\betab,\psib_{s_{m^*},e_{m^*}}^b \rangle - \psib_{s_{m^*},e_{m^*}}^{\tau_j} \langle\betab,\psib_{s_{m^*},e_{m^*}}^{\tau_j} \rangle \Big\rangle.
	\label{mzar:eq:main_th_proof_interval_rate2}
\end{align}
Here $\psib_{s,e}^b$ (with $1 \le s < b < e \le p$) is a $p$-dimensional vector, with its $s$-th to $b$-th component being $\sqrt\frac{e-b}{(e-s+1)()b-s+1}$, its $b+1$-th to $e$-th component being $\sqrt\frac{b-s+1}{(e-s+1)(e-b)}$, and the remaining elements being 0.
In the following, we assume that $b \ge \tau_j$. The case that $b < \tau_j$ can be handled in a similar fashion. By Lemma 4 in \cite{bcf19}, we  have
\begin{align*}
	\langle\betab, \psib_{s_{m^*},e_{m^*}}^{\tau_j} \rangle^2 - \langle\betab, \psib_{s_{m^*},e_{m^*}}^{b} \rangle^2 &= (\cont{s^*}{e^*}{\tau_j}{\betab})^2 -  (\cont{s_{m^*}}{e_{m^*}}{b}{\betab})^2 \\
	&= \frac{|b -\tau_j| \eta_L}{|b -\tau_j| +  \eta_L} (\alpha_{j}\tau_{j}^{-1})^2 =: \kappa.
\end{align*}
In addition, since we assume event $A_T$,
\begin{align*}
	&\langle \betahb -\betab, \psib_{s_{m^*},e_{m^*}}^{b} \rangle^2  -  \langle \betahb -\betab, \psib_{s_{m^*},e_{m^*}}^{\tau_j} \rangle^2  \le \lambda_T^2, \\
	&2 \Big\langle \betahb -\betab, \psib_{s_{m^*},e_{m^*}}^b \langle\betab,\psib_{s_{m^*},e_{m^*}}^b \rangle - \psib_{s_{m^*},e_{m^*}}^{\tau_j} \langle\betab,\psib_{s_{m^*},e_{m^*}}^{\tau_j} \rangle \Big\rangle  \\ 
	& \qquad \qquad  \qquad \le 2\| \psib_{s_{m^*},e_{m^*}}^b \langle\betab,\psib_{s_{m^*},e_{m^*}}^b \rangle - \psib_{s_{m^*},e_{m^*}}^{\tau_j} \langle\betab,\psib_{s_{m^*},e_{m^*}}^{\tau_j}\rangle\|_2 \lambda_T 
	= 2 \kappa^{1/2} \lambda_T,
\end{align*}
where the final equality is also implied by Lemma 4 in \cite{bcf19}. Consequently, \eqref{mzar:eq:main_th_proof_interval_rate2} can be deducted from the stronger inequality $\kappa - 2\lambda_T\kappa^{1/2} - \lambda_T^2 > 0$. This quadratic inequality is implied by $\kappa > (\sqrt{2}+1)^2\lambda_T^2$, and could be restricted further to  
\begin{align}
	\frac{2|b -\tau_j| \eta_L}{|b -\tau_j| +  \eta_L} \ge \min(|b -\tau_j|, \eta_L) > (4\sqrt{2}+6) (\alpha_{j} \tau_{j}^{-1})^{-2}\lambda_{T}^{2} = C_3  (\alpha_{j} \tau_{j}^{-1})^{-2}\underline{\lambda}_{T}^{2}.
	\label{mzar:eq:main_th_proof_interval_rate3}
\end{align}
But since 
\[
\eta_L \ge \eta_T = (C_1/c_3 -1)^2 (\alpha_{j} \tau_{j}^{-1})^{-2}\lambda_{T}^{2} = (2\sqrt{C_3}/c_3)^2  (\alpha_{j} \tau_{j}^{-1})^{-2}\lambda_{T}^{2} > C_3 (\alpha_{j} \tau_{j}^{-1})^{-2}\underline{\lambda}_{T}^{2},
\]
we see that (\ref{mzar:eq:main_th_proof_interval_rate3}) is implied by $|b -\tau_j| >C_3 (\alpha_{j} \tau_{j}^{-1})^{-2}\underline{\lambda}_{T}^{2}$. To sum up, 
$|b^{*} -\tau_j| > C_3 (\alpha_{j} \tau_{j}^{-1})^{-2}\underline{\lambda}_{T}^{2}$ would result in (\ref{mzar:eq:main_th_proof_interval_rate1}), a contradiction. So we have proved that $|b^{*} -\tau_j| \leq C_3 (\alpha_{j} \tau_{j}^{-1})^{-2}\underline{\lambda}_{T}^{2}$.

\vspace{10pt}

{\bf Step 4.} With the arguments above valid on the event $A_{T}\cap B_{T}\cap D_{T}^M$, we can now proceed with the proof of the theorem. At the start of Algorithm~\ref{mzar:alg:not_algorithm_for_mzar}, we have $s=1$ and $e=p$ and, provided that $q \ge 1$, condition \eqref{mzar:eq:main_th_proof_there exists interval} is satisfied. Therefore the algorithm detects a change-point $b^{*}$ in that interval such that $|b^{*}-\tau_{j}| \leq C_3 (\alpha_{j} \tau_{j}^{-1})^{-2}\underline{\lambda}_{T}^{2}$. By construction, we also have that $|b^{*}-\tau_{j}| < 2/3 \delta_{T}$. This in turn implies that for all $l=1,\ldots,q$ such that $\tau_{l}\in [s,e]$ and $l\neq j$ we have either $\Ic_{l}^{L}, \Ic_{l}^{R} \subset [s,b^{*}]$ or $\Ic_{l}^{L}, \Ic_{l}^{R} \subset [b^{*}+1,e]$. Therefore \eqref{mzar:eq:main_th_proof_there exists interval} is satisfied within each segment containing at least one change-point. Note that before all $q$ change points are detected, each change point will not be detected twice. To see this, we suppose that $\tau_j$ has already been detected by $b$, then for all intervals $[s_k,e_k] \subset [\tau_j -C_3 (\alpha_{j} \tau_{j}^{-1})^{-2}\underline{\lambda}_{T}^{2} + 1, \tau_j - C_3 (\alpha_{j} \tau_{j}^{-1})^{-2}\underline{\lambda}_{T}^{2} + 2/3\delta_T + 1] \cup [\tau_j + C_3 (\alpha_{j} \tau_{j}^{-1})^{-2}\underline{\lambda}_{T}^{2} - 2/3\delta_T, \tau_j + C_3 (\alpha_{j} \tau_{j}^{-1})^{-2}\underline{\lambda}_{T}^{2}]$, Lemma 2 in \cite{bcf19}, together with the definition of $A_T$, guarantee  that
\begin{align*}
	\max_{s_k \leq b < e}\cont{s_k}{e_k}{b}{\betahb} &\le \max_{s\leq b <  e}\cont{s_k}{e_k}{b}{\betab} + \lambda_{T} \\
	&\le \sqrt{C_3 (\alpha_{j} \tau_{j}^{-1})^{-2}\underline{\lambda}_{T}^{2} } \alpha_{j} \tau_{j}^{-1}+ \sqrt{C_3 (\alpha_{j+1} \tau_{j+1}^{-1})^{-2}\underline{\lambda}_{T}^{2}} \alpha_{j+1} \tau_{j+1}^{-1} +  \lambda_{T}\\
	& < (2\sqrt{C_3}/c_3 +1) \lambda_{T} = C_1 \underline{\lambda}_{T} < \zeta_{T}.
\end{align*}
Once all the change-points have been detected, we then only need to consider $[s_k,e_k]$ such that
\[
[s_k,e_k] \subset [\tau_j - C_3 (\alpha_{j} \tau_{j}^{-1})^{-2}\underline{\lambda}_{T}^{2}+1, \tau_{j+1} + C_3 (\alpha_{j+1} \tau_{j+1}^{-1})^{-2}\underline{\lambda}_{T}^{2}]
\]
for $j = 1,\ldots,q$. For such intervals, we have, by Lemmas 2 and 3 of \cite{bcf19}
\begin{align*}
	\max_{s_k \leq b < e_k}\cont{s_k}{e_k}{b}{\betahb} &\le \max_{s\leq b <  e}\cont{s_k}{e_k}{b}{\betab} + \lambda_{T} \\
	&\le \sqrt{ C_3 (\alpha_{j} \tau_{j}^{-1})^{-2}\underline{\lambda}_{T}^{2}} \alpha_{j} \tau_{j}^{-1} + \sqrt{ C_3 (\alpha_{j+1} \tau_{j+1}^{-1})^{-2}\underline{\lambda}_{T}^{2}} \alpha_{j+1} \tau_{j+1}^{-1} + \lambda_{T} \leq C_1 \underline{\lambda}_{T} <  \zeta_{T}.
\end{align*}

Hence  no further scales is detected and the algorithm terminates. \hfill$\square$

\subsection{Proof of Theorem~\ref{mzar:theorem:consistency_result_2}}
\label{mzar:sec:proof_consistency_2}
The proof of Theorem~\ref{mzar:theorem:consistency_result_2} is similar to that of Theorem~\ref{mzar:theorem:consistency_result}. In the following, we shall still divide the proof into four steps as before, but focus on the main differences. 

{\bf Step 1.} Let $\{\rho_h: h \in \mathbb{Z}\}$ the true  auto-correlation function of $\{X_t\}$ and  $\hat{\rho}_h$ be its sample version (without de-meaning). Let $\boldsymbol{\rho} = (\rho_0, \ldots,\rho_{p})'$. First, we note that for $\alpha > 2$, the distribution of the innovations has finite second moment. It then follows from \citet{Anderson1964} that $\hat{\boldsymbol{\rho}}- {\boldsymbol{\rho}} = O_p(T^{-1/2})$.  The least-square estimator for AR($p$) can be written as 
\[
\hat{\betab} = \begin{bmatrix} \sum_{i=p}^{T-1}{X_i^2}  & \ldots &  \sum_{i=p}^{T-1}{X_i X_{i-p+1}} \\
	& \ddots & \\
	\sum_{i=1}^{T-p}{X_i X_{i+p-1}}  & \ldots &  \sum_{i=1}^{T-p}{X_i^2}
\end{bmatrix}_{p \times p}^{-1}
\begin{bmatrix}
	\sum_{i=p}^{T-1}{X_i X_{i+1}}  \\
	\vdots\\
	\sum_{i=1}^{T-p}{X_i X_{i+p}}\\
\end{bmatrix}.
\]
This is asymptotically equivalent to 
\[
\begin{bmatrix} \hat{\rho}_0  & \ldots &  \hat{\rho}_{p-1}  \\
	& \ddots & \\
	\hat{\rho}_{p-1}  & \ldots &  \hat{\rho}_0
\end{bmatrix}^{-1}
\begin{bmatrix}
	\hat{\rho}_p \\
	\vdots\\
	\hat{\rho}_1 \\
\end{bmatrix},
\]
which converges to $\betab$ at $O_p(T^{-1/2})$ in view of the Yule--Walker equations. 
Now for $0 < \alpha \le 2$,  despite infinite second moment in the innovations thus the time series, the auto-correlation function is still well-defined, in the sense of \citet{davis1986}. It follows from \citet{hannan1977} that for any sufficiently small $\epsilon>0$, $T^{1/\alpha-\epsilon}\|\hat{\betab} - \betab\| \rightarrow 0$ in probability. See also \citet{yohai1977} and \citet{davis1986}.  In conclusion, we have that $T^{\max(1/2,1/\alpha)-\epsilon}\|\hat{\betab} - \betab\| \rightarrow 0$ in probability.

{\bf Steps 2 and 3.}  The following arguments are simpler, due to the fact that $p$ is fixed. Because we go through all the intervals $[s,e]$ over $\{1,\ldots,p\}$, we could see that under the event that $T^{\max(1/2,1/\alpha)-\epsilon}\|\hat{\betab} - \betab\|_\infty < 1$ (N.B. here the norm does not matter, as $p$ is fixed), for any $j = 1,\ldots, q$, and taking $C_1 = \sqrt{p}$ and $C_2 = 1/2$,
\[
\max_{\tau_j \le b < \tau_j+1} \mathcal{C}_{\tau_j,\tau_j+1}^b({\betab}) \ge \all_T/\sqrt{2} - 2 \|\hat{\betab} - \betab\|_\infty >  C_2 \all_T.
\]
On the other hand, for all the intervals $[s,e]$ that do not include any of the change-points $\{\tau_1, \ldots, \tau_q\}$,
\[
\max_{s \le b < e} \mathcal{C}_{s,e}^b({\betab}) \le \sqrt{p}  \|\hat{\betab} - \betab\|_\infty < C_1 T^{-\max(1/2,1/\alpha)+\epsilon}.
\]

{\bf Step 4.}  We shall now proceed with the proof under the event that $T^{\max(1/2,1/\alpha)-\epsilon}\|\hat{\betab} - \betab\|_\infty < 1$, which happens with probability one as $T\rightarrow \infty$. At the start of Algorithm~\ref{mzar:alg:not_algorithm_for_mzar}, we have $s=1$ and $e=p$. Since we pick the threshold $\zeta_T < C_2 \all_T$, and we consider only the narrowest intervals with the corresponding contrasts (i.e. CUSUM-type statistic) over the threshold, we would end up considering all $[\tau_j,\tau_j+1]$ for $j \in \{1,\ldots, q\}$. Notice that before all the $q$ change-points are detected, we would not consider other longer intervals, because of the nature of Algorithm~\ref{mzar:alg:not_algorithm_for_mzar}. In addition, we will not consider intervals without any change because their corresponding contrast values would be below the threshold, as proved in the previous step. Once all the changes are detected, we then only need to consider the intervals located in between consecutive change-points, which all have corresponding contrast values smaller than $C_1 T^{\epsilon-\max(1/2,1/\alpha)}$, thus the threshold $\zeta_T$. Hence the algorithm would terminate with no further scales detected.
\vspace{1cm}


\newpage
\bibliographystyle{chicago}
\bibliography{references}

\vskip 0.39cm
\noindent
\small
\hspace{5cm} Rafal Baranowski, London School of Economics and Political Science
\vskip 2pt
\noindent
\hspace{5cm} E-mail: r.baranowski@lse.ac.uk
\vskip 2pt

\noindent
\hspace{5cm} Yining Chen, London School of Economics and Political Science
\vskip 2pt
\noindent
\hspace{5cm} E-mail: y.chen101@lse.ac.uk
\vskip 2pt

\noindent
\hspace{5cm} Piotr Fryzlewicz, London School of Economics and Political Science
\vskip 2pt
\noindent
\hspace{5cm} E-mail: p.fryzlewicz@lse.ac.uk
\end{document}